\documentclass[prd,amsmath,amssymb,twocolumn,tightenlines,floatfix,notitlepage,preprintnumbers,superscriptaddress,nofootinbib,showpacs]{revtex4}

\usepackage{graphicx}
\usepackage{xspace}
\usepackage[small]{subfigure}
\usepackage{color}
\usepackage{multirow}

\def\nslash{n\!\!\!\slash}
\def\bnslash{\bar n\!\!\!\slash}

\newcommand{\nn}{\nonumber} 
\newcommand{\bn}{{\bar n}}
\newcommand{\nz}{n_z}
\newcommand{\bnz}{\bar{n}_z}
\newcommand{\mcdot}{\!\cdot\!}

\newcommand{\be}{\begin{equation}}
\newcommand{\ee}{\end{equation}}

\newcommand{\vect}[1]{\mathbf{#1}}
\newcommand{\abs}[1]{\left\lvert #1\right\rvert}
\newcommand{\bra}[1]{\left\langle #1\right\rvert}
\newcommand{\ket}[1]{\left\lvert #1\right\rangle}

\newcommand{\minus}{\!-\!}
\newcommand{\plus}{\!+\!}

\newcommand{\df}{\mathrm{d}}
\newcommand{\Lqcd}{\Lambda_{\text{QCD}}}

\newcommand{\GeV}{\text{ GeV}}

\newcommand{\as}{\alpha_s}
\newcommand{\MSbar}{\overline{\text{MS}}}

\newcommand{\kt}{${\rm k_T}$\xspace}
\newcommand{\cusp}{\mathrm{cusp}}

\newcommand{\Ga}{\Gamma}

\newcommand{\cO}{\mathcal{O}}
\newcommand{\cM}{\mathcal{M}}

\newcommand{\cI}{\mathcal{I}}

\newcommand{\cG}{\mathcal{G}}

\newcommand{\cH}{\mathcal{H}}

\newcommand{\cL}{\mathcal{L}}

\newcommand{\cB}{\mathcal{B}}
\newcommand{\cP}{\mathcal{P}}
\newcommand{\cY}{\mathcal{Y}}
\newcommand{\cK}{\mathcal{K}}
\newcommand{\cE}{\mathcal{E}}

\newcommand{\e}{\epsilon}

\newcommand{\eq}[1]{Eq.~\eqref{#1}}
\newcommand{\eqs}[2]{Eqs.~\eqref{#1} and \eqref{#2}}
\newcommand{\eqss}[3]{Eqs.~\eqref{#1}, \eqref{#2}, and \eqref{#3}}
\renewcommand{\sec}[1]{Sec.~\ref{sec:#1}}
\newcommand{\ssec}[1]{Sec.~\ref{ssec:#1}}
\newcommand{\sssec}[1]{Sec.~\ref{sssec:#1}}
\newcommand{\appx}[1]{App.~\ref{app:#1}}
\newcommand{\fig}[1]{Fig.~\ref{fig:#1}}
\newcommand{\figs}[2]{Figs.~\ref{fig:#1} and \ref{fig:#2}}
\newcommand{\tab}[1]{Table~\ref{tab:#1}}

\newcommand{\m}{{a}}
\newcommand{\B}{{b}}
\newcommand{\CM}{{c}}
\newcommand{\taun}{\tau_1}
\newcommand{\taum}{\tau_{1}^\m}
\newcommand{\tauB}{\tau_{1}^\B}
\newcommand{\tauCM}{\tau_{1}^\CM}
\newcommand{\taumB}{\tau_{1}^{\m,\B}}
\newcommand{\taumBCM}{\tau_{1}^{\m,\B,\CM}}
\newcommand{\tauBCM}{\tau_{1}^{\B,\CM}}
\newcommand{\hattaun}{\hat\tau_1}

\newcommand{\qB}{q_B}
\newcommand{\qJ}{q_J}
\newcommand{\qBJ}{q_{B,J}}
\newcommand{\qBm}{q_B^\m}
\newcommand{\qJm}{q_J^\m}
\newcommand{\qBB}{q_B^\B}
\newcommand{\qJB}{q_J^\B}
\newcommand{\qBCM}{q_B^\CM}
\newcommand{\qJCM}{q_J^\CM}

\newcommand{\qBJB}{q_{B,J}^\B}

\newcommand{\qBmB}{q_{B}^{\m,\B}}
\newcommand{\qperp}{\vect{q}_\perp}

\newcommand{\nB}{n_B}
\newcommand{\nJ}{n_J}
\newcommand{\nBJ}{n_{B,J}}

\newcommand{\nJm}{n_J^\m}
\newcommand{\nBB}{n_B^\B}
\newcommand{\nJB}{n_J^\B}
\newcommand{\nBCM}{n_B^\CM}
\newcommand{\nJCM}{n_J^\CM}

\newcommand{\nBJCM}{n_{B,J}^\CM}
\newcommand{\nJmBCM}{n_{J}^{\m,\B,\CM}}

\newcommand{\bnB}{{\bar n}_B}
\newcommand{\bnJ}{{\bar n}_J}

\newcommand{\wB}{\omega_B}
\newcommand{\wJ}{\omega_J}

\newcommand{\wJm}{\omega_J^\m}

\newcommand{\wJB}{\omega_J^\B}
\newcommand{\wBCM}{\omega_B^\CM}
\newcommand{\wJCM}{\omega_J^\CM}
\newcommand{\wBJCM}{\omega_{B,J}^\CM}

\newcommand{\Shemi}{S_\mathrm{hemi}}
\newcommand{\tShemi}{\widetilde{S}_\mathrm{hemi}}

\newcommand{\cumulant}{{\rm c}}
\newcommand{\sigmac}{\sigma_\cumulant}

\allowdisplaybreaks[1]

\DeclareMathOperator{\Tr}{Tr}
\DeclareMathOperator{\tr}{tr}

\DeclareMathOperator{\Li}{Li}
\DeclareMathOperator{\hyp}{{}_2F_1}

\begin{document}

\preprint{MIT-CTP 4375}
\preprint{LA-UR-13-20960}

\title{ \Large Using 1-Jettiness to Measure 2 Jets in DIS 3 Ways\vspace{0.3cm}}
\author{\large Daekyoung Kang}
\affiliation{\normalsize Center for Theoretical Physics,  Massachusetts Institute of
Technology, Cambridge, MA 02139, USA\smallskip}

\author{\large Christopher Lee}
\affiliation{\normalsize Theoretical Division, MS B283, Los Alamos National Laboratory, 
Los Alamos, NM 87545, USA \medskip }

\author{\large Iain W. Stewart\bigskip}
\affiliation{\normalsize Center for Theoretical Physics,  Massachusetts Institute of 
Technology, Cambridge, MA  02139, USA\smallskip}


\begin{abstract}
\bigskip 
\normalsize
  We predict cross sections in deep inelastic scattering (DIS) for the
  production of two jets---one along the proton beam direction created by
  initial state radiation (ISR) and another created by final state radiation
  after the hard collision.  Our results include fixed order corrections and a
  summation of large logarithms up to next-to-next-to-leading logarithmic (NNLL)
  accuracy in resummed perturbation theory. We make predictions for three
  versions of a DIS event shape 1-jettiness, each of which constrains hadronic
  final states to be well collimated into two jets along the beam and
  final-state jet directions, but which differ in their sensitivity to the
  transverse momentum of the ISR from the proton beam.  We use the tools of soft
  collinear effective theory (SCET) to derive factorization theorems for these
  three versions of 1-jettiness. The sensitivity to the ISR gives rise to
  significantly different structures in the corresponding factorization
  theorems---for example, dependence on either the ordinary or the generalized
  $k_\perp$-dependent beam function.  
  Despite the differences among 1-jettiness definitions, we show that the leading nonperturbative correction that shifts the tail region of their distributions is given by a single universal nonperturbative parameter $\Omega_1$, even accounting for hadron mass effects.
Finally,  we give numerical results for $Q^2$ and
  $x$ values explored at the HERA collider, emphasizing that the target of
  our factorization-based analyses is to open the door for higher-precision jet
  phenomenology in DIS.
\end{abstract}


\maketitle

\vspace*{40cm}
\clearpage

\tableofcontents

\newpage

\section{Introduction}
\label{sec:Intro}

Deep inelastic scattering (DIS) of an energetic lepton from a proton
target at large momentum transfer probes the partonic structure of the
proton and the nature of the strong interaction, and was an important
ingredient in the development of the theory of Quantum Chromodynamics
(QCD)~\cite{Taylor:1991ew,Kendall:1991np,Friedman:1991nq,Bjorken:1968dy,Gross:1973id,Politzer:1973fx}.
Modern DIS experiments at HERA and Jefferson Lab continue to
illuminate the internal partonic structure of hadrons, yielding
information on parton distribution functions of all types, as well as
the value of the strong coupling $\as$ itself (see \emph{e.g.}
\cite{Bethke:2011tr}). The precision of $\as$ extractions from DIS jet
cross sections is currently limited by the availability of theoretical
predictions only at next-to-leading order (NLO) \cite{Bethke:2011tr}.

Predicting the dependence of such cross sections on jet algorithms,
sizes, and vetoes to high accuracy currently presents a formidable
challenge. The dependence on more ``global'' observables
characterizing the jet-like structure of final states can often be
predicted to much higher accuracy.  Indeed, some of the most precise
extractions of $\as$ today come from hadronic event shapes in $e^+e^-$
collisions, for which theoretical predictions in QCD exist to N$^3$LL
accuracy in resummed perturbation theory matched to $\cO(\as^3)$
fixed-order results
\cite{GehrmannDeRidder:2007bj,GehrmannDeRidder:2007hr,Weinzierl:2008iv,Weinzierl:2009ms,Becher:2008cf,Chien:2010kc,Abbate:2010xh},
along with a wealth of data from LEP. Using event shapes to describe
jet-like final states in QCD in a global manner holds the potential to
improve the description of jet production in DIS to the same high
level of precision.

Thrust distributions in DIS were considered in \cite{Antonelli:1999kx}
and calculated to NLL accuracy in resummed perturbation theory, and were compared to
$\cO(\as^2)$ fixed-order results calculated numerically
\cite{Catani:1996vz,Graudenz:1997gv}.  Since then the improvement of
theoretical predictions for DIS event shapes beyond these orders of
accuracy has not received much attention.  The introduction of soft
collinear effective theory (SCET)
\cite{Bauer:2000ew,Bauer:2000yr,Bauer:2001ct,Bauer:2001yt,Bauer:2002nz}
has brought about a revolution in methods to achieve higher-order
resummation in a variety of applications in QCD, leading, for example,
to the N$^3$LL resummation of thrust \cite{Becher:2008cf} and heavy
jet mass \cite{Chien:2010kc} in $e^+e^-$ collisions mentioned above
. SCET has been used to predict a wide variety of event shapes in
$e^+e^-$ collisions \cite{Schwartz:2007ib,Hornig:2009kv,Hornig:2009vb}
and $pp$ collisions
\cite{Stewart:2010pd,Stewart:2010tn,Berger:2010xi,Jouttenus:2013hs},
going beyond the resummed accuracy previously available. A wealth of
data now exists on event shapes in DIS from measurements at HERA by
the ZEUS and H1 collaborations
\cite{Adloff:1997gq,Adloff:1999gn,Aktas:2005tz,Breitweg:1997ug,Chekanov:2002xk,Chekanov:2006hv}. To
take advantage of these data, for instance to achieve high-precision
extractions of $\as$, requires commensurate accuracy in theoretical
predictions. Thanks to advances already made in tools and calculations
for $e^+e^-$ and $pp$ event shapes, the time is ripe to extend the
accuracy of DIS event shape predictions beyond NLL. (DIS in the
endpoint region, $x\to 1$, has been studied with SCET
in~\cite{Manohar:2003vb,Chay:2005rz,Becher:2006mr,Chen:2006vd,Fleming:2012kb}.)

Traditional ways to define jet cross sections involve the use of a jet algorithm
(such as \kt-type recombination algorithms or infrared-safe cone algorithms
\cite{Catani:1991hj,Catani:1993hr,Ellis:1993tq,Dokshitzer:1997in,Salam:2007xv,Cacciari:2008gp}),
and often a jet veto as well. Predicting the dependence on jet algorithms,
sizes, and vetoes to high accuracy is currently a formidable theoretical problem
in QCD. In particular, non-global logarithms (NGLs)
\cite{Dasgupta:2001sh,Dasgupta:2002dc} can arise and complicate resummation
beginning at NLL order for observables that probe soft radiation with different
measures in sharply divided regions of phase space, as occurs with some jet
vetoes, for instance
\cite{Dasgupta:2002bw,Banfi:2002hw,Appleby:2002ke,Delenda:2006nf,Banfi:2010pa,Hornig:2011tg,Tackmann:2012bt}.
Similar clustering logs due to the way algorithms cluster soft gluons can also
spoil resummation beginning at NLL order
\cite{Banfi:2005gj,Delenda:2006nf,Banfi:2010pa,KhelifaKerfa:2011zu,Kelley:2012kj,Kelley:2012zs}.
NGLs and clustering logs limit the precision one can achieve in theoretical
predictions for jet cross sections in QCD. A great deal of progress has been
made to resum NGLs numerically in the large-$N_C$ limit
\cite{Dasgupta:2001sh,Dasgupta:2002dc}, to understand the origin and structure
of NGLs in the framework of effective field theory
\cite{Kelley:2011ng,Hornig:2011iu,Hornig:2011tg,Liu:2012sz}, and to find ways to
minimize their numerical impact (\emph{e.g.}
\cite{Delenda:2012mm,Jouttenus:2013hs}), but a generic approach to obtain NNLL
and higher order predictions does not yet exist. These complications due to
non-global methods of measuring jets provide a strong motivation to use
\emph{global} measurements of hadronic final states that still probe their
jet-like structure and are resummable to arbitrarily high accuracy in QCD
perturbation theory. The first steps needed for higher order resummation in DIS
are the derivations of appropriate factorization theorems.

Precisely such a global measure of jet-like structure of hadronic final states
is the $N$-jettiness introduced in \cite{Stewart:2010tn}. $N$-jettiness $\tau_N$
is global event shape that is a generalization of thrust \cite{Farhi:1977sg} and
can be used in any type of collision to constrain the final state to contain
$N+N_B$ jets, where $N_B$ is the number of initial-state hadronic ``beam''
directions. In $e^+e^-$ collisions, events with small $\tau_N$ contain $N$ jets
in the final state; in $pp$ collisions, they contain $N+2$ jets, with two along
the beam directions from initial state radiation (ISR). In DIS, small $\tau_N$
constrains events to have $N+1$ jets, with one jet along the beam direction from ISR from the
proton.

In this paper we will predict a special case of $N$-jettiness cross sections in
DIS, the 1-jettiness. We define a whole class of DIS 1-jettiness observables by
\be
\label{tau1intro}
\tau_1 = \frac{2}{Q^2}\sum_{i\in X} \min \{ q_B \cdot p_i, q_J\cdot p_i\}\,,
\ee
where $q_B$ is a four-vector along the incident proton beam direction and $q_J$
is another four-vector picking out the direction of the additional final-state
jet we wish to measure. Particles $i$ in the final state $X$ are grouped into
regions, according to which vector $q_{B,J}$ they are closer to as measured by
the dot products in \eq{tau1intro}. Different choices of $q_{B,J}$ give
different definitions of the 1-jettiness.  In this paper we consider three such
choices:
\begin{subequations}
\label{tauABC}
\begin{align}
\taum: \qquad \qBm &= xP \,, & \qJm &= \text{jet axis} \\
\tauB: \qquad \qBB &= xP \,,  & \qJB &= q+xP \\
\tauCM: \qquad \qBCM &= P \,, & \qJCM &= k\,,
\end{align}
\end{subequations}
where $P$ and $k$ are the initial proton and electron momenta, and $Q$ and $x$
are the usual DIS momentum transfer and the Bj\"{o}rken scaling variable.  The
three versions of $\tau_1$ in \eq{tauABC} are named for one of their distinctive
properties: $\taum$ \emph{aligns} the vector $\qJm$ with the physical jet axis
as identified by a jet algorithm or by minimization of the sum in \eq{tau1intro}
over possible directions of $\qJm$, see for example Ref.~\cite{Thaler:2011gf}.
This jet axis is almost but not quite equal to $q+xP$, which is used as the vector
$\qJB$ in $\tauB$.  The measurement of $\tauB$ groups final state particles in
\eq{tau1intro} into exact back-to-back hemispheres in the \emph{Breit} frame.
Finally, $\tauCM$ groups particles into exact back-to-back hemispheres in the
\emph{center-of-momentum} frame.

Note that the three $\tau_1$'s in \eq{tauABC} are physically distinct
observables. Each one of them can be defined in any reference frame, but the
definitions may be simpler in one frame versus another.  The DIS 1-jettiness
$\taum$ coincides with the version of 1-jettiness recently considered in
\cite{Kang:2012zr} at NLL order, and is closest in spirit to the original
N-jettiness event shape in \cite{Stewart:2010tn}. No factorization theorems so
far exist for either $\tauB$ or $\tauCM$.

There are in fact a number of DIS event shapes that have been measured by
experiments at HERA.  Two versions of thrust~\cite{Farhi:1977sg} were measured
by the H1 Collaboration \cite{Adloff:1997gq,Adloff:1999gn,Aktas:2005tz}, and by
the ZEUS collaboration \cite{Breitweg:1997ug,Chekanov:2002xk,Chekanov:2006hv}.
The DIS thrust variables $\tau_{n N}$ are all based on hemispheres in the
Breit frame where the axis $\hat n$ is either frozen to $\hat z$ (along
the virtual $\gamma$ or weak boson), or determined from a minimization. They
have been computed to NLL+${\cal
  O}(\alpha_s^2)$~\cite{Antonelli:1999kx,Dasgupta:2002dc}.  The $\tau_{n N}$
measure particles from only one hemisphere, and the choice of normalization $N$
determines whether they are global or non-global~\cite{Dasgupta:2002dc}
(where the non-global variables were used for the experimental measurements).
Our 1-jettiness event shapes defined in \eqs{tau1intro}{tauABC} are global
variables, avoiding NGLs by including information from all particles in the
final state.  We will demonstrate that our DIS 1-jettiness variable $\tau_1^b$
actually exactly coincides with the DIS thrust $\tau_Q\equiv \tau_{z Q}$,
computed in \cite{Antonelli:1999kx} at NLL.

It would be interesting to re-analyze HERA data to measure global 1-jettiness or
thrust variables. For such measurements, one may be concerned about the
contribution of the proton remnants to \eq{tau1intro}.  However, these
remain close to the $q_B$ axis, so their contributions to the sum giving
$\tau_1$ are exponentially suppressed~\cite{Stewart:2009yx}. (To see this
exponential written out explicitly see Eqs.~(\ref{tau1mY}) and (\ref{ftau1}).)
It is only the larger angle soft radiation and ISR in the beam region and the
collision products in the $q_J$ region that need to be measured.  In fact, we
will show below that one \emph{can} measure $\tau_1^{a,b,c}$ only from the
products in the $q_J$ region, obtaining the $q_B$-region contributions by
momentum conservation (however for $\taum$ this is true only in the two-jet
region $\taum\ll 1$).

We will give predictions for cross sections in the three versions of $\tau_1$ in
\eq{tauABC} accurate for small $\tau_1$. We will also prove factorization
theorems for all three variables $\taumBCM$. The structure of these factorization
theorems will differ because $\taumBCM$ each probe initial- and final-state
radiation in DIS differently. Besides grouping final-state hadrons into
different regions, each version has a different sensitivity to the transverse
momentum of ISR. For $\tauBCM$, the nonzero $k_\perp$ of ISR causes
the final-state jet momentum to deviate from the $q_J$ axis by an amount 
$\simeq k_\perp$ due to momentum conservation. This affects the measurement of $\tauB$
or $\tauCM$ at leading order. For $\taum$, $\qJm$ is always aligned with the
physical jet momentum and so is insensitive to the $k_\perp$ of ISR at leading
order.  This leads to different structures in the factorization theorems for
$\taumBCM$.

Before proceeding let us summarize the merits of the three versions of $\tau_1$.
$\tauBCM$ have the experimental advantage of being entirely measurable from just
the collision products in the so-called ``current'' hemisphere, while for
$\taum$ this is true only for $\taum\ll 1$. From a theoretical perspective,
since in this paper we give predictions for $\taumBCM$ at the same order of
accuracy (resummed to NNLL), currently they are equally preferred. However,
$\tauCM$ involves more nontrivial integrals over the transverse momenta of beam
and jet radiation, leading us to anticipate that $\taumB$ will be easier to
extend to higher accuracy. In addition, the factorization theorem we prove for
$\tauCM$ is valid only when the DIS variable $y\sim 1$, that is, for large
lepton energy loss in the CM frame producing a jet in a direction $q_J$ fairly
close to the initial electron direction. It is thus perhaps fair to say that
$\tauB$ possesses the best combination of advantages of experimental
measurability, theoretical calculability, and kinematic range of applicability.
Nevertheless, we emphasize that comparing $\taumBCM$ with each other can shed
light on the transverse recoil of ISR, and can test the universality of
nonperturbative effects which we will discuss below.

We will prove that the cross sections in all three variables factorize 
as special cases of the form:
\begin{align} \label{factorizationintro}
&\frac{d\sigma}{dx\,dQ^2\,d\tau_1} = \frac{d\sigma_0}{dx\,dQ^2} \sum_{\kappa}
   H_\kappa(Q^2,\mu)\! \int\!\! dt_J dt_B dk_S d^2\vect{p}_\perp 
   \nn \\
&\qquad \times J_q(t_J - (\vect{q}_\perp + \vect{p}_\perp)^2,\mu)\: 
     \cB_{\kappa/p}(t_B,x,\vect{p}_\perp^2,\mu)
  \nn \\
&\qquad \times  S_{\text{hemi}}(k_S,\mu) \: \delta\Bigl( \tau_1 - \frac{t_J}{s_J} 
   - \frac{t_B}{s_B} - \frac{k_S}{Q_R}\Bigr)
  \,,
\end{align}
where $\kappa$ runs over quark and antiquark flavors, $s_J,s_B,Q_R$ are
normalization constants given in \eqs{QR}{sJsB} that depend on the choice of
observable $\tau_1$ in \eq{tauABC}. $\sigma_0$ is the Born cross section,
$H_\kappa$ is a hard function arising from integrating out hard degrees of
freedom from QCD in matching onto SCET, $J_q$ is a quark jet function describing
collinear radiation in the final-state jet, and ${\cal B}_{\kappa/p}$ is a quark
beam function containing both perturbative collinear radiation in a function $\cI_{\kappa j}$ as well as the proton parton distribution function (PDF) $f_{j/p}$:
\begin{align} 
\cB_{\kappa/p}(t,x, \vect{p}_\perp^2,\mu)
  =\! \sum_j\!\!\int_x^1\! \frac{dz}{z}\:
\cI_{\kappa j}\Big(t,\frac{x}{z},\vect{p}_\perp^2,\mu\Big)\, f_{j/p}(z,\mu).  
\end{align}
This beam function depends on the transverse virtuality $t$ of the quark
$\kappa$ as well as the transverse momentum $\vect{p}_\perp$ of ISR.
$S_{\text{hemi}}$ in \eq{factorizationintro} describes soft radiation from both
the proton beam and the final-state jet. Despite the fact that the 1-jettiness
\eq{tau1intro} may not divide the final state into hemispheres, we will
nevertheless show that the soft function for any 1-jettiness in DIS is related
to the hemisphere soft function $\Shemi$.  Finally, $\vect{q}_\perp$ is the
transverse momentum of the momentum transfer $q$ in the DIS collision with
respect to the jet and beam directions.

We briefly discuss differences in the factorization theorem for
$\taumBCM$.  For $\taum$, the jet axis is aligned so that the argument
of the jet function $t_J-(\vect{q}_\perp+\vect{p}_\perp)^2 \to t_J$
with zero transverse momentum, and $\vect{p}_\perp$ then gets averaged
over in \eq{factorizationintro}, removing the dependence on this
variable in the beam function and yielding the ordinary beam function
of Ref.~\cite{Stewart:2009yx}. For $\tauB,\tauCM$, the convolution
over $\vect{p}_\perp$ remains and thus they are sensitive to
transverse momentum of ISR.  Thus for $\tauB,\tauCM$ results depend on
generalized $\vect{p}_\perp$-dependent beam function introduced in
Ref.~\cite{Mantry:2009qz}.  The final difference is that
$\vect{q}_\perp$ is identically zero for $\tauB$, while it is nonzero
for $\tauCM$, causing these observables to differ and inducing
additional complications in the convolution over $\vect{p}_\perp$ for
$\tauCM$. In particular the cross section for $\tau_1^c$ does not
start at $\tau_1^c=0$, but rather at $\tau_1^c=\vect{q}_\perp^2/Q^2$
due to the nonzero $\vect{q}_\perp$ injected into the collision and
the choice here for the jet axis.

The ingredients in the factorization theorem \eq{factorizationintro} depend on
an arbitrary scale $\mu$ that arises due to integrating out degrees of freedom
from QCD, matching onto a theory of collinear and soft modes, and then
integrating out collinear degrees of freedom and matching onto just soft modes.
The resulting hard, jet, beam, and soft functions each depend on logs of $\mu$
over physical variables. Renormalization group (RG) evolution allows us to
evolve each function from a scale $\mu_{H,J,B,S}$ where these logs are minimized
to the common scale $\mu$. This evolution resums logs of $\taumBCM$ to all
orders in $\as$, to a given order of logarithmic accuracy determined by the
order to which we know the anomalous dimensions for the RG evolution. We will
use this technology to resum logs of 1-jettiness in DIS to NNLL accuracy for
$\taumBCM$.

The factorized cross section in \eq{factorizationintro} accurately predicts the
$\tau_1$ distribution in the peak region and for the tail to the right of the
peak, where $\tau_1\ll 1$ and logs of $\tau_1$ are large. To be accurate for
larger $\tau_1$, the prediction of \eq{factorizationintro} must be matched
onto predictions of fixed-order QCD perturbation theory to determine the
``non-singular'' terms.  In this paper we do not perform the matching onto the
$\cO(\as)$ and $\cO(\as^2)$ tail of the $\tau_1$ distributions, leaving that to
future work.  However, by comparing the unmatched predictions of
\eq{factorizationintro} integrated over $\tau_1$ to the QCD total cross section
at $x,Q^2$ we can estimate the small size of these missing corrections at large
$\tau_1$. We emphasize that \eq{factorizationintro} accurately captures the
distribution for smaller $\tau_1$ near the peak region.

The factorization theorem \eq{factorizationintro} also allows us to account for
nonperturbative effects---not only in the parton distributions $f(x,\mu)$, but
also through a shape function that appears in the soft function $S$. In $e^+e^-$
collisions, the leading nonperturbative corrections from this shape function
have been shown to be universal for different event shapes and collision
energies~\cite{Salam:2001bd,Lee:2006fn,Lee:2006nr,Mateu:2012nk} (for earlier
work see~\cite{Dokshitzer:1995zt,Akhoury:1995sp,Korchemsky:1994is}). The same
conclusions hold for the soft shape function in \eq{factorizationintro},
endowing it with real predictive power. We will analyze the dominant effects of
the nonperturbative soft shape function on the DIS 1-jettiness. For the peak
region we include a simple nonperturbative model function to show the impact
these corrections have and how they modify the perturbatively calculated
distribution. For the tail region the leading shape function power correction is
a simple dimension-1 parameter $\Omega_1^{a,b,c}$ that induces a shift to
$\tau_1^{a,b,c}$, and is defined by a matrix element of a soft Wilson line
operator. For our observables we will prove that there is universality for this
correction, namely that $\Omega_1^a =\Omega_1^b = \Omega_1^c$. This follows from
a general analysis we carry out for how the direction of axes affect
nonperturbative matrix elements for two-jet soft Wilson line operators.

The paper is organized as follows: In \sec{kinematics} we review the kinematics
of DIS in several commonly used reference frames, laying out the notation for
our subsequent analyses.  In \sec{observables} we define the three versions of
1-jettiness in DIS that we will use in this paper and consider their physics in
some detail.  In \sec{QCD} we follow the usual formalism for calculating
the DIS cross section in QCD, and introduce an additional measurement of the
1-jettiness into the hadronic tensor that appears therein.  \sec{SCET} is the
technical heart of the paper. Here we present the elements of the SCET formalism
that we need and give a detailed proof of the factorization theorems for the
generic DIS 1-jettiness in \eq{tau1intro} and the three specializations we give in
\eq{tauABC} . In particular we derive in
each factorization theorem how the observable depends on the transverse momentum
of ISR through the beam function, and also show that by rescaling arguments we
can always use the hemisphere soft function for each version of 1-jettiness.

In \sec{NLO} we use the factorization theorems from \sec{SCET} to give
predictions for the singular terms in the $\tau_1$ distributions at fixed order
$\cO(\as)$, and also enumerate the results for the hard, jet, beam, and soft
functions that we will need to perform the RG evolution in the next section.  In
\sec{NNLL} we perform the RG evolution and give our resummed predictions to NNLL
accuracy. We compare our predictions for $\tauB$ to those of
\cite{Antonelli:1999kx} at NLL. We also explain the ``profiles'' for the
individual hard, jet, beam, and soft scales which we use to perform the RG
evolution~\cite{Ligeti:2008ac,Abbate:2010xh,Berger:2010xi}.  These profiles
allow for a smooth transition from the tail region into the peak region where
the soft scale becomes nonperturbative, and into the far tail region where the
resummation of logarithms must be turned off. Then we explain how we
incorporate nonperturbative hadronization corrections into our predictions
through a soft shape function and discuss the $\Omega_1$ parameters. We show that the shifts $\Omega_1^{a,b,c}$ to the tail region of all three versions of the 1-jettiness distributions obey universality.

In \sec{results} we present numerical results for our predictions to NNLL for
the $\taumBCM$ cross sections, including also their $x$ and $Q^2$ dependence. We
consider both integrated (cumulant) and differential cross sections. The
particular results we present are for $x,Q^2$ values studied at HERA
\cite{Aktas:2005tz,Chekanov:2006hv}.  However, the analytic results we give in
\sec{NNLL} can just as easily be used for other experiments at different
kinematics, such as at Jefferson Lab (JLab) \cite{Dudek:2012vr}, or for nuclear
states other than the proton, such as those at the future Electron-Ion Collider
(EIC) \cite{Accardi:2012hwp} and Large Hadron Electron Collider (LHeC)
\cite{AbelleiraFernandez:2012cc}.

In \sec{conclusions} we conclude. In several Appendices we collect various
technical details that are used in the main body of the paper. In particular, in
\appx{rge} we collect the anomalous dimensions we need to get to NNLL accuracy
in the $\tau_1$ cross sections, and in \appx{Laplace} we give the resummed cross
sections in an alternative formalism \cite{Becher:2006mr,Becher:2006nr} to that
used in \sec{NNLL} \cite{Ligeti:2008ac,Abbate:2010xh}. In \sec{NNLL} we use a
formalism that expresses the result of the RG evolution of
\eq{factorizationintro} entirely in momentum space, while in \appx{Laplace} we
use a formalism that expresses the RG evolution through Laplace space objects.
These two approaches give identical analytic results at each order in resummed
perturbation theory, but since both are commonly used in the SCET literature we
provide both results for people who prefer one or the other. Indeed, all of our
numeric results have been cross checked between two codes which each use one of
these two approaches.

The reader mainly interested in the phenomenology of DIS 1-jettiness and our
numerical predictions may read Secs.~\ref{sec:Intro}--\ref{sec:observables} and
then skip to \sec{results}. For those interested in details of the factorization
and resummation, we provide these in Secs.~\ref{sec:QCD}--\ref{sec:NNLL} and the
Appendices.

\section{Kinematics of DIS} \label{sec:kinematics}

In this section we define the kinematic variables in DIS that we will use
throughout the paper.  We also consider three reference
frames---center-of-momentum (CM), target rest frame, and Breit frame---and
describe the picture of the events in each of these frames.

\subsection{Kinematic variables}
\label{ssec:DISvariables}

In DIS, an incoming electron with momentum $k$ and a proton with momentum $P$ undergo hard scattering by exchange of a virtual boson (photon or $Z$) with
a large momentum $q$, and outgoing electron $k'$.  The boson momentum $q$ can be
determined from the initial- and final-state electron momenta,
\be\label{q}  
q = k - k' .
\ee

In inclusive DIS,
the final states from the hard scattering are inclusively denoted as $X$
and their total momentum is denoted as $p_X$.
Using \eq{q} momentum conservation $k+P= k'+p_X$ 
can be written as
\be
q + P= p_X .
\ee
The momentum scale $Q$ of the hard scattering is defined 
by the virtuality of the exchanged gauge boson.
Because the boson has a spacelike (negative) virtuality, 
one defines the positive definite quantity $Q^2$ by
\be
Q^2 \equiv -q^2.
\label{Q2}
\ee
where we will be interested in $Q\gg \Lqcd$.  Next one defines dimensionless
Lorentz invariant variables.  The Bj\"{o}rken scaling variable $x$ is defined by
\be
x = -\frac{q^2}{2P\mcdot q}
  = \frac{Q^2}{2P\mcdot q},
\label{x}
\ee
where $x$ ranges between $0\le x \le 1$.
Another Lorentz-invariant quantity $y$ is defined by 

\be
y = \frac{2P\cdot q}{2P\cdot k}  = \frac{Q^2}{x s},
\label{y}
\ee
where the total invariant mass $s=(P+k)^2=2P\mcdot k$ and $y$ ranges from $0\le
y\le 1$.  The variable $y$ measures the energy loss of the electron in the
target rest frame.  For a given $s$ \eq{y} relates $x$, $y$, and $Q^2$ to one
another, allowing one of the three variables to be eliminated.  The invariant
mass of the final state in terms of the above variables is
\be
p_X^2  =  \frac{1-x}{x}Q^2 = (1-x) y s\,.
\ee
In the classic DIS region one has $p_X^2\sim Q^2$ for generic $x$.  In the
endpoint region $1-x\sim \Lqcd/Q$, the final state is a single narrow jet with
momentum of order $Q$ in the virtual boson direction (and studied with SCET in
Refs.~\cite{Manohar:2003vb,Chay:2005rz,Becher:2006mr,Chen:2006vd,Fleming:2012kb}).
The resonance region where $1-x\sim \Lqcd^2/Q^2$ cannot be treated with
inclusive perturbative methods.

In this work we are interested in the classic region where $1-x\gg \Lqcd/Q$
i.e., $x\sim 1-x < 1$.  In this region one can have more than a single jet.
Below, we will make an additional measurement that picks out \emph{two-}jet-like
final states.

\subsection{Center-of-momentum frame}
\label{ssec:CMframe}

\begin{figure}[t]{
    \includegraphics[width=\columnwidth]{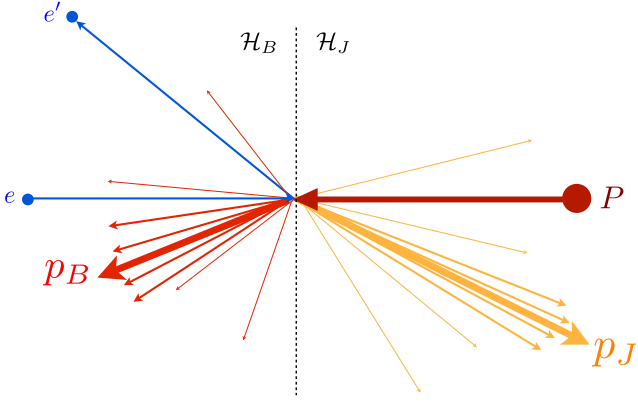}
    \vspace{-1em} 
    { \caption[1]{ Two-jet like event in center-of-momentum frame, in which one jet is produced by initial state radiation from the proton, and the other by the hard collision with the electron. Particles are grouped into two regions $\cH_{J,B}$ with total momenta $p_{J,B}$ in each region. Different choices of ``1-jettiness'' observables will give different boundaries for the two regions.}
  \label{fig:CM}} }
\end{figure}

A two-jet-like event in the CM frame is illustrated in \fig{CM}. 
An incoming electron and proton collide and produce in the final state an outgoing electron and hadrons.
The hadrons, mostly collimated into two jets with additional soft particles elsewhere, are grouped into two regions $\cH_B$ and $\cH_J$,
and $p_J$ and $p_B$ are the  total momenta of particles in the each region. The regions $\cH_{B,J}$ are not necessarily hemispheres in this frame, though we drew them as such in \fig{CM}.
The definitions of the regions are described in \ssec{Njettiness}.
As shown in \fig{CM}, the electron direction is defined to be the $+z$ direction
and the proton direction to be the $-z$ direction.  
In the CM frame
the initial electron and proton momenta are
\be
\label{kPCM}
 k^\mu = \sqrt{s}\,\frac{\nz^\mu}{2} \,, 
 \quad\quad 
 P^\mu = \sqrt{s}\,\frac{\bnz^\mu}{2} \,,
\ee
where the light cone vectors are
\be
 \nz=(1\,,0\,,0\,,1) \,,
 \quad\quad
 \bnz=(1\,,0\,,0\,,-1).
\ee
They satisfy $\nz\mcdot\bnz=2$ and $ \nz\mcdot \nz=\bnz\mcdot\bnz=0$.  An
arbitrary four vector $V^\mu$ can be written as 
\begin{align}
 V^\mu =V^+ \frac{\bnz^\mu}{2} + V^-
\frac{\nz^\mu}{2} +V_T^\mu \,, 
\end{align} 
where $V^+ \equiv V\mcdot \nz$ and $V^-\equiv V\mcdot\bnz$ and $V_T^2=-{\bf
  V}_T^2<0$.
In this frame $x,y$ take the values 
\be
\label{xyCM}
x = \frac{Q^2}{\sqrt{s} \, \bnz\mcdot q}\,,\quad y = \frac{\bnz\mcdot q}{\sqrt{s}}\,.
\ee
and so $q$ is given by
\be
\label{qCM}
q^\mu = y\sqrt{s}\, \frac{\nz^\mu}{2} - x\sqrt{s}\left( 1 - \frac{\vect{q}_T^2}{Q^2}\right) \frac{\bnz^\mu}{2} + q_T^\mu\,,
\ee
which satisfies $Q^2 = -q^2 = xys$. Here $q_T$ is a four-vector transverse
to $\nz,\bnz$ and satisfies $q_T^2 = -\vect{q}_T^2 <0$.
%

\subsection{Target rest frame}
\label{ssec:restframe}

\begin{figure}[t]{
    \includegraphics[width=\columnwidth]{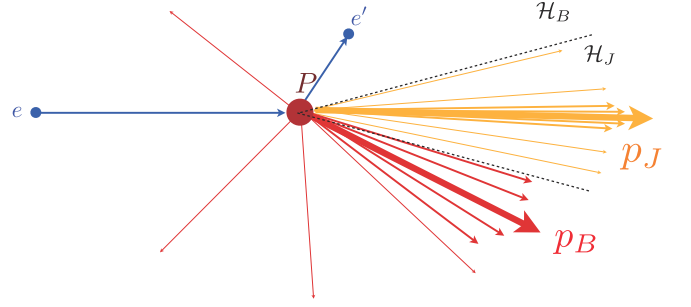}
    \vspace{-3em} 
    { \caption[1]{
Two-jet like event in target rest frame. 
The regions $\cH_{J,B}$ and directions of the total momenta $p_{J,B}$ in these regions are boosted from the CM frame in \fig{CM}. Both jets go forward, but those in $\cH_{J}$ are more highly collimated. 
 }
  \label{fig:rest}} }
\end{figure}

The same two-jet like event in \fig{CM} is illustrated as it would appear in the target rest frame in \fig{rest}. 
The proton is at rest.
The regions in \fig{rest} are transformed from those in \fig{CM} 
because of the boost along the proton direction.
In this frame, the initial electron and proton momenta are
\be
k^\mu = \frac{s}{M} \frac{\nz^\mu}{2} \, , \quad P^\mu = M\frac{\nz^\mu+\bnz^\mu}{2}\,,
\ee
satisfying $2 k\mcdot P = s$. Here $M$ is the proton mass.  We reach this frame
by a boost of momenta $p^\mu$ in the CM frame along the $z$ direction,
\be
\nz\mcdot p \to \frac{M}{\sqrt{s}} \ \nz\mcdot p \,,\quad 
\bnz\mcdot p \to \frac{\sqrt{s}}{M} \ \bnz\mcdot p\,.
\ee
Therefore, in this frame, $q^\mu$ in \eq{qCM} is boosted to become
\be
\label{qtarget}
q^\mu = \frac{Q^2}{xM}\frac{n_z^\mu}{2} - xM\left(1 - \frac{\vect{q}_T^2}{Q^2}\right)\frac{\bn_z^\mu}{2} + q_T^\mu\,,
\ee
and $x,y$ are given by  
\be
x = \frac{Q^2}{2M (E-E')}\,, \quad y =  \frac{E-E'}{E} \,.
\ee
Here $E$ and $E'$ are the energies of the incoming and outgoing electron,
respectively, measured in the target rest frame. Here $y$ is the
fractional electron energy loss.

%

\subsection{Breit Frame}
\label{ssec:Breit}

In the Breit frame, the virtual boson with momentum $q^\mu$ and proton with
momentum $P^\mu$ collide along the $z$ direction.  This frame is useful because
the proton initial state radiation moving along the proton direction can be
relatively well separated from other scattering products.  One might worry that
an ISR jet, which we want to measure in this paper, could be contaminated by the
proton remnants which are difficult to separate from ISR. However, the
$1$-jettiness observable in \ssec{Njettiness} that we use to measure the jets in
the final state is actually insensitive to this contamination since
contributions from the region of the beam remnant give exponentially suppressed
contributions to the variable. The contributions from the beam region are by far
dominated by the initial state radiation at larger angles.  The picture of the
two-jet like event in the Breit frame is similar to \fig{CM} with incoming
electron replaced by virtual boson and with the outgoing electron removed.

The Breit frame is defined as that in which the momentum transfer $q$ is purely
spacelike:
\be
q^\mu = Q\frac{\nz^\mu-\bnz^\mu}{2}\,,
\ee
where we align $\bnz$ to be along the proton direction:
\be
P^\mu = \frac{Q}{x}\frac{\bnz^\mu}{2}\,.
\ee
The incoming electron has momentum
\be
k^\mu = \frac{Q}{y}\frac{\nz^\mu}{2} + Q\frac{1-y}{y} \frac{\bnz^\mu}{2} + k_T^\mu\,,
\ee
where $\vect{k}_T^2 = Q^2(1-y)/y^2$. The outgoing electron then has momentum
\be
k'^{\mu} = Q\frac{1-y}{y} \frac{\nz^\mu}{2} + \frac{Q}{y}\frac{\bnz^\mu}{2} + k_T^\mu\,.
\ee
Unlike the CM and target rest frames, where for a fixed $s$ the incident momenta
are fixed, in the Breit frame the incident momenta are functions of $x,y$. Thus
each point in the differential cross section in $x,y$ corresponds to a different
Breit frame.

\section{Hadronic Observables}
\label{sec:observables}

\subsection{$N$-jettiness}
\label{ssec:Njettiness}

To restrict final states to be two-jet-like, we must make a measurement on the
hadronic state and require energetic radiation to be collimated along two
light-like directions. An observable naturally suited to this role is the
$N$-jettiness \cite{Stewart:2010tn}. In our case, with one proton beam, 
$1$-jettiness $\taun$ can be used to restrict final states to those that have \emph{two}
jets: one along the original proton direction (beam) from ISR and another
produced from the hard scattering. Recalling the definition of $\taun$ in \eq{tau1intro}:
\be \label{tau1def}
\taun = \frac{2}{Q^2}\sum_{i\in X}\min\{ \qB\mcdot p_i,\qJ\mcdot p_i\}\,,
\ee
where $\qB,\qJ$ are massless four-vectors chosen to lie along the beam and jet
directions.

The minimum operator in \eq{tau1def} groups particles in $X$ with the
four-vector to which they are closest (in the sense of the dot product). We will
call the region in which particles are grouped with the beam $\cH_B$ and the
region in which particles are grouped with the jet $\cH_J$.  We denote the total
momentum in the beam region as $p_B$ and total momentum in the jet region as
$p_J$:
\be\label{pBpJ}
 p_B = \sum_{i\in \cH_B} p_i \,, 
 \quad\qquad
 p_J = \sum_{i\in \cH_J} p_i \,.
\ee
These regions are illustrated for two examples in the CM and target rest frames
in \fig{CM} and \fig{rest}.

The 1-jettiness $\taun$ can be expressed as  the sum
\be \label{tau1tauBJ}
 \taun=\tau_B + \tau_J \,,
\ee
where $\tau_B$ and $\tau_J$ are defined by
\be\label{tauBJ}
 \tau_B= \frac{2\qB\mcdot p_B}{Q^2} \,,
 \quad\qquad
 \tau_J = \frac{2\qJ\mcdot p_J}{Q^2} \,.
\ee
The variables $\tau_{B,J} $ are projections of $p_{B,J}$ onto the references vector $\qBJ$.
They can be thought as two independent observables, and $\taun$ is one possible
combination of them. Another combination gives a generalized rapidity gap and is
discussed in \appx{observables}.

The reference vectors $\qB$ and $\qJ$ can be expressed as
\be  \label{qBqJ}
 \qB^\mu =\wB\, \frac{\nB^\mu}{2} \,, 
 \quad\qquad 
 \qJ^\mu =\wJ\, \frac{\nJ^\mu}{2} \,,
\ee
for light-like vectors $\nBJ$ given by $\nBJ=(1\,,\vect{n}_{B,J})$, where
$\vect{n}_{B,J}$ are unit 3-vectors satisfying $\vect{n}_{B,J}^2=1$. Below we
will use the vectors $\nBJ$ to define the directions of the collinear fields in
SCET which we use for the degrees of freedom that describe fluctuations
collimated in the beam and jet regions.
Refs.~\cite{Stewart:2010tn,Thaler:2011gf} discussed the possibility of also
minimizing over possible vectors $\qBJ$ to give the smallest possible $\taun$ in
\eq{tau1def}, and Ref.~\cite{Thaler:2011gf} developed a fast algorithm to carry
out this minimization. Here we will take $\qBJ$ to be fixed vectors. We will discuss
several possible choices for $\qBJ$ below, each giving a different definition of
$\taun$.

Measuring $\taun$ to be small means the final state has at most two collimated
jets, one in the $\qB$ direction and one in the $\qJ$ direction (irrespective of
the exact definition of $\qB$ and $\qJ$).  For power counting purposes we will
use $\taun\sim\lambda^2$ which defines a small parameter $\lambda\ll 1$ in which we
will perform the expansion to obtain the leading-order factorization theorem
for DIS 1-jettiness cross sections.

\subsubsection{$\taum$: 1-jettiness aligned with the jet axis}

\begin{figure*}[t]{
 \includegraphics[width=.32\textwidth]{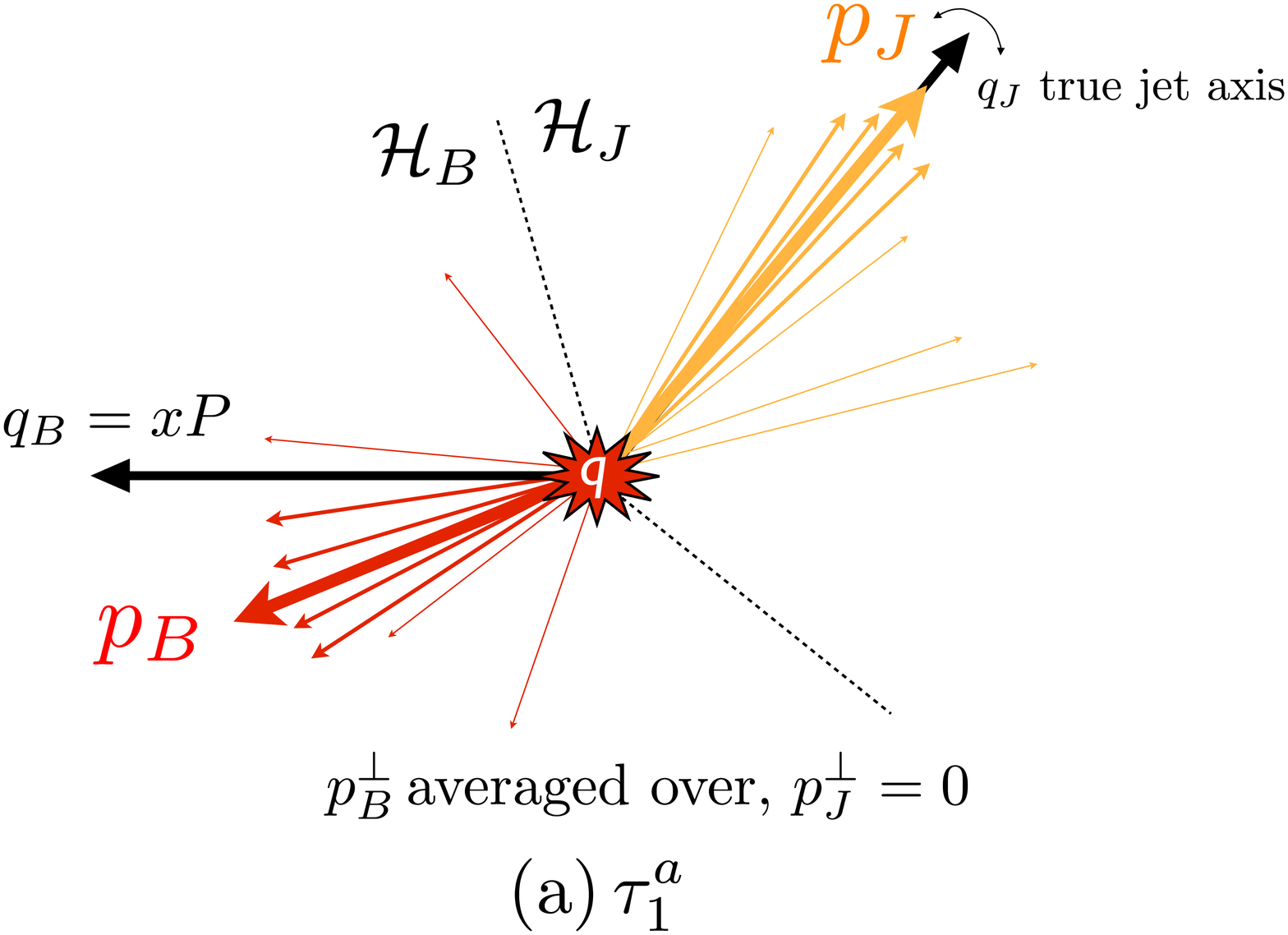}
 \includegraphics[width=.32\textwidth]{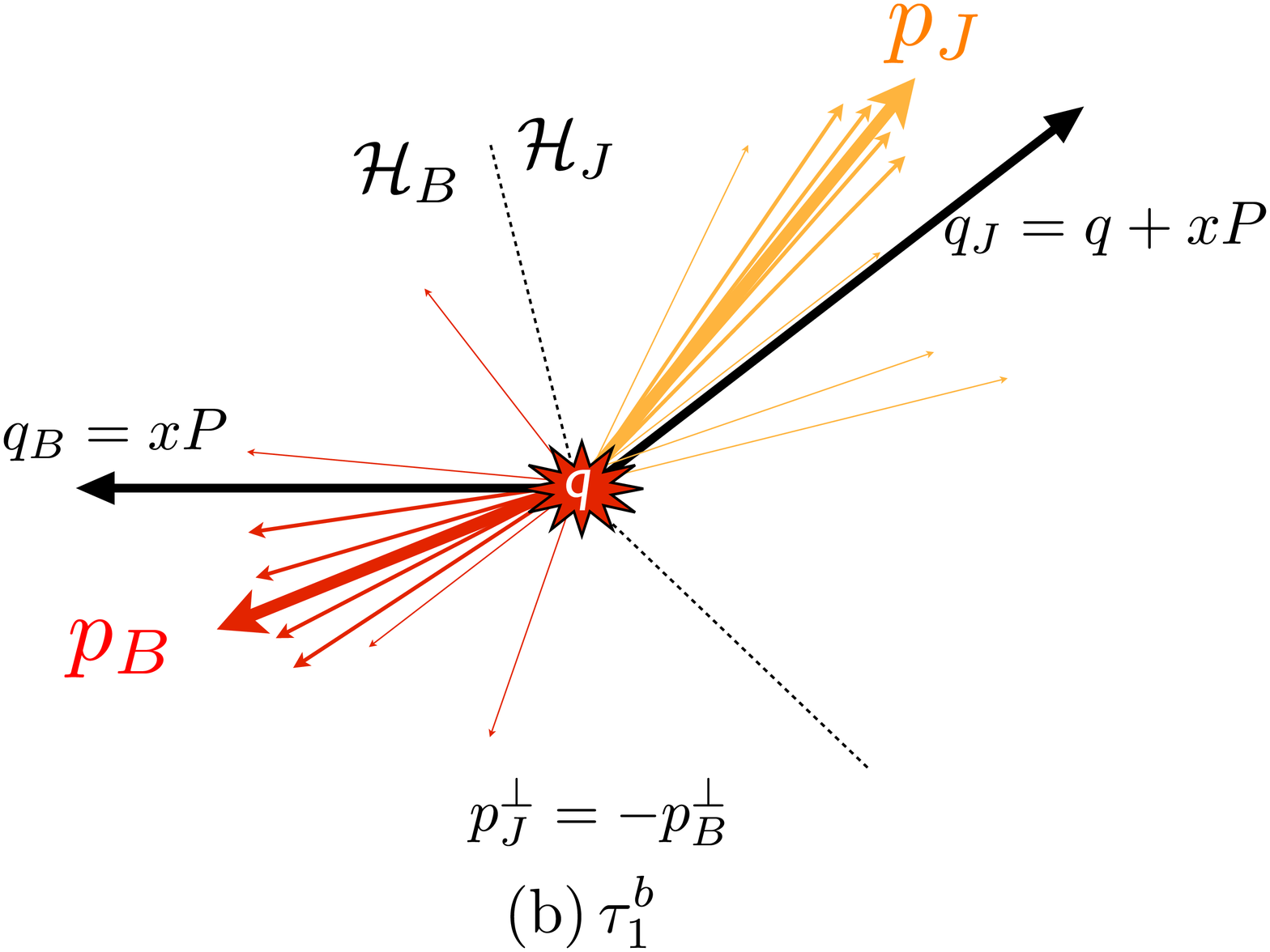}
 \includegraphics[width=.32\textwidth]{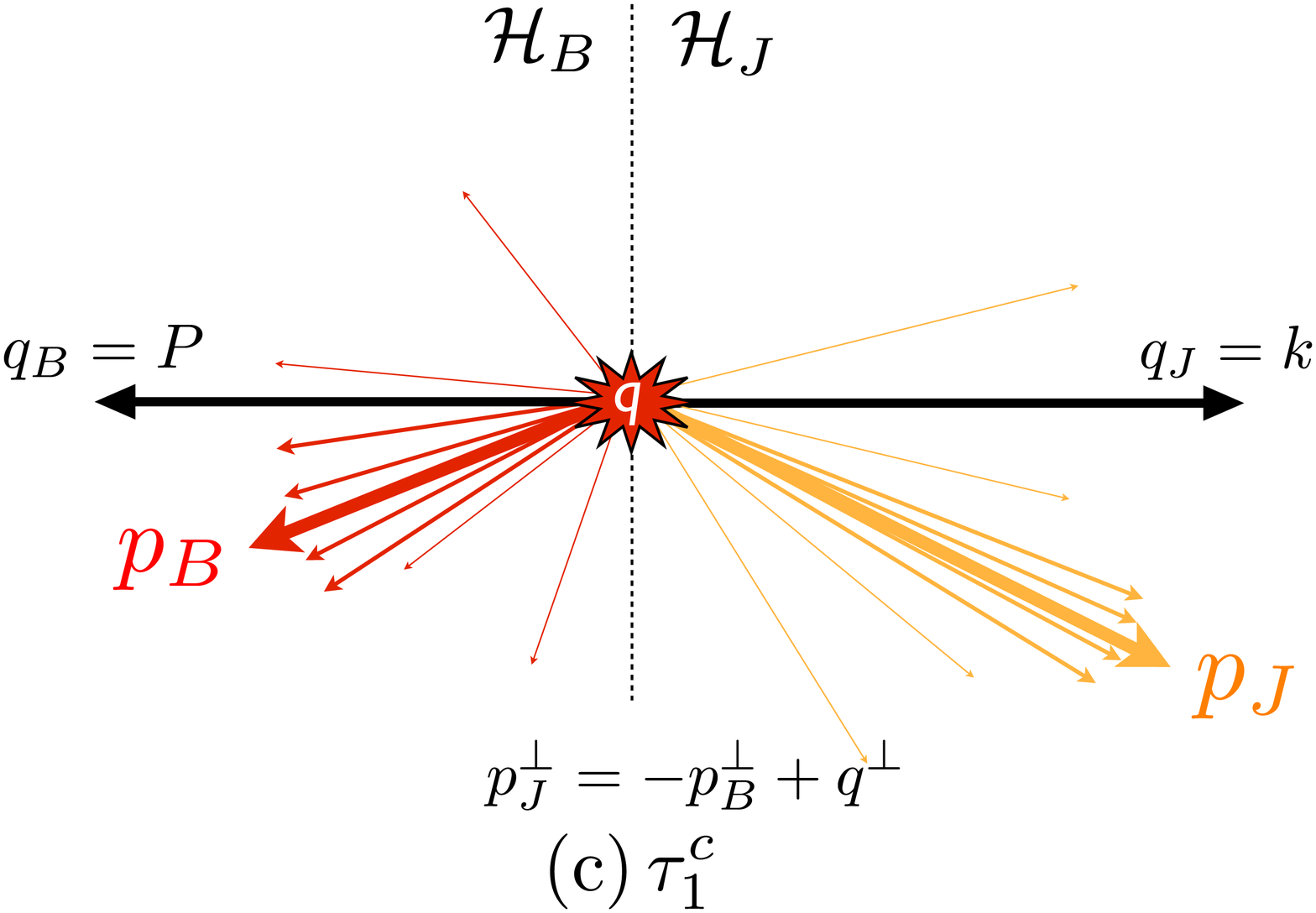} 
\vspace{-1ex}{ 
  \caption[1]{(a) 1-jettiness $\taum$ measures the small light-cone component of
    the momentum in the jet region $\cH_J$ along the ``true'' jet axis $\qJm$,
    which is proportional to the jet invariant mass and is thus insensitive at
    leading order in $\lambda$ to the transverse momentum $p_B^\perp$ of ISR.  Thus
    $p_B^\perp$ gets averaged over in calculating the $\taum$ cross section. (b)
    1-jettiness $\tauB$ measures the small light-cone component of $p_J$ along
    the fixed axis $\qJB=q+xP$. This projection \emph{is} sensitive to and
    balances the transverse momentum $p_B^\perp$ of ISR. The transverse momenta of
    $p_B$ and $p_J$ get convolved together in calculating the cross section.
    Both $\taum$ and $\tauB$ divide the final state into hemispheres in the
    Breit frame.  (c) 1-jettiness $\tauCM$ divides event into back-to-back
    hemispheres in the CM frame and projects beam and jet momenta onto
    $\nz,\bnz$ axes. These projections are sensitive to the transverse momentum
    $p_B^\perp$ of ISR. The momentum transfer $q$ has a nonzero transverse component
    in these coordinates, and the jet and beam momenta are convolved in $p_B^\perp$
    in calculating the cross section.}
\label{fig:tau}} }
\end{figure*}

The first version of 1-jettiness that we consider is $\taum$, which is defined by choosing the beam reference vector $\qBm$ in \eq{tau1def} to be proportional to the proton momentum, and the jet reference vector $\qJm$ to be the jet momentum as given by a jet algorithm such as anti-\kt\ \cite{Cacciari:2008gp}:
\be
\label{truetau}
\taum = \frac{2}{Q^2} \sum_{i\in X} \min\{ \qBm\cdot p_i,\qJm\cdot p_i\}\,.
\ee 
These reference vectors are given by the values
\be\label{qBJm}
 {\qBm}^\mu = xP^\mu \,,
 \quad\qquad
 {\qJm}^\mu = q^\mu +xP^\mu + q_J^{\perp\,\mu}
\,,\ee
where $q_J^\perp$ is $\cO(Q\lambda)$. This is because $xP$ is the longitudinal
momentum of the parton that hard scatters from the virtual photon of momentum
$q$, which would produce a jet of momentum $q+xP$, but the colliding parton may
also have a transverse momentum of order $Q\lambda$. It cannot be larger,
otherwise it would cause $\taun$ to be larger than $\cO(\lambda^2)$. Various jet
algorithms give the same value of $\qJm$ up to negligible power corrections of
$\cO(Q\lambda^2)$, and the cross section does not actually depend on which of
these algorithms is used. Here it would also be equivalent to leading power to
define $\taum$ by minimizing the sum in \eq{truetau} with respect to $\hat n_J$
in $\qJm$.  The total momentum of particles in the jet region $\cH_J$ is $p_J =
\qJm + k$ for a soft momentum $k$ of $\cO(Q\lambda^2)$. Thus, to the order we
are working, the sum over particles in the jet region $\cH_J$ in \eq{truetau}
gives the total invariant mass of those particles, $2\qJm\mcdot p_J = p_J^2 =
m_J^2$ (for more discussion of this see
Refs.~\cite{Jouttenus:2011wh,Jouttenus:2013hs}).

We will show below that in deriving the correct factorization theorem for the
$\taum$ cross section, we must use the fact that $q_J^\perp$ is chosen to make
the relative transverse momentum between $\qJm$ and the actual jet momentum
$p_J$ be zero (technically the dominant $\cO(Q\lambda)$ part must be zero and a
small $\cO(Q\lambda^2)$ part is still allowed). That is, $\qJm$ is \emph{aligned}
with the jet, hence the name $\taum$. This is also important for experimentally
measuring $\taum$.  Nevertheless, once this factorization theorem is known,
$q_J^\perp$ is not directly required for calculating the objects such as hard
and soft functions that appear in the factorization theorem.  For the other
versions of 1-jettiness we consider below, the reference vector $\qJ$ is not
aligned exactly with the jet, and the transverse momentum between $\qJ$ and the
jet momentum $p_J$ will be nonzero, as illustrated in \fig{tau}. This will
change the structure of the corresponding factorization theorems, introducing
convolutions over the transverse momenta of radiation from the beam and from the
final-state jet.

\subsubsection{$\tauB$: hemisphere 1-jettiness in the Breit frame} \label{sssec:tauB}
A second way to define 1-jettiness in DIS is  
\be
\label{tau1Bdef}
\tauB = \frac{2}{Q^2}\sum_{i\in X}\min\{ \qBB\cdot p_i , \qJB\cdot p_i \}\,,
\ee 
where
\be
\label{Breitvectors}
 {\qBB}^\mu = xP^\mu \,, \quad {\qJB}^\mu = q^\mu +xP^\mu \,.
\ee
In this case, $\qJB$ is given exactly by the quantity $q+xP$ which can be
constructed from the electron and proton momenta $k,k',P$, and needs no
information about the jet momentum given by any jet-finding algorithm. Thus in
general $\qJB$ differs by a transverse momentum $q_J^\perp\sim Q\lambda$ from
the vector $\qJm$ used in the $\taum$ definition of 1-jettiness we introduced
above in \eq{truetau}. Note that since $q = \qJB - \qBB$, $q$ itself has zero
tranverse momentum $q_\perp$ with respect to the directions $\nJB,\nBB$ of
$\qJB,\qBB$.

This choice of vectors is natural in the \emph{Breit} frame (hence the name
$\tauB$), in which it divides the final state into back-to-back hemispheres. In
the Breit frame,
\be
\tauB\stackrel{\text{Breit}}{=} 
 \frac{1}{Q}\sum_{i\in X}\min\{ \bnz\mcdot p_i, \nz\mcdot p_i\}\,.
\ee
This definition directly corresponds to the thrust $\tau_Q$ in DIS defined in
\cite{Antonelli:1999kx}\,.

We will often work in the CM frame in intermediate stages of calculation below.
Expressing $\qBJB$ in the CM frame, we find
\begin{align} \label{BreitvectorsCM}
 {\qBB}^\mu &= x\sqrt{s}\, \frac{\bnz^\mu}{2}
 \,, \\
 {\qJB}^\mu &= y\sqrt{s}\,\frac{\nz^\mu}{2} + x(1-y)\sqrt{s}\, \frac{\bnz^\mu}{2} 
  + q_T^\mu
 \,, \nn
\end{align}
where $\vect{q}_T^2 = (1-y)Q^2$ and $\qJB$ is a massless vector.  $\qJB$ in \eq{BreitvectorsCM} can
also be written in the form
\begin{align}
\label{qJpTY}
 {\qJB}^\mu &= P_T e^Y \frac{\nz^\mu}{2} + P_T e^{-Y}\frac{\bnz^\mu}{2} 
  + P_T\, \hat n_T^\mu\,,
\end{align}
where the jet transverse momentum and rapidity are
\be \label{pTxyrelations}
 P_T = Q\sqrt{1-y} \,,
 \quad\qquad
 Y = \frac{1}{2}\ln\frac{y}{x(1-y)}\,,
\ee
and $\hat n_T$ is a unit vector in the direction of $q_T$. These
relations can be inverted to give
\be
\label{xypTrelations}
x = \frac{P_T\, e^{-Y}}{\sqrt{s}-P_T e^Y} \,,
 \quad\qquad 
y = \frac{P_T\, e^Y}{\sqrt{s}}\,.
\ee
Equating the 0th components of  \eqs{qBqJ}{qJpTY}, we find that
\be
\label{omegaJ}
\wJB = 2P_T\cosh Y = [y+x(1-y)]\sqrt{s}\,.
\ee

Calculating $\tauB$ in the CM frame groups particles into non-hemisphere-like
regions. Particles with momenta $p$ are grouped into the beam or jet regions
according to which dot product is smaller:
\begin{align}
 & \cH_B: \frac{x\sqrt{s}\,  \nBB\mcdot p}{2} <  \frac{\wJB \nJB\mcdot p}{2}  \,,
  \nn\\
 & \cH_J:\frac{x\sqrt{s}\,  \nBB\mcdot p}{2} > \frac{\wJB \nJB\mcdot p}{2} \,.
\end{align}
Using \eq{omegaJ}, we can write these conditions as
\be \label{hemitauB}
 \cH_B: \quad \frac{\nBB\mcdot p}{\nJB\mcdot p} <  1 - y + \frac{y}{x} \,,
 \quad \cH_J: \frac{\nBB\mcdot p}{\nJB\mcdot p} >  1 - y + \frac{y}{x} \,.
\ee

In order to understand the regions defined by \eq{hemitauB}, let us consider
simple case $y\sim 1$ and $x<y$. For this case $q_J^\B$ in \eq{qJpTY} is
$\nz$-collinear because in \eq{pTxyrelations} $P_T$ and $Y$ are small and large,
respectively. We can replace $\nJB$ and $\nBB$ in \eq{hemitauB} by $\nz$ and
$\bnz$ and set $\bnz\mcdot p/\nz\mcdot p= 1/(\tan^2\theta/2)$ where $\theta$ is
the polar angle of massless particle $p$. Then, the jet region is a symmetric
cone around the $\nz$ direction of opening angle given by 
\be
\label{coneangle}
\tan^2\frac{R}{2} \approx \frac{x}{y}\,,
\ee
and the beam region is everything outside. For generic $x$ and $y$, the jet
region is not symmetric around the $\nJB$.

As mentioned above in the description of $\taum$, the vector $\qJB=q+xP$ is the
4-momentum of a jet produced by scattering at momentum $q$ on an
incoming parton with momentum exactly equal to $xP$. In general the colliding
incoming parton will have a nonzero transverse momentum due to ISR, causing the
produced jet momentum to deviate by ${\cal O}(Q\lambda)$ from $\qJB$. The scale
${\cal O}(Q\lambda)$ is perturbative and this transverse momentum is much larger
than the intrinsic transverse momentum of partons in the proton. 
The observable $\taum$ differs from $\tauB$ in that $\taum$ measures the true
invariant mass $m_J^2$ of the jet while $\tauB$ simply projects the jet momentum
onto the fixed axis $\qJB = q+xP$ which does not vary with the exact direction
of the jet. The jet axis varies from $\qJB$ due to ISR from the beam before the
hard collision. This subtle difference leads to a different structure in the
factorization theorems for $\taum$ and $\tauB$.

For the 1-jettiness for DIS studied in \cite{Kang:2012zr}, the procedure for
determining the $q_J$ was described as determining the jet axis from a jet
algorithm. This makes their $q_J$ correctly correspond with our $q_J^a$.
However, they also used the formulas \eqs{qJpTY}{pTxyrelations} to describe
their $\qJ$, which yields $\qJ=q+xP$, and this would correspond to our $\tauB$.
This choice neglects the $\cO(Q\lambda)$ transverse momentum between $\qJ$ and
the jet momentum $p_J$, which taken literally would lead to an incorrect
factorization theorem for the observable $\taum$.  However, after the correct
form of the factorization theorem for $\taum$ is known (which was written in
\cite{Kang:2012zr}), this approximation is valid for calculating the objects in
that theorem to leading order in $\lambda$.  Thus, the $\taun$ in
\cite{Kang:2012zr} is the same as our $\taum$ defined above in \eq{truetau},
where $\qJ$ is aligned along $p_J$.

\subsubsection{$\tauCM$: hemisphere 1-jettiness in the CM frame}

\begin{table}[t]
\begin{ruledtabular}
\renewcommand{\arraystretch}{1.5}
\begin{tabular}{c|ccc}
 1-jettiness   & axis $\qJ$ &  axis $\qB$ &   \\ \hline
 generic $\taun$ & $\displaystyle \wJ\frac{\nJ}{2}$  
                  & $\displaystyle \wB\frac{\nB}{2}$  & \\
 $\taum$ & $xP+q+q^\perp_J$ & $xP$  &  \\
 $\tauB$ & $xP+q$  & $xP$    &  \\
 $\tauCM$& $k$   & $P$     &  
\end{tabular}
\end{ruledtabular} 
\caption{Reference vectors $\qJ$ and $\qB$
  defining the axes for various versions of 1-jettiness.  For $\taum$ the
  $\qJ$ axis is defined to be the jet momentum $\qJm$ given by, e.g., the
  anti-\kt algorithm. This axis is given by $q+xP$ up to transverse momentum
  corrections of order $q^\perp_J\sim\cO(Q\lambda)$. The exact value of $q^\perp_J$
  will not be needed for our calculation, only the fact that there is no relative
  transverse momentum larger than $\cO(Q\lambda^2)$ between the momentum $p_J$
  in the jet region $\cH_J$ and the axis $\qJm$.  This is in contrast to
  $\tauB$, for which the cross section will depend on the transverse momentum
  between $p_J$ and $\qJB = q+xP$, but where $q_\perp=0$. Finally for $\tauCM$
  we also have $q_\perp\ne 0$. \label{tab:axes} }
\end{table}

A third way to define the 1-jettiness in DIS is with the proton and electron
momenta
\be
\label{CMvectors}
 {\qBCM}^\mu = P^\mu\,, 
  \quad\qquad 
 {\qJCM}^\mu = k^\mu\,.
\ee
We use the superscripts $\CM$ because this choice naturally divides the final
state into hemispheres in the CM frame, mimicking the thrust defined in the CM
frame for $e^+e^-$ collisions~\cite{Farhi:1977sg}.

In the CM frame the momenta $k$ and $P$ are along the $z$ and $-z$ directions as
in \eq{kPCM}. In this frame the reference vectors $q_{J,B}$ are given by the
light-cone directions $\nBJCM$ and normalizations $\wBJCM$:
\be \label{nBnJCM}
 {\nBCM} ^\mu = \bnz^\mu\, , 
 \quad\qquad 
 {\nJCM}^\mu = \nz^\mu \, ,
\ee
and 
\be
\label{omegaBJCM}
\wBCM = \sqrt{s} \, , \quad\qquad \wJCM  = \sqrt{s} \, .
\ee
In this frame, $\taun$ is then given by
\be
\tauCM = \frac{1}{xy\sqrt{s}}\sum_{i\in X}\min\{ \bnz\mcdot p_i,\nz\mcdot p_i\}\,.
\ee
The minimum here assigns particles to either the hemisphere containing the
proton or electron. States with small $\taun$ thus have two nearly back-to-back
jets in this frame.

\bigskip

The essential differences among $\taum, \tauB, \tauCM$ are illustrated in
\fig{tau} drawn in the CM frame and summarized in \tab{axes}. $\tauB$ and
$\tauCM$ project the jet momentum onto a fixed axis, and are sensitive at
leading order to the transverse momentum of initial state radiation from the
incoming proton, while $\taum$ always projects the jet momentum onto the axis
with respect to which it has no transverse momentum, and so measures the
invariant mass of the jet which is insensitive at leading order to the
transverse momentum of ISR.
\tab{axes} summarizes the choices of reference vectors $q_{J,B}$ for the three
versions of 1-jettiness defined in this section.

\subsection{Versions of DIS Thrust} \label{sec:versions}

Several thrust DIS event shapes have been considered in the
literature~\cite{Dasgupta:2003iq}, and some of them have been measured by
experiments.  One version, called $\tau_Q$ in \cite{Antonelli:1999kx} but not
yet measured, is defined in the Breit frame by
\be
\label{tauQdef}
\tau_Q\overset{\text{Breit}}{=} 1 - \frac2Q \sum_{i\in\cH_C} p_{z\, i} \,,
\ee
where $\cH_C$ is the ``current hemisphere'' in the direction set by the virtual
boson $q$. We will show below in section~\ref{sec:BeamRegion} that $\tau_Q$ is
equivalent to our $\tauB$.

Another version of thrust, used in \cite{Adloff:1997gq,Breitweg:1997ug} and
called $\tau_{tE}$ in \cite{Dasgupta:2002dc}, is defined using a thrust axis
whose definition involves a maximization procedure over particles in the current
hemisphere $\cH_C = \cH_J$ in the Breit frame:
\be
\label{tautE}
\tau_{tE} = 1 - \max_{\vect{n}} \frac{\sum_{i\in\cH_C} 
  \abs{\vect{p}_i\cdot\vect{n}}}{\sum_{i\in\cH_C} \abs{\vect{p}_i}}\,.
\ee
The maximization aligns the vector $\vect{n}$ with the direction of the jet in
the current hemisphere, just like the $\qJm$ vector in our definition of
$\taum$. However, because the sums in both the numerator and denominator are limited to
$\cH_C$, the observable is actually non-global \cite{Dasgupta:2002dc}, cutting
out radiation from the remnant hemisphere.\footnote{The variable $\tau_{tE}$ is
  also not IR safe without a minimal energy constraint on the ${\cal H}_c$
  hemisphere.} Thus it differs from our $\taum$ which sums over both
hemispheres. It cannot be simply related to a global version of 1-jettiness as
above.  A global thrust event shape, $\tau_{tQ}$, can be obtained by replacing
the denominator in \eq{tautE} by $Q/2$, but this version of the thrust event
shape is also not related to our $\tau_1^a$.

Yet another variation is $\tau_{zE}$ \cite{Adloff:1997gq,Dasgupta:2002dc} which
is like \eq{tautE} with the same normalization, but with respect to the
$\vect{z}$-axis in the Breit frame. It is also not global
\cite{Dasgupta:2002dc}. H1 and ZEUS have measured $\tau_{zE}=\tau_c^{\rm
  H1}=1-T_\gamma^{\rm ZEUS}$ and $\tau_{tE}=\tau^{\rm
  H1}=1-T_T^{\rm ZEUS}$ \cite{Aktas:2005tz,Chekanov:2006hv}. It
would be interesting to reanalyze the data to measure the global observables
$\taumBCM$ we predict in this paper at NNLL order.

\subsection{Jet and Beam Momenta}
\label{ssec:momenta}

\subsubsection{Jet and beam contributions to 1-jettiness}

\begin{table*}[tbh]
\begin{ruledtabular}
\renewcommand{\arraystretch}{2.5}
\begin{tabular}{cc|ccccccc}
1-jettiness   & frame  & $Q_J$ & $Q_B$ & $R_J$ & $R_B$ & $Q_R$ & $s_J$ & $s_B$ \\ \hline
generic $\taun$ &   & $\displaystyle \frac{Q^2}{\wJ}$
                    & $\displaystyle \frac{Q^2}{\wB}$ 
                    & $\displaystyle \sqrt{\frac{\wB \nJ\mcdot\nB}{2\wJ}}$  
                    & $\displaystyle \sqrt{\frac{\wJ \nJ\mcdot\nB}{2\wB}}$ 
                    & $\displaystyle \frac{Q^2}{\sqrt{2\qJ\mcdot \qB}}$ 
                    & $\displaystyle \frac{q_B\mcdot q}{q_B\mcdot q_J} Q^2 $
                    & $\displaystyle -\frac{q_J\mcdot q}{q_B\mcdot q_J} Q^2$ \\ \hline
\multirow{3}{*}{$\taumB$} & CM 
                & $\displaystyle \frac{\sqrt{xy}Q}{y+x(1-y)}$  &$\displaystyle\sqrt{\frac{y}{x}}Q$ 
                & $\displaystyle \frac{\sqrt{xy}}{y+x(1-y)}$  &$\displaystyle\sqrt{\frac{y}{x}}$ 
                & \multirow{3}{*}{$Q$} & \multirow{3}{*}{$Q^2$} & \multirow{3}{*}{$Q^2$} \\
                 & Breit & $Q$ & $Q$ & $1$ & $1$ &  \\
                 & Target-rest & $xM$ & $\displaystyle \frac{Q^2}{xM}$    
                               & $\displaystyle \frac{xM}{Q}$ & $\displaystyle \frac{Q}{xM}$ &\\
\hline
\multirow{3}{*}{$\tauCM$} & CM & $\sqrt{xy}Q$  & $\sqrt{xy}Q$ 
                               & $1$  & $1$ & \multirow{3}{*}{$\sqrt{xy}Q$} & \multirow{3}{*}{$yQ^2$} & \multirow{3}{*}{$xyQ^2$} \\
                          & Breit & $\displaystyle \frac{yQ}{2-y}$  & $xQ$ 
                                  & $\displaystyle \frac{\sqrt{y/x}}{2-y}$  & $\displaystyle \sqrt{\frac{x}{y}}$ & \\
                  & Target-rest & $xyM$ & $\displaystyle \frac{Q^2}{M}$ 
                                & $\displaystyle \frac{M}{\sqrt{s}}$ & $\displaystyle \frac{\sqrt{s}}{M}$ &\\
\end{tabular}
\end{ruledtabular}
\caption{Kinematic variables characterizing 1-jettiness. Normalizations $Q_J$ and $Q_B$ in the expression \eq{tauQJQB}  and sizes $R_{J,B}$ of the jet and beam regions $\cH_{J,B}$ in \eq{RJRBmeaning} for the different versions of 1-jettiness, in three different reference frames described in \sec{kinematics}, and the Lorentz invariant combinations $Q_R \equiv Q_{J}/R_{J} = Q_B/R_B$ in \eq{QR} and $s_{J,B}$ given in \eq{sJsB}.
  \label{tab:QJQB} }
\end{table*}

The cross sections for the different versions of 1-jettiness in
\ssec{Njettiness} will all be expressed in terms of \emph{beam}, \emph{jet}, and
\emph{soft} functions that depend on the projections of the total momenta in the
regions $\cH_B$ and $\cH_J$ onto the reference vectors $\qBJ$ in the definition
of the 1-jettiness \eq{tau1def}. These vectors point in the direction of
light-cone vectors $\nB = \bnz$ and $\nJ$, which varies for the three different
versions of 1-jettiness $\taumBCM$.  The expression $\taun$ in \eq{tau1tauBJ}
can be written in terms of $\nJ\cdot p_J$ and $\nB\mcdot p_B$ as
\be
\label{tauQJQB}
\taun = \frac{\nJ\mcdot p_J}{Q_J} + \frac{\nB\mcdot p_B}{Q_B}\,,
\ee
where $Q_J$ and $Q_B$ are given by
\be
\label{QJQBdefs}
Q_J=\frac{Q^2}{\wJ} \,,\quad Q_B=\frac{Q^2}{\wB}.
\ee
\tab{QJQB} lists explicit expressions for $Q_{B,J}$ in the CM, Breit, and target rest frames
for the three versions of 1-jettiness $\taum, \tauB, \tauCM$.

For the three different cases $\taumBCM$ of \eq{tauQJQB},  the contributions $\nJ\cdot p_J$ will be with respect to different
vectors $\nJmBCM$, and  $\nJ\cdot p_J,\nB\cdot p_B$ will include momenta of
particles in different regions $\cH_{J,B}$ in the three cases.  For
$\taum$, the differences between energies $\wJ^\B$ and $\wJ^\m$ and between unit
vectors $\nJB$ and $\nJm$ are of order $\lambda$ since the vectors $\qJB$ and
$\qJm$ differ due to the transverse momentum of ISR of order $Q\lambda$.  So
using the same expression $\wJ^\B$ in \eq{omegaJ} for $\wJ^\m$ is correct up to
corrections suppressed by $\lambda$ that can be neglected in computing $\taum$.
Nevertheless, the values of $\nJ\mcdot p_J$ in the equations for $\tauB$ and
$\taum$ do differ at leading power ($\nJ\mcdot p_J\sim Q\lambda^2$) because the
$\cO(\lambda)$ difference in the axes $\nJB$ and $\nJm$ is dotted the into transverse
momentum in $p_J$ which is of $\cO(Q\lambda)$. This difference is reflected in
the different factorization theorems for $\taum$ and $\tauB$.

The discussion on the jet and beam regions $\cH_{J,B}$ in \sssec{tauB} can be
done for a generic $\taun$.  For particles with momenta $p$ grouped into the
beam or jet region, the criteria $\qJ\cdot p < \qB\cdot p$ and $\qB\cdot p <
\qJ\cdot p$ that define the regions $\cH_{J,B}$, respectively, can be written
\begin{subequations}
\label{RJRBmeaning}
\begin{align}
p\in \cH_J: &\quad  \frac{\nJ\mcdot p}{\bn_J\mcdot p} < \frac{\wB \nJ\mcdot \nB}{2\wJ} \equiv R_J^2
\,,\\ 
p\in \cH_B: &\quad \frac{\nB\mcdot p}{\bn_B\mcdot p} < \frac{\wJ \nJ\mcdot \nB}{2\wB} \equiv R_B^2\,.
\end{align}
\end{subequations}
Here $\bn_J$ and $\bn_B$ are the normalized conjugate vectors to $\nJ$ and
$\nB$, respectively.  Their definitions are 
\be
\label{nBbarnJbar}
\bn_{J}^\mu \equiv \frac{2\nB^\mu}{\nJ\mcdot
\nB} \, ,\quad \bn_B^\mu \equiv \frac{2\nJ^\mu}{\nJ\mcdot \nB}\,,
\ee
chosen so that $\nJ\cdot\bnJ= \nB\cdot\bnB=2$.
The parameters $R_{J,B}$
characterize the sizes of the regions $\cH_{J,B}$ into which the 1-jettiness
\eq{tau1def} partitions final-state particles.  The variables on the left-hand
sides are analogous to the ratio of momenta related to rapidity: $n\cdot
p/\bn\cdot p=e^{-2Y}$ for back-to-back directions $n,\bn$.  They can be
interpreted as a generalized rapidity, $e^{-2Y_{\nJ\bn_J}}$ or $e^{-2
  Y_{\nB\bn_B}}$ as defined by \eq{YnBnJ}.  These rapidities are defined in
terms of 4-vectors $\bn_{J,B}$ and $n_{J,B}$, which are not in general
back-to-back.  $R_{J,B}$ in \eq{RJRBmeaning} characterizes the range of these
generalized rapidities that are included in each of the regions $\cH_{J,B}$.

\subsubsection{Invariants for 1-jettiness}

For later purposes we will express \eq{tauQJQB} in terms of separate
$n_J$-collinear, $n_B$-collinear, and soft contributions:
\be
\label{tauQJQBcs}
\tau_1 = \frac{n_J\mcdot (p_J^c + k_J) } {Q_J} + \frac{n_B\mcdot (p_B^c + k_B) } {Q_B}\,,
\ee
where $p_J^c$ is the total momentum of all $n_J$-collinear modes, $p_B^c$ is the
total momentum of all $n_B$-collinear modes, and $k_{J,B}$ are the total momenta
of soft modes in regions $\cH_{J,B}$, respectively. These modes are defined by
the scaling of their light-cone components of momentum:
\begin{align}
& \text{$n_J$-collinear}:   &(n_J\mcdot p,\bn_J\mcdot p,p_\perp) 
  \sim  &\: Q(\lambda^2,1,\lambda) 
 \nn \\
& \text{$n_B$-collinear}:  & (n_B\mcdot p,\bn_B\mcdot p,p_\perp) 
  \sim &\: Q(\lambda^2,1,\lambda) 
  \nn\\
& \text{soft}:   & k \sim &\:  Q\lambda^2\,.
\end{align}

The normalization constants $Q_{J,B}$ in \eq{tauQJQBcs} are not Lorentz
invariant (which for SCET corresponds to a reparameterization
invariance~\cite{Chay:2002vy,Manohar:2002fd}), but by combining them with other
kinematic quantities we can form invariants in terms of which we can express
\eq{tauQJQBcs}.  One set of such combinations uses $R_{J,B}$ in
\eq{RJRBmeaning}. The sizes $R_{J,B}$ of the regions $\cH_{J,B}$ are not
Lorentz-invariant---they depend on the choice of frame.  However, the ratios
$Q_J/R_J$ and $Q_B/R_B$ \emph{are} Lorentz/reparameterization invariant and, in
fact, are equal:
\be \label{QR}
Q_R\equiv\frac{Q_J}{R_J}=\frac{Q_B}{R_B}=\frac{Q^2}{\sqrt{2\qJ\mcdot \qB}}
\,.\ee
Expressions for $R_{B,J}$ and $Q_R$ for each case $\taumBCM$ are given in
\tab{QJQB}.  (Strictly speaking, dot products with $\qJm$ are not
Lorentz-invariant due to dependence on the jet algorithm, but for calculating
$Q_R$ and $s_{J,B}$ we can use the approximation $\qJm=\qJB = q+xP$ to leading
order in $\lambda$, which does give Lorentz-invariant dot products.)

It is useful to re-express the soft contribution in \eq{tauQJQBcs} by rescaling
the vectors $n_{J,B}$ by $n'_{J,B} = n_{J,B}/R_{J,B}$, which gives us
\be
\label{tauS}
\tau_S \equiv \frac{n_J\mcdot k_J}{Q_J} + \frac{n_B\mcdot k_B}{Q_B} 
  = \frac{n_J'\mcdot k_J + n_B'\mcdot k_B}{Q_R}\,.
\ee
This relation will help us simplify the soft function in the factorized $\taun$
cross sections later on.  This is because rewriting the particle grouping in
\eq{RJRBmeaning} in terms of $n'_{J,B}$ absorbs the factor $R_{J,B}$ giving
$\nJ'\cdot p/\bn_J'\cdot p <1$ and $\nB'\cdot p/\bn_B'\cdot p>1$.  Hence
with these variables the hemispheres $\cH_{J,B}$ are symmetric, which makes it
possible to connect our soft function to the usual hemisphere soft
function.

We can also re-express the $n_{J,B}$ collinear contributions to $\taun$ in
\eq{tauQJQBcs} in terms of another set of Lorentz-invariant combinations
involving $Q_{J,B}$. In the $\taun$ factorization theorems we derive below, the
arguments of the collinear jet and beam functions appearing therein will
naturally depend on ``transverse virtualities'' $\bn\cdot p \, n\cdot p$ of the
$n_J$-collinear jet and of the struck parton in the proton, respectively.
Relating the $n_J$-collinear contribution to $\taun$ to the transverse
virtuality $t_J$ of the jet,
\be \label{tauJcfromtJ}
\tau_J^c \equiv \frac{n_J\mcdot p_J^c}{Q_J} 
  = \frac{\bn_J\mcdot p_J \,n_J\mcdot p_J^c}{\bn_J\mcdot p_J \,Q_J} 
  = \frac{t_J}{\bn_J\mcdot q\,Q_J} + \cO(\lambda^4) \,,
\ee
where in the middle step we simply multiplied top and bottom by the large
component $\bn_J\mcdot p_J$ of the total collinear momentum in region $\cH_J$,
and in the last step we used in the denominator $\bnJ\cdot p_J=\bnJ\cdot q+\cO(Q\lambda^2)$. The large component of the jet momentum can only come from
the momentum transferred into the collision by the virtual boson of momentum
$q$---the proton with which it collides only has a large component in the
$n_J\cdot p$ component.  Similarly, the $n_B$-collinear contribution to
$\taun$ is
\be
\label{tauBcfromtB}
\tau_B^c \equiv \frac{n_B\mcdot p_B^c}{Q_B} 
 = \frac{-\bn_B\mcdot p_x \,n_B\mcdot p_x}{\bn_B\mcdot p_x \,Q_B} 
 = \frac{t_B}{-\bn_B\mcdot q \,Q_B} + \cO(\lambda^4) \,,
\ee
where $p_x$ is the momentum of the parton that is struck by the virtual boson of
momentum $q$. In the middle step we used that $n_B\mcdot p_B^c = -n_B\mcdot p_x$
since the struck parton recoils against the ISR and balances the small component
of momentum in the $n_B$ direction. In the last step, we defined the positive
virtuality $t_B\equiv -\bn_B\mcdot p_x \,n_B\mcdot p_x$ of the spacelike struck
parton and in the denominator used that $\bn_B\mcdot p_x = -\bn_B\mcdot q +
\cO(Q\lambda^2)$. This is because the collision of the virtual boson and struck
parton is the $n_J$-collinear jet which has no large momentum in the $n_B\cdot
p$ component. Thus momentum conservation requires that the large components of
$\bn_B\cdot q$ and $\bn_B\cdot p_x$ cancel.

The quantities in the denominators of the relations
\eqs{tauJcfromtJ}{tauBcfromtB} are Lorentz invariant:
\begin{subequations} \label{sJsB}
\begin{align}
s_J&\equiv \bnJ\mcdot q\, Q_J=\frac{q_B\mcdot q }{q_B\mcdot q_J}\, Q^2 
\,,\\
s_B&\equiv -\bnB\mcdot q\, Q_B=\frac{-q_J\mcdot q }{q_B\mcdot q_J}\, Q^2
\,,
\end{align}
\end{subequations}
where the minus sign in $s_B$ makes it positive since $\bnB\cdot q<0$.
For the cases $\taumBCM$, $s_{J}$ and $s_B$ take the special values given in \tab{QJQB}.

Using the definitions of $Q_R$ and $s_{J,B}$ in \eqs{QR}{sJsB} these factors can
be combined to give the transverse virtuality of the exchanged boson $q$:
\be\label{ssQR2}
\frac{s_J s_B}{Q_R^2}=-\bnB\mcdot q\bnJ\mcdot q \frac{\nB\mcdot\nJ}{2}=Q^2(1-\qperp^2/Q^2)\,,
\ee
where we used
\be\label{qnJnB}
q=\bnB\mcdot q \frac{\nB}{2} +\bnJ\mcdot q \frac{\nJ}{2} +q_\perp\,,
\ee
and $q^2=-Q^2$.  The transverse momentum $q_\perp$ is orthogonal to $\nBJ$.
The relation \eq{ssQR2} will be useful in evaluating the fixed-order $\taun$ cross section in \appx{NLO}. 
We will use that $\qperp^2/Q^2\sim\lambda^2$ when 1-jettiness is measured to be small, $\taun\sim \lambda^2$. A larger $\vect{q}_\perp$ cannot be transferred into the final state for this to be true, since particles have to be collimated along $q_{J,B}$ or be soft.

\subsection{Momentum Conservation and the Beam Region}
\label{sec:BeamRegion}

We noted earlier that the contribution of proton remnants to $\tau_1$ is
exponentially suppressed, by a factor $e^{-\abs{\Delta Y}}$ of their rapidity
with respect to $\qB$. Only the energetic ISR and soft radiation at larger
angles in $\cH_B$ contribute to $\tau_1$. Although these contributions are
easier to measure, one may still prefer to measure particles only in the $\cH_J$
jet region in the direction of $\qJ$. In general, such a restriction in the final
state is non-global, and leads to NGLs. However, by momentum conservation, we
can show that each of the global $\taumBCM$ observables we consider can be
rewritten in terms of momenta of particles only in the $\cH_J$ region (for case
$a$ this is true only in the 2-jet region $\taum\ll 1$).

First, consider $\tauB$. In the Breit frame,
\begin{align}
\tauB &\overset{\text{Breit}}{=}    \frac1Q \sum_{i\in X} \text{min}\{\nz\cdot p_i,\bnz\cdot p_i\} \\
&=  \frac1Q \biggl[   \sum_{i \in \cH_J^\B}  ( E_i  - p_{z\,i} )    
  +   \sum_{i \in \cH_B^\B}   (E_i + p_{z\,i})  \biggr] \nn \\
&= \frac1Q \biggl[   \sum_{i \in X}  ( E_i  + p_{z\,i} )    
  - 2  \sum_{i \in \cH_J^\B}   p_{z\,i}  \biggr]
\,,\nn
\end{align}
where $X = \cH_J^b + \cH_B^b$ denotes the entire final state. Note that in the
Breit frame,
\begin{align}
p_X = P +  q =  \Bigl( \frac{Q}{2x} , 0, 0,   Q - \frac{Q}{2x} \Bigr)\,,
\end{align}
where $p_X^\mu \equiv \sum_{i\in X}  p_i^\mu$. 
Thus,  $E_X + p_{zX} = Q$, and we obtain
\be
\label{tauBtauQ}
\tau_1^b \overset{\text{Breit}}{=} 1 - \frac2Q \sum_{i\in\cH_J^b} p_{z\, i}
\equiv \tau_Q\,, 
\ee
where in the last equality we recall that \eq{tauBtauQ} is precisely the
definition in \eq{tauQdef} of the DIS thrust variable called $\tau_Q$ in
\cite{Antonelli:1999kx}, where the hemisphere $\cH_J^b$ in the Breit frame was
called the ``current hemisphere'' $\cH_C$. We will comment further on the
relation between the results of \cite{Antonelli:1999kx} for $\tau_Q$ and our
results for $\tauB$ in \ssec{tauQ} below.  \eq{tauBtauQ} shows that $\tauB$ can always be
computed just in terms of the measurements of momenta of particles in the
current hemisphere $\cH_C = \cH_J^b$.

The same arguments as for $\tauB$ in the Breit frame apply to $\tauCM$ in the CM
frame. In the CM frame,
\begin{align}
\tauCM & \overset{\text{CM}}{=} \frac{1}{xy\sqrt{s}} \sum_{i\in X} \min\{n_z\mcdot p_i , \bnz\mcdot p_i\} \\
&= \frac{1}{xy\sqrt{s}} \biggl[ \sum_{i\in X} (E_i+p_{z\,i}) - 2\sum_{i\in\cH_J^\CM} p_{z\,i} \biggr]\,. \nn
\end{align}
In this frame, we have that
\begin{align}
p_X &= P+q \\
&= \frac{\sqrt{s}}{2}\biggl(y \plus 1 \minus x \Bigl( 1\minus \frac{\vect{q}_T^2}{Q^2}\Bigr),\frac{2\vect{q}_T}{\sqrt{s}}, y \minus 1 \plus x\Bigl( 1 \minus \frac{\vect{q}_T^2}{Q^2}\Bigr)\biggr)   \nn \,,
\end{align}
so
\be
\tauCM \overset{\text{CM}}{=}  \frac{1}{x} \biggl( 1 - \frac{2}{y\sqrt{s}}\sum_{i\in\cH_J^\CM} p_{z\,i}\biggr)\,.
\ee
Thus, $\tauCM$ also can be measured just from momenta of particles in the
$\cH_J$ hemisphere in the CM frame. 

Finally, the above argument can be extended to apply also to the 1-jettiness $\taum$, but
only for the region where $\taum\ll 1$. $\taum$ can be written
\be
\label{taumhemi}
\taum = \frac{2}{Q^2} \biggl[\sum_{i\in\cH_J^a} q_J^a\mcdot p_i + \sum_{i\in\cH_B^a} q_B^a\mcdot p_i\biggr]\,.
\ee
Now, $q_B^a = q_B^b$, while $q_J^a = q_J^b+\cO(Q\lambda)$. Thus the regions
$\cH_{J,B}^a$ differ from those for $\tauB$, $\cH_{J,B}^b$, by a change in the
region boundary of $\cO(\lambda)$. This does not affect the assignment of
collinear particles to the two regions, since none of them change regions under
this small change in boundary. An $\cO(\lambda)$ fraction of the soft particles
switch from one region to the other, but this then produces a correction
suppressed by $\lambda$ to the soft contribution $\tau_S$ in \eq{tauS}. Thus,
\eq{taumhemi} can be expressed
\begin{align}
\label{taumtauB}
\taum &= \! \frac{2}{Q^2} \biggl[\sum_{i\in\cH_J^b}\! 
 (q_J^a \minus q_J^b) \mcdot p_i +\!\!\!  \sum_{i\in\cH_J^b}\!\!  
 q_J^b \mcdot p_i +\!\!\! \sum_{i\in\cH_B^b} \!\!q_B^b\mcdot p_i\biggr] \! 
 \plus  \cO(\lambda^3) \nn \\
&\!  = \tauB  + \frac{2}{Q^2}\sum_{i\in\cH_J^b} (q_J^a - q_J^b)\mcdot p_i 
 + \cO(\lambda^3)\,,
\end{align}
in the regime where $\taun\sim \lambda^2\ll 1$. This is the regime we aim to
predict accurately in this paper.  Thus, in this limit $\taum$ can also be computed
just by measuring particles in the ``current hemisphere'' $\cH_J^b= \cH_C$ in
the Breit frame, as long as both axes $\qJm$ and $\qJB$ are measured.  For
larger $\tau_1^a$, both regions $\cH_{J,B}^\m$ would need to be measured, and we
emphasize that the contribution of proton remnants is still exponentially
suppressed.

In summary, for small $\tau_1$ none of the three versions of 1-jettiness
$\taumBCM$ require direct measurement of particles from initial state radiation
in the beam region.  Furthermore, for larger $\tau_1$ values the variables
$\tau_1^{b,c}$ still do not require such measurements (though $\tau_1^a$ does).
All three $\taun$'s are global observables since measurement of $\taun$ by
summing over the particles only in the $\cH_J$ region is still affected by ISR
from the proton beam through momentum conservation.

\section{Cross section in QCD}
\label{sec:QCD}

In this Section we organize the full QCD cross section into the usual leptonic
and hadronic tensors, but with an additional measurement of 1-jettiness inserted
into the definition of the hadronic tensor. We express it in a form that will be
easily matched or compared to the effective theory cross section we consider in
the following section.

\subsection{Inclusive DIS cross section}
\label{ssec:inclusiveDIS}

We begin with the inclusive DIS cross section in QCD, differential in the
momentum transfer $q$,
\be
\label{CMcs}
\begin{split}
\frac{d\sigma}{d^4 q} &=  \frac{1}{2s} \int d\Phi_L \sum_X \left\langle\abs{\cM(eP\to LX)}^2\right\rangle    \\
 & \quad \times (2\pi)^4\delta^4(P + q - p_X) \delta^4(q - k + k')\,,
\end{split}
\ee 
where $L$ is the final lepton state with momentum $k'$, and $X$ is the final
hadronic state with momentum $p_X$. $d\Phi_L$ is the phase space for the lepton
states, and the $\sum_X$ includes the phase space integrals for hadronic states.
The squared amplitude $\abs{\cM}^2$ is averaged over initial spins, and summed
over final spins. Recall that $q$ (and $x,y$) can be determined entirely by
measurements of the lepton momenta. Later in \ssec{1jettiness} we will insert
additional measurements such as 1-jettiness on the state $X$.

We wish to express the cross section differential in the Lorentz-invariant
variables $Q^2,x$ using \eqs{Q2}{x}.  Although $Q^2,x$ are Lorentz-invariant, at
intermediate stages of integration we can work in a particular frame. In either
the CM or Breit frame, the proton momentum is of the form $P = n_z\mcdot P\,
\bnz/2$.  So we decompose $q$ along the $\nz,\bnz$ directions, $q = n_z\cdot q\,
\bnz/2 + \bnz\cdot q\, \nz/2 + q_T$. Then the delta functions defining
$Q^2,x$ take the form
\be
\delta\biggl(x - \frac{Q^2}{n_z\mcdot P\, \bn_z\mcdot q}\biggr)
\delta\left(Q^2 + n_z\mcdot q\,\bnz\mcdot q - \vect{q}_T^2\right) \,.
\ee
Inserting these into \eq{CMcs} and integrating over $q^+$ and $q^-$, we obtain
\be
\label{xQ2cs}
\begin{split}
\frac{d\sigma}{dx \,dQ^2} 
&= \frac{1}{4xs}\int d^2q_T^{\phantom{2}} \int d\Phi_L  \delta^4(q - k + k') \\
&\quad\times\sum_X (2\pi)^4 \delta^4(P + q - p_X)\langle\abs{\cM}^2\rangle \,,
\end{split}
\ee
where $q$ is now given by the value
\be \label{qcomponents}
q^\mu = \frac{Q^2}{xn_z\mcdot P}\, \frac{\nz^\mu}{2}
   - xn_z\mcdot P \left( 1 - \frac{\vect{q}_T^2}{Q^2}\right) \frac{\bnz^\mu}{2} 
   + q_T^\mu\,.
\ee

For a single electron final state $L=e(k')$ (which is all we have at the leading
order in $\alpha_{em}$ at which we are working), the integral over $\Phi_L$ in
\eq{xQ2cs} takes the form
\be
\int\frac{d^3 k'}{(2\pi)^3 2E_{k'}} = \int\frac{ d^4 k'}{(2\pi)^3}\delta(k'^2)\,,
\ee
so, performing the $k'$ integral, we obtain
\be
\label{xQ2cs2}
\begin{split}
\frac{d\sigma}{dx \,dQ^2} &= \frac{1}{4(2\pi)^3\,xs}\int d^2q_T^{\phantom{x}} 
 \delta((q - k)^2) \\
&\quad\times\sum_X (2\pi)^4 \delta^4(P + q - p_X)\langle\abs{\cM}^2\rangle \,.
\end{split}
\ee
To use the first delta function, we need to pick a particular frame in which to
complete the $q_T$ integration. In the CM frame,
\be
\delta((q-k)^2) = \delta(Q^2 + 2q\cdot k )  
 = \frac{Q^2}{xs}\,\delta\left(\!\vect{q}_T^2
  - \left(1-\frac{Q^2}{xs}\right)Q^2\!\right),
\ee
where we use \eq{qcomponents}, $k = \sqrt{s}\,\nz/2$ and $P=\sqrt{s}\,\bnz/2$.
We use this delta function to perform the $\vect{q}_T^2$ integral in
\eq{xQ2cs2}, and then use that the spin-averaged squared amplitude is independent of
$\varphi_q$, to obtain
\be
\label{xQ2cs3}
\begin{split}
\frac{d\sigma}{dx \,dQ^2} &= \frac{Q^2}{8(2\pi)^2\,x^2s^2} 
 \sum_X (2\pi)^4 \delta^4(P + q - p_X)\langle\abs{\cM}^2\rangle \,.
\end{split}
\ee
Here the integrand is evaluated in the CM frame with $q$ now given by 
\be \label{qCMy}
q^\mu = y\sqrt{s}\,\frac{\nz^\mu}{2} 
      - xy\sqrt{s}\,\frac{\bnz^\mu}{2} 
       + \sqrt{1-y} \, Q\, \hat n_T^\mu\,,
\ee
where $\hat n_T = (0,1,0,0)$ in $(n^0,n_1,n_2,n_3)$ coordinates.

The matrix element $\cM$ is given by
\be
\label{factoredM}
\cM(eP\to e'X) = \!\! \sum_{I=\gamma,Z} \! 
  \langle e'X\lvert J^{\mu}_{I,EW}(0) D^I_{\mu\nu}J^{\nu}_{I,QCD}(0)\rvert eP\rangle,
\ee
where the sum over $I$ is over photon and $Z$ exchange, $J_{I,EW}$ is the
appropriate electron electroweak current, $J_{I,QCD}$ is the quark electroweak
current, and $D^I_{\mu\nu}$ is the $\gamma$ or $Z$ propagator. There is an
implicit sum over quark flavors. The matrix element can be factored,
\be
\cM(eP\to e'X) = \sum_{I=\gamma,Z}
 \langle e'\lvert J^{\mu}_{I,EW} D^I_{\mu\nu}\rvert e\rangle 
 \langle X\lvert J^{\nu}_{I,QCD}\rvert P\rangle
 \,.
\ee
More conveniently, we can express the sum over $i$ as being over the vector and
axial currents in QCD,
\be
 J^{\mu}_{Vf} = \bar q_f \gamma^\mu q_f \, , \quad\qquad
 J^{\mu}_{Af} = \bar q_f \gamma^\mu \gamma_5 q_f\,.
\ee
The sum in \eq{factoredM} can then be expressed as
\be
\label{factoredMVA}
 \cM = \sum_{I=V,A} \sum_f L_{I\mu} \bra{X} J^\mu_{If} \ket{P} \,,
\ee
defining the leptonic vector $L_{I\mu}$, which contains the electron matrix
element, electroweak propagator, and electroweak charges of the quarks implicit
in \eq{factoredM}. The sum over $f$ in \eq{factoredMVA} is over quark flavors.

Now the cross section in \eq{xQ2cs3} can be written
\be
\frac{d\sigma}{dx\,dQ^2} = 
  \sum_{I,I'=V,A} \!\!\! L^{II'}_{\mu\nu}(x,Q^2)W^{II'\mu\nu}(x,Q^2)\,,
\ee
where
\begin{subequations}
\label{LWtensors}
\begin{align}
\begin{split}
\label{Ltensor}
 L_{II'}^{\mu\nu}(x,Q^2) &= 
  \frac{Q^2}{32\pi^2 x^2 s^2} L_I^{\mu\dag}(x,Q^2) L_{I'}^\nu(x,Q^2)
  \,,
\end{split}
\\[5pt]
\begin{split}
\label{Wtensor}
 W_{II'}^{\mu\nu}(x,Q^2) &= 
   \sum_X \bra{P}J_{I}^{\mu \dag} \ket{X}\bra{X}J_{I'}^\nu\ket{P}  \\
 &\quad \times  (2\pi)^4\delta^4(P + q - p_X) 
 \,.
\end{split}
\end{align}
\end{subequations}
Here $L_{II'}^{\mu\nu},W_{II'}^{\mu\nu}$ depend on $x,Q^2$ through the
components of $q$ given in \eq{qcomponents}. The average over initial electron
and proton spins is implicit in \eq{LWtensors}, as is the sum over quark flavors
in \eq{Wtensor}.

\subsubsection{Leptonic tensor}
\label{ssec:leptontensor}

The leptonic tensor in \eq{Ltensor} is given by
\be
\label{leptontensor}
L_{\mu\nu ff' \!}^{II'}(x,\!Q^2)  =
  -\frac{\alpha_{em}^2}{2x^2s^2} \Bigl( L_{gff'}^{II'\!}(Q^2) g^{T}_{\mu\nu} 
 + iL_{\epsilon ff'}^{II'\!}(Q^2) \epsilon^{T}_{\mu\nu}\Bigr),
\ee
where
\be
\label{metrictensor}
 g^{T}_{\mu\nu} = g_{\mu\nu} - 2\frac{k_\mu k'_\nu + k'_\mu k_\nu}{Q^2} 
 \,,\quad 
 \epsilon^{T}_{\mu\nu} = \frac{2}{Q^2}\epsilon_{\alpha\beta\mu\nu}k^\alpha k'^\beta\,,
\ee
where $k'=k-q$, with $q$ given in \eq{qCMy}\,
and
\begin{align} \label{Lrslts}
L_{gff'}^{VV} &= Q_f Q_{f'} - \frac{(Q_f v_{f'} + v_f Q_{f'})v_e }{1+m_Z^2/Q^2}
+ \frac{ v_f v_{f'}(v_e^2 + a_e^2)}{(1+m_Z^2/Q^2)^2}
 \,,\nn \\
L_{\epsilon ff'}^{VV} &= -a_e\frac{Q_f v_{f'} + v_f Q_{f'}}{1+m_Z^2/Q^2} +
\frac{2a_e v_f v_{f'} v_e}{(1+m_Z^2/Q^2)^2} 
 \,,\\
L_{gff'}^{AA} &= \frac{a_f a_{f'} (v_e^2 + a_e^2)}{(1+m_Z^2/Q^2)^2} \,,\quad
L_{\epsilon ff'}^{AA} = \frac{2a_f a_{f'} v_e a_e}{(1+m_Z^2/Q^2)^2}  
 \,,\nn\\
L_{gff'}^{AV} &= L_{g f'f}^{VA} = \frac{a_f}{1+m_Z^2/Q^2}\left[ Q_{f'} v_e -
  \frac{v_{f'}(v_e^2+a_e^2)}{1+m_Z^2/Q^2}\right]
  \,,\nn \\
L_{\epsilon ff'}^{AV} &= L_{\epsilon f'f}^{VA} =
  \frac{a_f a_e}{1+m_Z^2/Q^2}\left( Q_{f'} - \frac{2v_{f'}v_e}{1+m_Z^2/Q^2}\right),\nn
\end{align}
where we have made explicit the flavor indices $f,f'$. $Q_f$ is the electric
charge of the quark $q_f$ in units of $e$; $v_f,a_f$ are the weak vector and
axial charges of $q_f$; and $v_e,a_e$ the weak vector and axial charges of the
electron. The vector and axial charges are given by:
\begin{align} 
\label{VAcharge}
&a_f=\frac{T_f}{\sin 2\theta_W}, &v_f=\frac{T_f-2Q_f\sin^2\theta_W}{\sin 2\theta_W}
\,,\nn\\
&a_e=-\frac{1}{2\sin 2\theta_W}, &v_e=\frac{-1+4\sin^2\theta_W}{2\sin 2\theta_W}
\,,
\end{align}
where $T_f=1/2$ for $f=u,c,t$ and $-1/2$ for $f=d,s,b$.

\subsection{1-jettiness cross section}
\label{ssec:1jettiness}

To form the cross section differential in the 1-jettiness $\taun$, we insert a
delta function measuring $\taun$ into the hadronic tensor \eq{Wtensor}:
\be
\label{tauCS}
\frac{d\sigma}{dx \, dQ^2 \, d\taun} = \!\!  \sum_{I,I'=V,A} \!\!\! L^{II'}_{\mu\nu}(x,Q^2)W^{II'\mu\nu}(x,Q^2,\taun)\,,
\ee
where the $\taun$-dependent hadronic tensor is
\begin{align}
\label{Wtau}
W_{II'}^{\mu\nu}(x,Q^2,\taun) &=  \sum_X \bra{P}J_{I}^{\mu \dag} \ket{X}\bra{X}J_{I'}^\nu\ket{P}  \\
& \times  (2\pi)^4\delta^4(P + q - p_X)\delta(\taun - \taun(X))\,. \nn
\end{align}
Here the 1-jettiness $\taun(X)$ of state $X$ is defined by \eq{tau1def}. The
definition depends on the choices of reference vectors $\qBJ$.

The sum over states $X$ in \eq{Wtau} can be removed by using an operator
$\hattaun$ which gives $\taun(X)$ when acting on the state $X$:
\be
\hattaun\ket{X} = \taun(X)\ket{X}\,.
\ee

This operator can be constructed from a momentum-flow operator as in
\cite{Bauer:2008dt}.  Explicitly,
\be
\label{tau1operator}
\hattaun = \hattaun^J + \hattaun^B\,,
\ee
where
\be
\label{tau1JBoperators}
\hattaun^{J,B} = \frac{2}{Q^2}\int_{Y_{J,B}}^\infty dY_{J,B}'\,\, q_{J,B}\mcdot \widehat P(Y_{J,B}')\,.
\ee
Here $\widehat P(Y_{J,B}')$ is a momentum flow operator that can be defined and
explicitly constructed in terms of the energy-momentum tensor, which can be
obtained for massless partons
using~\cite{Sveshnikov:1995vi,Korchemsky:1997sy,Belitsky:2001ij,Bauer:2008dt}
and for massive hadrons using~\cite{Mateu:2012nk}. It measures the momentum flow
in the generalized rapidity direction $Y_{J,B}'$, which we define as we did
below \eq{nBbarnJbar} by
\begin{align}
 e^{-2Y_{J}} &=\frac{ n_{J}\mcdot p }{ \bn_{J}\mcdot p} 
 \,,
 & e^{-2Y_{B}} &=\frac{ n_{B}\mcdot p }{ \bn_{B}\mcdot p} 
 \,.
\end{align}
The lower limits $Y_{J,B}$ on the integral in \eq{tau1JBoperators} are given
according to \eq{RJRBmeaning} by $Y_{J,B} = \frac{1}{2}\ln (1/R_{J,B})$. These
values depend on the frame of reference and choice of 1-jettiness $\taun$. For
example, for the choice $\tauCM$ of \eq{CMvectors} in the CM frame for $y$ near
1, the beam and jet regions are hemispheres and $Y_{J,B}=0$. For the choice
$\tauB$ of \eq{Breitvectors} in the CM frame, the jet region is given by the
lower limit $Y_J = \frac{1}{2}\ln(y/x)$, and the beam region is given by the
lower limit $Y_B = \frac{1}{2}\ln(x/y)$.

In the massless limit the generalized rapidities $\exp(-2Y_{J,B})\rightarrow
(1-\cos\theta_{J,B})/(1-\cos\theta_{B,J}) {n_J\mcdot n_B}/{2}$ defining
generalized ``pseudorapidities''. They depend only on angles $\theta_{J,B}$ from
the $\vect{n}_{J,B}$ directions and $n_J\cdot n_B$, and so simply characterize
angular directions in space over which we integrate in \eq{tau1JBoperators}.

Using \eq{tau1JBoperators} the hadronic tensor \eq{Wtau} can be written
\be \label{WxQ2tau}
 W_{II' \!}^{\mu\nu}(x,\! Q^2, \! \taun) 
 = \!\!\int\!\! d^4 x\,  e^{iq\cdot x} \!\bra{P} \! J_I^{\mu\dag} (x) 
 \delta(\taun - \hattaun) J_{I'}^\nu(0)\! \ket{P},
\ee
recalling that $q$ is given by \eq{qCMy}. $\hattaun$ can also be expressed in terms of momentum operators in the regions $\cH_{J,B}$, using \eq{tauQJQB}:
\be
\label{tau1hat}
\hattaun = \frac{\nJ\mcdot \hat p_J}{Q_J} + \frac{\nB\mcdot \hat p_B}{Q_B}\,,
\ee
where $\hat p_{J,B}$ measures the total 4-momentum in region $\cH_{J,B}$.

\section{Factorization in SCET}
\label{sec:SCET}

Soft-collinear effective theory (SCET)
\cite{Bauer:2000ew,Bauer:2000yr,Bauer:2001ct,Bauer:2001yt,Bauer:2002nz} is a
systematic expansion of QCD in a small parameter $\lambda$ which characterizes
the scale of collinear and soft radiation from energetic massless partons. Soft
and collinear modes are defined by the scaling of their momenta in light-cone
coordinates with respect to light-like vectors $n,\bn$ (not necessarily
back-to-back) satisfying $n^2=\bn^2=0$ and $n\cdot \bn=2$. We express the
components of a vector $p$ in $n,\bn$ light-cone coordinates as $p=(\bn\cdot p,
n\cdot p, p_\perp)$, where
\be
\label{pnnbar}
p = \bn\mcdot p \frac{n}{2} + n\mcdot p \frac{\bn}{2} + p_\perp\,,
\ee
with $p_\perp$ being orthogonal to $n,\bn$, defined as
\be
\label{gperpdef}
p_\perp^\mu = g_\perp^{\mu\nu}p_\nu\,,\quad g_\perp^{\mu\nu} = g^{\mu\nu} - \frac{n^\mu \bn^\nu + n^\nu \bn^\mu}{2}\,.
\ee
In these light-cone coordinates, $n$-collinear and soft momenta scale as:
\begin{subequations}
\begin{align}
\text{collinear:} &\qquad  p_n \sim Q(1,\lambda^2,\lambda) \\
\text{soft:} &\qquad p_s \sim Q(\lambda^2,\lambda^2,\lambda^2)\,.
\end{align}
\end{subequations}
The parameter $\lambda$ is determined by the virtuality of the modes $p_n^2\sim
Q^2\lambda^2$ that contribute to the observable in question. Collinear momenta
will be expressed as the sum of a large ``label'' piece and a small ``residual''
piece: $p_n = \tilde p_n + k$, where $\tilde p_n = \bn\cdot\tilde p \, n/2 +
\tilde p_\perp$ contains the $\cO(Q)$ longitudinal and $\cO(Q\lambda)$
transverse pieces, and $k$ is the residual $\cO(Q\lambda^2)$ piece.

\subsection{Matching onto SCET}

Now we are ready to match the currents in \eq{WxQ2tau} onto operators in SCET.
The QCD current
\be
\label{JQCD}
J_{If}^\mu(x) = \bar q_f (x) \Gamma_I^\mu q_f(x)\,,
\ee
with $\Gamma_V^\mu = \gamma^\mu$ and $\Gamma_A^\mu = \gamma^\mu\gamma_5$,
matches onto operators in SCET,
\be
\label{matching}
\begin{split}
J_{If}^\mu(x)& = \sum_{n_1 n_2}\int d^3\tilde p_1 d^3\tilde p_2 
  e^{i(\tilde p_1 - \tilde p_2)\cdot x}  \\
&\qquad \times \Big[ C_{Ifq\bar q \alpha\beta}^{\mu}(\tilde p_1,\tilde p_2)
  \cO_{q\bar q}^{\alpha\beta}(\tilde p_1,\tilde p_2;x) \\
&\qquad\  +  C_{If gg\lambda\rho}^{\mu}(\tilde p_1,\tilde p_2) 
  \cO_{gg}^{\lambda\rho}(\tilde p_1,\tilde p_2;x) \Big] \,,
\end{split}
\ee
neglecting power corrections of $\cO(\lambda^2)$. The quark and gluon SCET operators are
\begin{subequations}
\label{SCEToperators}
\begin{align}
\label{Oqq}
\cO_{q\bar q}^{\alpha\beta}(\tilde p_1,\tilde p_2;x) &= 
  \bar\chi_{n_1,\tilde p_1}^{\alpha j}(x) \chi_{n_2,\tilde p_2}^{\beta j}(x) \\
\label{Ogg}
\cO_{gg}^{\lambda\rho} (\tilde p_1,\tilde p_2;x) &= \sqrt{\omega_1\omega_2}\,
  \cB^{\perp\lambda c}_{n_1,\tilde p_1}(x) \cB^{\perp\rho c}_{n_2,-\tilde p_2}(x)\,
\end{align}
\end{subequations}
where we sum over fundamental color indices $j$ and adjoint color indices $c$,
but fix the spin indices $\alpha\beta$ and $\lambda\rho$. We leave implicit that
$\chi\equiv\chi_q$ carries flavor $q$. Below we will also leave the flavor index
$f$ on the current $J_I$ implicit. The collinear fields $\chi_{n_i,\tilde p_i}$
and $\cB^\perp_{n_i,\tilde p_i}$ carry label momenta
\be
\tilde p_i  = \frac{\omega_i n_i}{2} + \tilde p_i^\perp\,,
\ee 
where $i=1,2$. The momentum of each collinear field can be written in
$n_i,\bn_i$ light-cone coordinates as in \eq{pnnbar}, with the residual $x$
dependence of the SCET fields being conjugate to momenta $k$ of order
$Q\lambda^2$. In \eq{matching}, the integrals over $\tilde p_{1,2}$ are
continuous versions of discrete sums over the label momenta, and the measures
are given  by $d^3\tilde p_i \equiv d\omega_i d^2\tilde p_\perp^i$.

The quark jet fields $\chi_{n_1,\tilde p}(x)$ are products of collinear quark
fields with collinear Wilson lines,
\be
\chi_{n,\tilde p} = \big[ \delta(\omega - \bn\mcdot \cP)
  \delta^2(\tilde p_\perp - \cP_\perp)W_{n}^\dag \xi_{n} \big]\,,
\ee
where $\cP^\mu$ is a label momentum operator \cite{Bauer:2001ct} which acts on
collinear fields and conjugate fields as:
\be
\cP^\mu\phi_{n,p} = \tilde p^\mu \phi_{n,p}\,,\quad \cP^\mu\phi^\dag_{n,p} 
  = -\tilde p^\mu\phi^\dag_{n,p}
\ee
and $W_n$ is the Wilson line 
\be
  W_{n}(x) = \sum_{\text{perms}}
   \exp\left[-\frac{g}{\bn\mcdot\cP}\bn\mcdot A_{n}(x)\right]\,,
\ee
where $A_n^\mu(x) = \sum_{\tilde p}A_{n,\tilde p}^\mu(x)$ is a $n$-collinear
gluon field. The gluon jet fields $\cB^\perp_n$ are collinear gauge-invariant
products of gluon fields and Wilson lines,
\be
\cB^\perp_{n,\tilde p} = \frac{1}{g}
[\delta(\omega + \bn\mcdot \cP)\delta^2(\tilde p_\perp + \cP_\perp)
W_{n}^\dag(\cP_\perp +g A_{n}^\perp)W_{n}]\,.
\ee
The matching coefficients $C_{q\bar q},C_{gg}$ in \eq{matching} are calculated
order-by-order in $\as$ by requiring that matrix elements of both sides of
\eq{matching} between collinear states in QCD and in SCET be equal.

Collinear fields are decoupled from soft fields by the field redefinitions
\cite{Bauer:2001yt}
\be
\label{BPS}
\chi_n = Y_n\chi_n^{(0)} \,, \quad 
A_n^a T^a =  \cY_n^{ab} A_n^{(0)\, b}  T^a= Y_n A_n^{(0)b} T^b Y_n^\dag\,,
\ee
where $Y_n$ is a Wilson line of soft gluons in the fundamental representation.
For $n=n_B$ we have
\be
\label{Ydef}
Y_{n_B}(x) = P\exp \left[ig\int_{-\infty}^0 ds \,n_B\cdot A_s(n_B s+x)\right]\,,
\ee
and $\cY_{n_B}$ is defined similarly but in the adjoint representation. Soft
gluons carry momenta scaling as $\lambda^2$ in all components. Additional
factors accompanying outgoing states turn the path in \eq{Ydef} into $x$ to
$\infty$ \cite{Arnesen:2005nk} for outgoing collinear particles, see also
\cite{Chay:2004zn}. So for $n=n_J$ we have
\begin{align} \label{Ydef2}
  Y^\dagger_{n_J}(x) = P\exp \left[ig\int_0^{+\infty} ds \,n_J\cdot A_s(n_J s+x)\right]\,.
\end{align}
After the field redefinition \eq{BPS}, the operators in \eq{SCEToperators} become
\begin{align} \label{SCEToperators0}
\cO_{q\bar q}^{\alpha\beta}(\tilde p_1,\tilde p_2;x) &=
  \bar\chi_{n_1,\tilde p_1}^{(0)\alpha j}(x) T[Y_{n_1}^\dag Y_{n_2}]^{jk}(x)
  \chi_{n_2,\tilde p_2}^{(0)\beta k}(x) \,, \nn \\
\cO_{gg}^{\lambda\rho} (\tilde p_1,\tilde p_2;x) &= 
  \sqrt{\omega_1\omega_2}\, \cB^{(0)\perp\lambda c}_{n_1,\tilde p_1}(x)  \\
 &  \qquad\times T[\cY_{n_1}^\dag \cY_{n_2}]^{cd}(x) 
  \cB^{(0)\perp\rho d}_{n_2,-\tilde p_2}(x)\, . \nn
\end{align}
The directions $n_1$ and $n_2$ will each get set equal to either $n_J$ or $n_B$ later on, replacing $Y_{n_i}$ with $Y_{n_B}$ in \eq{Ydef} for $n_i=n_B$ or with $Y_{n_J}^\dag$ in \eq{Ydef2} for $n_i=n_J$. Henceforth we use only the decoupled collinear fields and drop the ${}^{(0)}$
superscripts. 

The measurement operators in \eq{tau1hat} also split up into collinear and soft
pieces. Since $\hat p$ is linear in the energy-momentum tensor, which itself
splits linearly into decoupled collinear and soft components after the field
redefinition \eq{BPS}~\cite{Bauer:2008dt}, $\hat p$ splits up as
\be
\label{phatsplit}
\hat p= \hat p^{n_1} + \hat p^{n_2}  + \hat p^s\,
\ee
where $\hat p^{n_1,n_2,s}$ is built only out of the $n_1$-collinear,
$n_2$-collinear, or soft energy-momentum tensor of SCET, respectively.

After matching the product of currents $J^\dag_\mu J'_\nu$ in the hadronic
tensor in \eq{WxQ2tau} onto SCET, there will be products of the quark and gluon
operators $(\cO_{q\bar q}+\cO_{gg})(\cO_{q\bar q}'+\cO_{gg}')$. The $\cO_{q\bar
  q}\cO'_{gg}$ and $\cO_{gg}\cO'_{q\bar q}$ cross terms will vanish inside the
proton-proton matrix element by quark-number conservation (only one of the
fields $\bar\chi_{n_1}$ or $\chi_{n_2}$ in $\cO_{q\bar q}$ will
create/annihilate a quark in the collinear proton). Thus only the $\cO_{q\bar
  q}\cO'_{q\bar q}$ and $\cO_{gg}\cO'_{gg}$ operator products can contribute.

In fact, for DIS, only the quark operator product contributes, just as in
Drell-Yan (DY) \cite{Stewart:2009yx}. Following the arguments in
\cite{Stewart:2009yx}, we know that the matching coefficients
$C_{Igg}^\mu(\tilde p_1,\tilde p_2)$ must be a linear combination of $\tilde
p_1^\mu$ and $\tilde p_2^\mu$, and obey the symmetry
\be
C_{Igg\mu}^{\lambda\rho}(\tilde p_1,\tilde p_2) 
  = C_{Igg\mu}^{\rho\lambda}(-\tilde p_2,-\tilde p_1)\,,
\ee
due to the structure of the operator \eq{Ogg}. This requires $C_{Igg}^\mu$ to be
proportional to $(\tilde p_2 - \tilde p_1)^\mu$, which the $x$ integration in
\eq{WxQ2tau} will eventually set equal to $q^\mu$. Vector current conservation
in QCD requires $q_\mu C_{Vgg}^\mu = 0$, which requires that $C_{Vgg}^\mu$ be
identically zero. The axial current matching coefficient $C_{Agg}^\mu$ can be
nonzero, but still proportional to $\tilde p_2^\mu - \tilde p_1^\mu = q^\mu$,
which gives zero contribution when contracted with the lepton tensor
\eq{leptontensor}. Thus for DIS we need only consider the quark operator
contribution as in DY.

\begin{widetext}

\subsection{Factorization of the Hadronic Tensor}

The hadronic tensor \eq{WxQ2tau} can now be written in SCET as
\begin{align}
\label{Wlong}
 W_{\mu\nu}^{II'}(x,Q^2,\taun) 
 &=  \int d^4x \,e^{iq\cdot x} \sum_{\stackrel{\mbox{\scriptsize $n_1n_2$}}{n_1'n_2'}} 
\int d^3\tilde p_1 d^3\tilde p_2 d^3\tilde p_1' d^3 \tilde p_2' \, 
  e^{i(\tilde p_2 - \tilde p_1)\cdot x} \int d\tau_J d\tau_B d\tau_s 
  \delta(\taun - \tau_J - \tau_B - \tau_s) 
 \nn \\
&\times \bra{P_{\nB}}  \bar C_{I q\bar q \mu}^{\beta\alpha}(\tilde p_1,\tilde p_2)
  \bar\chi_{n_2,\tilde p_2}^{\beta k}\overline T [Y_{n_2}^\dag Y_{n_1}]^{kj} 
  \chi_{n_1,\tilde p_1}^{\alpha j}(x)  
  \delta\Bigl(\tau_J - \frac{\nJ\mcdot \hat p^{\nJ}}{Q_J}\Bigr) 
  \delta\Bigl(\tau_B - \frac{\nB\mcdot \hat p^{\nB}}{Q_B} \Bigr) 
 \nn \\
& \times \delta\Bigl(\tau_s - \frac{\nJ\mcdot \hat p^s_J}{Q_J} 
 -   \frac{\nB\mcdot \hat p^s_B}{Q_B}\Bigr) 
  C_{I'q\bar q \nu}^{\alpha'\beta'}(\tilde p_1',\tilde p_2') 
 \bar\chi_{n_1',\tilde p_1'}^{\alpha' j'} T [Y_{n_1'}^\dag Y_{n_2'}]^{j'k'} 
  \chi_{n_2',\tilde p_2'}^{\beta' k'}(0)  \ket{P_{\nB}}
\,.
\end{align}
We have explicitly specified that the proton is an $\nB$-collinear state. The
conjugate quark matching coefficient is given by $\bar C_{Iq\bar
  q\mu}^{\beta\alpha}(\tilde p_1,\tilde p_2) = [\gamma^0 C^{\dag}_{Iq\bar
  q}(\tilde p_1,\tilde p_2)\gamma^0]^{\beta\alpha}$. We have used that the
measurement operator $\hat\tau_1$ can be written in the form \eq{tau1hat}, and
that the momentum operators $\hat p_{J,B}$ split up linearly into purely $\nJ$-
and $\nB$-collinear and soft operators as in \eq{phatsplit}. We dropped the
subscripts $J,B$ on the collinear momentum operators restricting them to the jet
or beam regions $\cH_{J,B}$ determined by the definition of $\taun$, since all
$\nB$ collinear particles are grouped in region $B$ and all $\nJ$-collinear
particles are grouped in region $J$. 
In the soft sector, the restrictions of the operators $\hat\tau^s_{J,B}$ to the $\cH_{J,B}$ regions remain.

Since the $n_1$-collinear, $n_2$-collinear and the soft sectors are all
decoupled from one another, the proton matrix element in \eq{Wlong} can be
factored,
\begin{align}
\label{Wfactored}
W_{\mu\nu}^{II'}(x,Q^2,\taun) &=  \int d^4 x\int d^3\tilde p_1 d^3\tilde p_2 
  e^{i(q + \tilde p_2 - \tilde p_1)\cdot x}  \int d\tau_J d\tau_B d\tau_s 
  \delta(\taun - \tau_J - \tau_B - \tau_s) 
  \bar C_{Iq\bar q \mu}^{\beta\alpha}(\tilde p_1,\tilde p_2) 
  C_{I'q\bar q \nu}^{\alpha'\beta'}(\tilde p_1,\tilde p_2) 
 \nn\\
&\quad \times \bra{0}[Y_{\nB}^\dag Y_{\nJ}]^{kj}(x) 
 \delta\Bigl(\tau_s - \frac{\nJ\mcdot \hat p^s_J}{Q_J} -   
  \frac{\nB\mcdot \hat p^s_B}{Q_B}\Bigr)  
  [Y_{\nJ}^\dag Y_{\nB}]^{j'k'}(0)\ket{0}
 \\
&\quad \times \! \biggl\{ \! \bra{P_{\nB}} \bar\chi_{\nB,\tilde p_2}^{\beta k}(x) 
 \delta\Bigl(\tau_B - \frac{\nB\mcdot \hat p^{\nB}}{Q_B}\Bigr)
 \chi_{\nB}^{\beta' k'}(0) \ket{P_{\nB}} \bra{0}\chi_{\nJ,\tilde p_1}^{\alpha j}(x)  
  \delta\Bigl(\tau_J - \frac{\nJ\mcdot \hat p^{\nJ}}{Q_J}\Bigr) 
  \bar\chi_{\nJ}^{\alpha' j'}(0)\ket{0} 
 \nn\\
&\quad \ \  +  \bra{P_{\nB}}  \chi_{\nB,\tilde p_1}^{\alpha j} (x) 
 \delta\Bigl(\tau_B - \frac{\nB\mcdot \hat p^{\nB}}{Q_B}\Bigr) 
  \bar\chi_{\nB}^{\alpha' j'}(0) \ket{P_{\nB}}  \bra{0}  
  \bar\chi_{\nJ,\tilde p_2}^{\beta k} (x)  
  \delta\Bigl(\tau_J - \frac{\nJ\mcdot \hat p^{\nJ}}{Q_J}\Bigr) 
  \chi_{\nJ}^{\beta' k'}(0)\ket{0}\! \biggr\} 
 \,.\nn
\end{align}
The last two lines account for the two ways to choose a pair of collinear fields
in the proton matrix element.  We have performed the sums over
$n_{1,2},n_{1,2}'$ sums using that the fields within each collinear matrix
element must all be in the same collinear sector. We also require that the
fields in the proton matrix element must be in the same collinear sector as the
proton, and those in the vacuum matrix element with the direction $\nJ$ in the
definition of $\taun$. The integrals over $\tilde p_{1,2}'$ have been absorbed
into the definition of the unlabeled fields $\chi_{n_{1,2}}$.  In the soft
matrix element we have used the fact that $T[Y_{\nJ}^\dagger Y_{\nB}] =
Y_{\nJ}^\dagger Y_{\nB}$ and $\overline T[Y_{\nB}^\dagger
Y_{\nJ}]=Y_{\nB}^\dagger Y_{\nJ}$ since the two Wilson lines are space-like
separated and the time ordering is the same as the path
ordering~\cite{Bauer:2003di,Fleming:2007qr}. For the soft Wilson line matrix
element corresponding to antiquarks in the beam and jet functions, we have used
charge conjugation to relate it to the matrix element shown in \eq{Wfactored}.

It is measuring $\taun$ to be small that enforces that the direction $\nJ$ on
the collinear fields in the vacuum matrix element be equal to the direction of
the vector $\qJ$ in the definition of the 1-jettiness $\taun$. We are free to
choose any vector $\qJ$ to define the observable $\taun$. Requiring that the
final-state jet $J$ be close to the direction of $\qJ$ may, in general, impose
additional kinematic constraints on $x,y,Q^2$ to ensure this. We will find below
that for $\taumB$, $\qJ$ is already chosen to be close to the final-state jet
and so imposes no additional constraints, while for $\tauCM$ requiring the jet
be close to $\qJCM=k$ requires $y$ to be near 1.

Next we wish to perform the $x$ integral in \eq{Wfactored} to enforce label momentum conservation.
Before doing so, we consider the residual momentum dependence conjugate to the
coordinate $x$ in the SCET matrix elements.  The collinear field $\chi_{n,\tilde
  p}(x)$ with a continuous label momentum $\tilde p$ depends only on single
spatial component $\bn\mcdot x$ because the residual momenta (conjugate to the
spatial components $n\mcdot x,x_\perp$) are reabsorbed into $\tilde p$ when the
discrete label is made continuous. Then, the matrix element of $n$-collinear
fields are $\cM_n=\cM_n(\bn\mcdot x)$. For convenience the soft matrix element
with $Y_n(x)$ and $Y_\bn(x)$ will be defined as $\cM_s(x)$.  Their Fourier
transforms take the form
\be \label{Fourier}
 \cM_{n}(\bn\cdot x) = \int\frac{d n\mcdot k}{2\pi}
  e^{i n\cdot k \, \bn\cdot x/2} \widetilde\cM_n(n\cdot k) 
  \,,\quad\qquad
 \cM_s(x)  = \int\! \frac{d^4 k_s}{(2\pi)^4} e^{ik_s\cdot x}\widetilde\cM_s(k_s)
 \,,
\ee
where $k,k_s$ is a residual or soft momentum of order $Q\lambda^2$.  When
combined with the exponentials containing $q$ or label momenta $\tilde p_{1,2}$,
we can expand the exponents using $q + \tilde p + k = (q+\tilde p)[1 +
\cO(\lambda^2)]$, and drop the terms of order $\lambda^2$. Then the remaining
integrals over $n\mcdot k,k_s$ are simply the Fourier transforms of
the position space matrix elements evaluated at $x=0$. So, we can set $x=0$ in
the SCET matrix elements, and perform the $x$ integral in \eq{Wfactored} to
enforce label momentum conservation.

In performing the $x$ integration, we have a choice to write $x$ and momenta in
$\nB,\bnB$ coordinates or $\nJ,\bnJ$ coordinates. In fact, we have freedom to
define the vectors $\bn_{J,B}$ as long as we choose them such that
$\bnB^2=\bnJ^2=0$ and $\nJ\mcdot \bnJ = \nB\mcdot \bnB = 2$. Since the
measurement of $\taun$ involves measurements of both $\nJ\mcdot p$ and
$\nB\mcdot p$ components of particles' momenta, it is convenient to choose
$\bnB$ to be proportional to $\nJ$ and $\bnJ$ to be proportional to $\nB$, as we did in \eq{nBbarnJbar}, a choice we will continue to use in what follows.

For the first pair of collinear matrix elements in \eq{Wfactored}, the $x$
integral and accompanying phase factor for label momentum conservation take the
form
\begin{align}
\label{delta1}
\int d^4 x \, e^{i(q+\tilde p_2 - \tilde p_1)\cdot x} &= 
 \int \frac{d\nB\mcdot x\; d\bnB\mcdot x}{2} d^2 x_\perp 
 \exp\Bigg(i\biggl\{(\bnB\mcdot q + \omega_2)\frac{\nB\mcdot x}{2} 
  + \nB\mcdot q \frac{\bnB\mcdot x}{2} - \omega_1\frac{\nJ\mcdot x}{2}
  + (q_\perp +\tilde p_2^\perp - \tilde p_1^\perp)\mcdot x_\perp \biggr\}\Bigg) 
 \nn \\
&= 2(2\pi)^4\delta(\bnB\mcdot q + \omega_2) 
 \delta\Bigl(\nB\mcdot q - \frac{\nJ\mcdot \nB\,\omega_1}{2}\Bigr) 
  \delta^2(q_\perp + \tilde p_2^\perp - \tilde p_1^\perp) 
 \nn \\
&= \frac{4}{\nJ\mcdot \nB}(2\pi)^4\delta(\bnB\mcdot q + \omega_2) 
 \delta(\bnJ\mcdot q -\omega_1)\delta^2(q_\perp +\tilde p_2^\perp -\tilde p_1^\perp) 
 \,,
\end{align}
where we used \eq{nBbarnJbar} to rewrite $\nJ\mcdot x$ in terms of $\bnB\mcdot x$ in the first line and to rewrite $\nB\mcdot q$ in terms of $\bnJ\mcdot q$ in the last line.  Exchanging $\omega_2$ and $-\omega_1$ in \eq{delta1} gives us the label momentum-conserving delta functions for the second pair of collinear matrix elements in \eq{Wfactored}.
Using these delta functions to perform the $\omega_{1,2}$ and $\tilde p_1^\perp$
integrals in \eq{Wfactored}, we obtain
\be
\label{Wfixedlabels}
\begin{split}
W_{\mu\nu}^{II'}(x,Q^2,\taun) &=   2(2\pi)^4(Q_JQ_B)^2\int d^2\tilde p_\perp \frac{2}{\nJ\mcdot \nB} \int d\tau_J d\tau_B d\tau_s^J\,d\tau_s^B \delta(\taun - \tau_J - \tau_B - \tau_s^J - \tau_s^B)   \\
&\quad \times \bigl[\bar C_{Iq\bar q \mu}^{\beta\alpha}C_{I'q\bar q \nu}^{\alpha'\beta'}\bigr]\Bigl(\bnJ\mcdot q\frac{\nJ}{2} + q_\perp + \tilde p_\perp,-\bnB\mcdot q\frac{\nB}{2} + \tilde p_\perp \Bigr)  \\
&\quad \times \bra{0}[Y_{\nB}^\dag Y_{\nJ}]^{kj}(0) \delta(Q_J\tau_s^J - \nJ\mcdot\hat p_J^s) \delta(Q_B\tau_s^B - \nB\mcdot\hat p_B^s)  [Y_{\nJ}^\dag Y_{\nB}]^{j'k'}(0)\ket{0} \\
&\quad \times \biggl\{ \bra{P_{\nB}}   \bar\chi_{\nB}^{\beta k} (0) \delta(Q_B\tau_B - \nB\mcdot \hat p^{\nB}) \Bigl[\delta(\bnB\mcdot q + \bnB\mcdot\cP) \delta^2(\tilde p_\perp - \cP_\perp)  \chi_{\nB}^{\beta' k'}(0)\Bigr] \ket{P_{\nB}}  \\
&\qquad\qquad \times \bra{0}  \chi_{\nJ}^{\alpha j} (0)  \delta(Q_J\tau_J - \nJ\mcdot \hat p^{\nJ}) \delta(\bnJ\mcdot q + \bnJ\mcdot \cP)\delta^2(q_\perp + \tilde p_\perp + \cP_\perp)\bar\chi_{\nJ}^{\alpha' j'}(0)\ket{0} \\
&\qquad +  \bra{P_{\nB}}   \chi_{\nB}^{\alpha j} (0) \delta(Q_B\tau_B - \nB\mcdot \hat p^{\nB}) \bigl[\delta(\bnB\mcdot q + \bnB\mcdot\cP) \delta^2(\tilde p_\perp - \cP_\perp)  \bar\chi_{\nB}^{\alpha' j'}(0)\bigr] \ket{P_{\nB}}  \\
&\qquad\qquad\times \bra{0} \bar\chi_{\nJ}^{\beta k} (0)  \delta(Q_J\tau_J - \nJ\mcdot \hat p^{\nJ}) \delta(\bnJ\mcdot q + \bnJ\mcdot \cP)\delta^2(q_\perp + \tilde p_\perp + \cP_\perp)\chi_{\nJ}^{\alpha' j'}(0)\ket{0}  \biggr\} \,,
\end{split}
\ee 
where we use the change of variables $\tilde p^\perp_2=\tilde p_\perp$ in 3rd
and 4th lines and $\tilde p^\perp_2=-\tilde p_\perp-q_\perp$ in 5th and 6th
lines.  Recall that $\bnB$ and $\bnJ$ are now fixed by \eq{nBbarnJbar}. The
collinear fields without labels implicitly contain a sum over all labels, with
the delta functions then fixing the labels to a single value (it is important to
recall that label operators $\cP^\mu$ acting on fields $\bar\chi_{n,\tilde p}$
give minus the label momentum, $-\tilde p^\mu$ \cite{Bauer:2001ct}).  The vector
$\nJ$ (may) implicitly depend on the integration variable $\tilde p_\perp$, at
least for the case of the $\taum$ distribution, which we will deal with below.
For $\tauB$ and $\tauCM$ the vector $\nJ$ is independent of $\tilde p_\perp$. We
have also indicated that the arguments of the matching coefficients $\bar C,C$ are
both set equal to the label momenta of the fields in the collinear proton and
vacuum matrix elements.

The result in \eq{Wfixedlabels} is organized in terms of
factorized matrix elements that can now be related to known functions in SCET.

\subsection{SCET Matrix Elements}

\subsubsection{Beam Functions}

The proton matrix elements in \eq{Wfixedlabels} can be expressed in terms of
\emph{generalized beam functions} \cite{Mantry:2010bi,Jain:2011iu} in SCET. In
covariant gauges (for discussion of similar matrix elements in
light-cone gauges see~\cite{Ji:2002aa,Collins:2003fm,Idilbi:2010im})  they are defined by
\begin{align} \label{genbeamdef}
\cB_q\Bigl(\omega k^+ \!\!,\frac{\omega}{P^-},k_\perp^2,\mu \Bigr) 
&= \frac{\theta(\omega)}{\omega}\!\!\int \!\frac{dy^-}{4\pi} e^{i k^+y^- \! /  2}\!  
 \bra{P_n(P^- \!)} \bar\chi_n\Bigl(y^-\frac{n}{2}\Bigr) \frac{\bnslash}{2} 
 \Bigl[\delta(\omega - \bn\cdot\cP)\frac{1}{\pi}
  \delta(k_\perp^2 - \cP_\perp^2)\chi_n(0)\Bigr]\!  \ket{P_n(P^-\!)} ,
 \\
\cB_{\bar q}\Bigl(\omega k^+ \!\!,\frac{\omega}{P^-},k_\perp^2,\mu \Bigr) 
&= \frac{\theta(\omega)}{\omega}\!\!\int\!\frac{dy^-}{4\pi} e^{ik^+y^- \!/2}\! 
\bra{P_n(P^-\!)} \tr\frac{\bnslash}{2}\chi_n\Bigl(y^-\frac{n}{2}\Bigr)
\Bigl[\delta(\omega - \bn\cdot\cP)\frac{1}{\pi}
 \delta(k_\perp^2 - \cP_\perp^2)\bar\chi_n(0)\Bigr]\!
\ket{P_n(P^-\!)}
 \,, \nn
\end{align}
where the light-cone components of vectors are given by $V^+\equiv n\cdot V$ and $V^- \equiv \bn\cdot V$. Note the dependence of the beam functions on the transverse label momentum $k_\perp$ is only on the squared magnitude $k_\perp^2$. The matrix
elements in \eqs{genbeamdef}{ordinarybeam} are similar to those that define
parton distribution functions, but the separation of the collinear fields in the
$n$ direction means there is energetic collinear radiation from the proton with
virtuality $\sim\omega k^+\gg \Lqcd^2$ (assuming we are measuring $k^+$ to be
large enough), which must be integrated out to match \eq{genbeamdef} onto
nonperturbative PDFs (where the separation of $\bar\chi_n,\chi_n$ fields is
zero). The generalized beam functions \eq{genbeamdef} are related to the
ordinary beam functions originally defined in \cite{Stewart:2009yx} by
integrating over all $\vect{k}_\perp$:
\be
\label{ordinarybeam}
B_{q,\bar q}\Bigl(\omega k^+,\frac{\omega}{P^-},\mu\Bigr) 
 = \int d^2\vect{k}_\perp \cB_{q,\bar q}
  \Bigl(\omega k^+ \!\!,\frac{\omega}{P^-},k_\perp^2,\mu \Bigr)\,.
\ee
This relationship would be subtle for PDFs, where it is true for the bare matrix
elements, but where after renormalization the two objects may no longer be
simply related. In the beam function case both sides have the same anomalous
dimension which is independent of $k_\perp$ and there is no such subtelty.

The proton matrix elements in \eq{Wfixedlabels} can now be expressed as
\begin{align}
\label{protonmatrixelements}
&\bra{P_{\nB}}   \bar\chi_{\nB}^{\beta k} (0) 
 \delta(Q_B\tau_B - \nB\mcdot \hat p^{\nB}) \bigl[\delta(\bnB\mcdot q +\bnB\mcdot\cP) 
 \delta^2(\tilde p_\perp - \cP_\perp)  \chi_{\nB}^{\beta' k'}(0)\bigr] \ket{P_{\nB}}  
 \\
&\qquad\qquad = -\bnB\mcdot q \frac{\nslash_B^{\beta'\beta}}{4}
 \frac{\delta^{kk'}}{N_C}\cB_q\Bigl(s_B\tau_B, -\frac{\bnB\mcdot q}{\bnB\mcdot
   P},\tilde p_\perp^2,\mu\Bigr),
 \nn \\
&\bra{P_{\nB}}   \chi_{\nB}^{\alpha j} (0) \delta(Q_B\tau_B - \nB\mcdot \hat
p^\bn) \bigl[\delta(\bnB\mcdot q + \bnB\mcdot\cP) \delta^2(\tilde p_\perp -
\cP_\perp)  \bar \chi_{\nB}^{\alpha' j'}(0)\bigr] \ket{P_{\nB}}  
 \nn \\
&\qquad \qquad = -\bnB\mcdot q\frac{\nslash_B^{\alpha\alpha'}}{4}
  \frac{\delta^{jj'}}{N_C}\cB_{\bar q}
  \Bigl(s_B\tau_B, -\frac{\bnB\mcdot q}{\bnB\mcdot P}, \tilde p_\perp^2,\mu\Bigr)
 \nn \,,
\end{align}
where $s_B$ is defined in \eq{sJsB}.  Now, to simplify the second argument of
the beam functions, we note that
\be
x = -\frac{q^2}{2q\mcdot P} = - \frac{\bnB\mcdot q \, \nB\mcdot q + q_\perp^2}{\nB\mcdot q\, \bnB\mcdot P} = -\frac{\bnB\mcdot q}{\bnB\mcdot P} + \cO(\lambda^2)\,,
\ee
where in the second equality we used that the proton momentum $P$ is exactly
along the $\nB$ direction, and in the last step used that $q_\perp$ is no bigger than $\cO(Q\lambda^2)$. (The directions $\nJ$ and $\nB$ will always be
chosen so that this is true, according to \eqs{delta1}{gperp}. In other words,
for events with small 1-jettiness, all the large momentum $q$ transferred into
the final state is collimated along $n_J$ and $n_B$ with no $\cO(Q)$ momentum
going in a third direction.) Thus to leading order in $\lambda$ the second
argument of the beam functions in \eq{protonmatrixelements} is always just $x$.

\subsubsection{Jet Functions}

The vacuum collinear matrix elements in \eq{Wfixedlabels} can be
written in terms of jet functions in SCET \cite{Bauer:2001yt}, defined with
transverse displacement of the jet in \cite{Stewart:2009yx} by
\begin{align} \label{jetfunction}
J_q(\omega k^+ + \omega_\perp^2,\mu) &= \frac{(2\pi)^2}{N_C}\int
\frac{dy^-}{2\abs{\omega}} e^{ik^+y^-/2}
\tr \Big\langle 0\Big| \frac{\bnslash}{2}\chi_n\Bigl(y^-\frac{n}{2}\Bigr)\delta(\omega +
\bn\cdot\cP) \delta^2(\omega_\perp + \cP_\perp)\bar\chi_n(0)
 \Big| 0 \Big\rangle, 
  \\
J_{\bar q}(\omega k^+ + \omega_\perp^2,\mu) &= \frac{(2\pi)^2}{N_C}
  \int \frac{dy^-}{2\abs{\omega}} e^{ik^+y^-/2} \Big\langle 0 \Big| \bar\chi_n
  \Bigl(y^-\frac{n}{2}\Bigr)\delta(\omega + \bn\cdot\cP) 
  \delta^2(\omega_\perp + \cP_\perp)\frac{\bnslash}{2}\chi_n(0)\Big| 0\Big\rangle
 \,. \nn
\end{align}
Thus the vacuum collinear matrix elements in \eq{Wfixedlabels} can be expressed
\begin{align} \label{jetME}
& \bra{0} \chi_{\nJ}^{\alpha j} (0)  \delta(Q_J\tau_J - \nJ\mcdot \hat p_n ) 
 \delta( \bnJ\mcdot q + \bnJ\cdot\cP) \delta^2(q_\perp +  \tilde p_\perp +\cP_\perp)
  \bar\chi_{\nJ}^{\alpha' j'}(0) \ket{0} 
 \\
& \qquad\qquad  = \frac{\bnJ\mcdot q}{(2\pi)^3} \frac{\nslash_J^{\alpha\alpha'}}{4} 
 \delta^{jj'}  J_q( s_J\tau_J +   (q_\perp+\tilde p_\perp)^2,\mu) \,,
\nn \\
&\bra{0}\bar \chi_{\nJ}^{\beta k} (0) \delta(Q_J\tau_J - \nJ\mcdot \hat p_n)
\delta( \bnJ\mcdot q + \bnJ\cdot\cP) \delta^2( q_\perp + \tilde p_\perp +
\cP_\perp) \chi_{\nJ}^{\beta' k'}(0) \ket{0} 
 \nn\\
&\qquad \qquad = \frac{\bnJ\mcdot q}{(2\pi)^3} \frac{\nslash_J^{\beta'\beta}}{4}
\delta^{kk'}  J_{\bar q}(s_J\tau_J + (q_\perp + \tilde p_\perp)^2,\mu)  
 \nn \,,
\end{align}
where $s_J$ is defined in \eq{sJsB} and $(q_\perp+\tilde p_\perp)^2= -(\vect{q}_\perp+\vect{\tilde{p}}_\perp)^2$.

\subsubsection{Hard and Soft Functions}

Using the above definitions of beam and jet functions, the hadronic tensor in \eq{Wfixedlabels} can be written as
\be
\label{Wcompact}
\begin{split}
W_{\mu\nu}^{II'}(x,Q^2,\taun) &=   -2(2\pi)\bnB\mcdot q\, \bnJ\mcdot q(Q_JQ_B)^2\int d^2\tilde p_\perp \frac{2}{\nJ\mcdot \nB} \int d\tau_J d\tau_B d\tau_s^J\,d\tau_s^B \delta(\taun - \tau_J - \tau_B - \tau_s^J - \tau_s^B)   \\
&\quad \times S(Q_J\tau_s^J,Q_B\tau_s^B,\nJ\mcdot \nB,\mu) J_q(s_J\tau_J + (q_\perp + \tilde p_\perp)^2, \mu)  \\
&\quad \times \Bigl[H^{II'}_{q\bar q\mu\nu}(q^2,\nJ,\nB ) \cB_q(s_B\tau_B, x,\tilde p_\perp^2,\mu)    +    H^{II'}_{q\bar q\mu\nu} (q^2,\nB,\nJ)  \cB_{\bar q}(s_B\tau_B, x,\tilde p_\perp^2,\mu)\Bigr]    \,.
\end{split}
\ee
where the hard function is defined
\be
\label{hardfunctiondef}
H^{II'}_{q\bar q \mu\nu}((\tilde p_1-\tilde p_2)^2,n_a,n_b) = \Tr\Bigl[\bar C_{Iq\bar q \mu}(\tilde p_1,\tilde p_2) \frac{\nslash_a}{4}C_{I'q\bar q \nu}(\tilde p_1,\tilde p_2) \frac{\nslash_b}{4}\Bigr]\,,
\ee
and the soft function is defined
\be
\label{softfunction}
S(k_J,k_B,\qJ, \qB,\mu) = \frac{1}{N_C}\tr\bra{0}[Y_{\nB}^\dag Y_{\nJ}](0) \delta(k_J - \nJ\mcdot\hat p_J^s) \delta(k_B - \nB\mcdot\hat p_B^s)  [Y_{\nJ}^\dag Y_{\nB}](0)\ket{0}\,.
\ee
To write \eq{Wcompact} we used the equality of the quark and antiquark jet functions $J_{q,\bar q}$ in QCD.

\paragraph{Structure of the hard functions}

In \eq{hardfunctiondef}, the matching coefficients $C,\bar C$ in the hard function \eq{hardfunctiondef} for the vector and axial currents $I=V,A$ take the form
\be
\label{CVCA}
 C_{Vf\,q\bar q}^\mu (\tilde p_1,\tilde p_2) =
  C_{Vfq}((\tilde p_1-\tilde p_2)^2) \gamma_\perp^\mu 
 \,,\quad
 C_{Af\,q\bar q^\mu} (\tilde p_1,\tilde p_2) = 
  C_{Afq}((\tilde p_1-\tilde p_2)^2) \gamma_\perp^\mu\gamma_5\,,
\ee
where $\gamma_\perp^\mu$ is transverse to the directions $n_{1,2}$ of the label
momenta $\tilde p_{1,2}$. We have shown the index $f$ for the quark flavor in
the current explicitly. In \eq{Wcompact} these directions are $n_{J,B}$. The
scalar coefficients $C_{Vf\,q\bar q},C_{Af\,q\bar q}$ depend only on the
symmetric Lorentz-invariant combination $(\tilde p_1-\tilde p_2)^2$. Using the
momentum-conserving delta function in \eq{Wfixedlabels}, this combination takes
the value $(\tilde p_1 - \tilde p_2)^2 =q^2$. Inserting \eq{CVCA} into
\eq{hardfunctiondef} we obtain
\be \label{hardfunction}
H^{II'}_{q\bar q \mu\nu}(q^2,\nJ,\nB,\mu) =
  C_{Ifq}(q^2,\mu) C_{I'f'q}(q^2,\mu) 
 \Tr \Bigl( \Gamma^{I}_\mu \frac{\nslash_J}{4}\Gamma^{I'}_\nu\frac{\nslash_B}{4}\Bigr)
 \,,
\ee
where $\Gamma^{V}_\mu = \gamma_\perp^\mu$ and $\Gamma^{A}_\mu =
\gamma_\perp^\mu\gamma_5$. Thus, there are two distinct traces to take in
\eq{hardfunction}:
\begin{align} \label{hardVA}
H^{VV,AA}_{q\bar q \mu\nu}(q^2,\nJ,\nB,\mu) &= -\frac{\nJ\mcdot
  \nB}{4}C_{V,A\,fq}(q^2,\mu) C_{V,A\,f'q}(q^2,\mu) g_\perp^{\mu\nu} 
  \,, \\
H^{VA,AV}_{q\bar q \mu\nu}(q^2,\nJ,\nB,\mu) &=-i\frac{\nJ\mcdot
  \nB}{4}C_{V,A\,fq}(q^2,\mu) C_{A,V\,f'q}(q^2,\mu) \epsilon_\perp^{\mu\nu}
 \,, \nn
\end{align}
where $g_\perp^{\mu\nu}$ and $\epsilon_\perp^{\mu\nu}$ are symmetric and
antisymmetric tensors orthogonal to $\nJ$ and $\nB$ given in \eq{geperp}.
Hence, $H^{VV,AA}$ and $H^{VA,AV}$ are symmetric and antisymmetric, respectively, under exchanging $\nJ$ and $\nB$.

\paragraph{Structure of the soft function}

The soft function \eq{softfunction} depends on the momenta $k_{B,J}$ projected
onto the $n_{B,J}$ directions in the regions $\cH_{B,J}$, respectively. The
shape of these regions in turn depends on the vectors $\qBJ =
\omega_{B,J}n_{B,J}/2$ in the definition of the 1-jettiness $\taun$ in
\eq{tau1def}. Indicating this dependence explicitly, we express the soft
function \eq{softfunction} as
\begin{align} \label{softdef}
S(k_J,k_B,\qJ, \qB,\mu) &= \frac{1}{N_C}\tr \sum_{X_s}
  \abs{\bra{X_s}[Y_{\nJ}^\dag Y_{\nB}](0)\ket{0}}^2 
  \: \delta\Bigl(k_J - \sum_{i\in X_s} 
    \theta(\qB\mcdot k_i - \qJ\mcdot k_i)\nJ\mcdot k_i\Bigr) 
 \nn \\
&\quad \times \delta\Bigl(k_B - \sum_{i\in X_s}
  \theta(\qJ\mcdot k_i - \qB\mcdot k_i)\nB\mcdot k_i\Bigr) 
 \,.
\end{align}
Note that the soft function for DIS involves the square of one incoming and one
outgoing Wilson line, and hence differs from that for $e^+e^-\to {\rm dijets}$
that has two outgoing lines, and for $pp\to L+0{\rm -jets}$ which has two
incoming lines.  We can relate \eq{softdef} to the usual hemisphere soft
function for DIS by generalizing an argument given in \cite{Feige:2012vc}. Note
that the Wilson lines $Y_n$ are invariant under rescaling of $n$ (boost
invariance):
\be
\label{Yrescaling}
Y_{\beta n_B} = P\exp\left[ ig\int_{-\infty}^0 ds\,\beta n_B\mcdot A_s(\beta n_B s)\right] = P\exp\left[ ig\int_{-\infty}^0 ds\, n_B\mcdot A_s( n_B s)\right] = Y_{n_B}\,,
\ee
and similarly for the lines extending from $0$ to $+\infty$, $Y_{\beta n_J}=Y_{n_J}$. 
Recall from \eq{RJRBmeaning} that
\be
\label{RJRBdefs}
 R_J = \sqrt{\frac{\wB \nB\mcdot \nJ}{2\wJ}}
 \,,\quad\qquad 
 R_B = \sqrt{\frac{\wJ \nJ\mcdot \nB}{2\wB}} 
 \,,
\ee
so defining $\nJ' = \nJ/R_J$ and $\nB' = \nB/R_B$ we have $(q_B-q_J)\cdot k_i =
\frac12 \omega_B R_B\, (n_B'-n_J')\cdot k_i$ since $\omega_J R_J=\omega_B R_B$.
This implies that the same partitioning defined in \eq{softdef} can be
expressed with $\theta(n_B'\cdot k_i - n_J'\cdot k_i)$ and $\theta(n_J'\cdot k_i
- n_B'\cdot k_i)$. Furthermore $\nB'\mcdot \nJ' = 2$. Thus expressing
\eq{softdef} in terms of the rescaled vectors, $\nJ'$ and $\nB'$, we obtain
\begin{align} \label{softrescaled}
S(k_J,k_B,\qJ, \qB,\mu) &= \frac{1}{N_C R_J R_B}\tr \sum_{X_s}
 \abs{\bra{X_s}[Y_{\nJ'}^\dag Y_{\nB'}](0)\ket{0}}^2  
 \delta\Bigl(\frac{k_J}{R_J} - \sum_{i\in X_s} 
 \theta(\nB'\mcdot k_i - \nJ'\mcdot k_i)\nJ'\mcdot k_i\Bigr)
  \nn \\
&\qquad \times \delta\Bigl(\frac{k_B}{R_B} - \sum_{i\in X_s}
  \theta(\nJ'\mcdot k_i - \nB'\mcdot k_i)\nB'\mcdot k_i\Bigr)  
 \nn \\
 &= \frac{1}{R_J R_B}\: \Shemi\Bigl(\frac{k_J}{R_J},\frac{k_B}{R_B},\mu\Bigr)
\,.
\end{align}
In the last equality we have expressed the fact that the expression in
\eq{softrescaled} is the same as the hemisphere soft function (up to the overall
$1/(R_J R_B)$ in front), with momentum arguments rescaled by $R_{J,B}$ as
indicated. Therefore from here on we will write all the $\taun$ factorization
theorems in terms of the DIS hemiphere soft function. Note that the vectors
$n_{J,B}'$ have been rescaled from $n_{J,B}$ such that they no longer have
timelike components equal to 1 nor spacelike magnitudes equal to each other, and
therefore do not partition the final states $X_s$ into hemispheres as viewed in
the original $n_{J,B}$ frame of reference. However, the soft function in
\eq{softrescaled} depends on $n_{J,B}'$ exactly like the hemisphere soft
function depends on $n_{J,B}$ and depends on the dot product $\nJ'\mcdot \nB'$,
which is 2, making it equal to the hemisphere soft function. Physically, there
exists a frame where $n_J'$ and $n_B'$ are back-to-back with equal time-like
components, so that the partitioning in this frame gives hemispheres.

In the 1-jettiness cross sections below, the soft function \eq{softrescaled} will always be projected symmetrically onto a function of a single variable $k_S$, following from \eq{Wcompact}:
\be
\label{Shemiprojection}
\Shemi(k_S,\mu) = \int dk_S^J\,dk_S^B\, \delta(k_S - k_S^J - k_S^B) \Shemi(k_S^J,k_S^B,\mu)\,,
\ee
We will use the same name $\Shemi$ for the hemisphere soft function of two variables in \eq{softrescaled} and its one-variable projection \eq{Shemiprojection}, distinguishing them by the number of arguments we write.

\subsubsection{Final Form of Factorization Theorem for Hadronic Tensor}

Changing variables in the arguments of the beam, jet, and soft functions in
\eq{Wcompact} gives
\begin{align}
\label{Wfinal}
W_{\mu\nu}^{II'}(x,Q^2,\taun) &=  
  \int d^2 p_\perp \frac{8\pi}{\nJ\mcdot \nB}   \int dt_J dt_B dk_s^J\,dk_s^B \,
  \delta\left(\taun - \frac{t_J}{s_J} - \frac{t_B}{s_B} -\frac{k_s^J+k_s^B}{Q_R} \right)      
 J_q\big(t_J - (\vect{q}_\perp + \vect{p}_\perp)^2, \mu\big) 
 \:
\\
&\quad \times  \Shemi\bigl({k_s^J},{k_s^B},\mu\bigr) 
 \Bigl[H^{II'}_{q\bar q\mu\nu}(q^2,\nJ,\nB,\mu) \cB_q(t_B,x,\vect{p}_\perp^2,\mu) 
 + H^{II'}_{q\bar q\mu\nu}(q^2,\nB,\nJ,\mu) \cB_{\bar q}(t_B,x,\vect{p}_\perp^2,\mu)
 \Bigr]  \,.\nn
\end{align}
We have written the arguments of the jet and beam function in terms of dimension
2 variables $t_{J,B}$,  the arguments of the soft function in terms of the
total light-cone momentum $k_s^J \equiv \nJ\mcdot k_s^J$ in region $J$ and
$k_s^B \equiv \nB\mcdot k_s^B$ in region $B$, and have rewritten the transverse
momentum arguments of the jet and beam functions in terms of two-vectors
$\vect{q}_\perp,\vect{p}_\perp$ instead of the the four-vectors $q_\perp,\tilde
p_\perp$. The constant $Q_R$ is defined in \eq{QR} and $s_{J,B}$ are defined in
\eq{sJsB}, and their special values for $\taumBCM$ are given in \tab{QJQB}.

\subsubsection{Factorization Theorem for Cross Section}

In the cross section \eq{tauCS}, the hard function \eq{hardfunction} gets
contracted with the leptonic tensor $L_{\mu\nu}$ in \eq{leptontensor}.  The
contraction of the leptonic tensor and the hard function can be performed using
the tensor contractions in \eqs{gg}{ee}, and can be expressed in terms of
Born-level cross section and scalar hard coefficients as
\begin{subequations}
\label{leptonhard}
\begin{align}
\sum_{II'} L^{II'}_{\mu\nu f f'}(x,Q^2) \frac{8\pi}{\nJ\mcdot \nB} H^{II'}_{q\bar q \mu\nu}(q^2,\nJ,\nB,\mu)
&= \frac{d\sigma_0}{dx\,dQ^2} H_{q}(\qJ,\qB,Q^2,\mu)\, ,
\\
\sum_{II'} L^{II'}_{\mu\nu f f'}(x,Q^2) \frac{8\pi}{\nJ\mcdot \nB} H^{II'}_{q\bar q \mu\nu}(q^2,\nB,\nJ,\mu)
&= \frac{d\sigma_0}{dx\,dQ^2} H_{\bar q}(\qJ,\qB,Q^2,\mu)\, ,
\end{align}
\end{subequations}
where the Born-level cross section is given by
\be
\label{Born}
\frac{d\sigma_0}{dx\,dQ^2} = \frac{4\pi \alpha_{em}^2}{x^2 s^2 Q^2} 
\frac{\qJ\mcdot k' \, \qB\mcdot k + \qJ\mcdot k \, \qB\mcdot k'}{\qJ\mcdot \qB}\,.
\ee
The hard coefficients of the quark and antiquark beam functions are
\be
\label{Hcoefficient}
H_{q\, , \bar q}(\qJ,\qB,Q^2,\mu) 
= \sum_{ff'}\left[
\left(C_{Vfq}^* C_{Vf'q} L_{gff'}^{VV} + C_{Afq}^* C_{Af'q} L_{gff'}^{AA} \right) 
\mp r(\qJ,\qB) 
\left( C_{Vfq}^* C_{Af'q} L_{\epsilon ff'}^{VA}  + C_{Afq}^* C_{Vf'q} L_{\epsilon ff'}^{AV}\right) 
\right],
\ee
where the relative minus signs for $H_{\bar q}$ come from the interchange of
$n_{J,B}$ in \eq{leptonhard}. The coefficients $C_{V,A}\equiv C_{V,A}(q^2,\mu)$
are functions of $q^2$ and $\mu$ and the leptonic coefficients
$L_{g,\epsilon}\equiv L_{g,\epsilon}(Q^2)$ given in \eq{Lrslts}. The coefficient
$r(\qJ,\qB)$ is given by
\be \label{r}
r(\qJ,\qB)= \frac{ \qJ\mcdot k' \, \qB\mcdot k- \qJ\mcdot k\, \qB\mcdot k'}
                { \qJ\mcdot k' \, \qB\mcdot k+\qJ\mcdot k\, \qB\mcdot k'} 
  \:.
\ee
Because the coefficient $r$ is a function of scalar products of $\qBJ$ and $k$ and $k'$
it becomes a function of $y$ and $Q^2$ once $\qBJ$ are specified as in \ssec{Njettiness}.
So, the hard coefficient $H_{q\, \bar{q}}$ also is a function of $y$ and $Q^2$ through the coefficient $r$.

Contracting \eq{leptontensor} with \eq{Wfinal} then gives for the cross section \eq{tauCS}, 
\be
\label{taucsfinal}
\begin{split}
\frac{d\sigma}{dx\,dQ^2\,d\taun} 
&= \int d^2\vect{p}_\perp \, \frac{d\sigma_0}{dx\,dQ^2}\int dt_J dt_B dk_S
\delta\left(\taun - \frac{t_J}{s_J} - \frac{t_B}{s_B} 
- \frac{k_S }{Q_R}\right)    J_q(t_J - (\vect{q}_\perp + \vect{p}_\perp)^2, \mu) \, 
        \Shemi(k_S,\mu)  \\
&\quad \times \bigl[ H_{q}(\qJ,\qB,Q^2,\mu) \cB_q(t_B, x,\vect{p}_\perp^2,\mu)   +   H_{\bar q}(\qJ,\qB,Q^2,\mu) \cB_{\bar q}(t_B, x, \vect{p}_\perp^2,\mu)\bigr] \,,
\end{split}
\ee
where we used the projection \eq{Shemiprojection} of the soft function onto a single variable.

\subsection{Results for three versions of 1-jettiness $\taum,\tauB,\tauCM$}
\label{ssec:csresults}

Now we will specialize the generic factorization theorem for 1-jettiness in
\eq{taucsfinal} to the specific cases $\taumBCM$. The discussion will be most
efficient if we begin with $\tauB$.

\subsubsection{1-jettiness $\tauB$}

The reference vectors $\qBB = xP$ and $\qJB = q+xP$ in \eq{Breitvectors} are used to define the 1-jettiness $\tauB$. 
In any frame $q$ can be written as $q = \qJB- \qBB$, 
so with respect to the directions $\nBB,\nJB$,  the transverse component $q_\perp = 0$ so that the  argument of the jet function in \eq{taucsfinal} is $(\vect{q}_\perp+\vect{p}_\perp)^2=\vect{p}_\perp^2$.  
Meanwhile, the coefficient $r(\qJ,\qB)$ in \eq{r} is given by
\be \label{rB}
r(q+xP,xP)=\frac{y(2-y)}{1+(1-y)^2} \: .
\ee
Note that $r$ is a function only of $y$. So, the hard coefficients $H_{q\, \bar
  q}$ in \eq{Hcoefficient} depend on $y$ and $Q^2$, and we define the hard
coefficients for $\tauB$ by $H_{q\,,\bar q}^b (y,Q^2,\mu) \equiv H_{q\,,\bar
  q}(q+xP,P,Q^2,\mu)$.  Therefore, using \eq{taucsfinal} the final factorization
theorem for $\tauB$ is given by
\be
\label{tauBcs}
\begin{split}
\frac{d\sigma}{dx\,dQ^2\,d\tauB} &=   
\frac{d\sigma_{0}^\B}{dx\,dQ^2} \int dt_J dt_B dk_S \:
\delta\left(\tauB - \frac{t_J}{Q^2} - \frac{t_B}{Q^2} 
  - \frac{k_S}{Q}\right)\, 
  \Shemi \left(k_S,\mu\right)   \\
&\quad\times \int d^2 \vect{p}_\perp   J_q(t_J -  \vect{p}_\perp^2, \mu)  \Bigl[ H_q^\B(y,Q^2,\mu) \cB_q(t_B, x,\vect{p}_\perp^2,\mu) + H^\B_{\bar q}(y,Q^2,\mu) \cB_{\bar q}(t_B, x,\vect{p}_\perp^2,\mu)\Bigr]     \,,
\end{split}
\ee
where we used \tab{QJQB} to substitute for $s_{J,B},Q_R$ in \eq{taucsfinal}, and
where the Born-level cross section is given by
\be
\label{Bornxy}
\frac{d\sigma_{0}^\B}{dx\,dQ^2} = \frac{2\pi \alpha_{em}^2}{Q^4}\left[(1-y)^2+1 \right]\,.
\ee

\subsubsection{1-jettiness $\taum$}

For the 1-jettiness $\taum$ defined in \eq{truetau}, the minimization inside the
sum over final state particles $i$ groups particles with the reference vector to
which they are closest.
The reference vector $\qJm$ with which the jet particles are grouped is aligned
with the jet momentum $p_J$, so that the jet has zero transverse label momentum
with respect to $\nJm$. This direction $\nJm$ is the one which would minimize
$\taum$ (to leading $\cO(\lambda^2)$)with respect to variations of $\qJm$. A jet with momentum $p_J = \wJ
\nJ/2 + \tilde p_{J \perp} + k$, where $k$ is residual, has a mass $m^2 = \wJ
\nJ\mcdot k+ \tilde p_{J \perp}^2$, so $\nJ\mcdot k = (m^2 - \tilde p_{J
  \perp}^2) /\wJ$. The choice of $\nJ$ which makes $\tilde p_{J \perp}^2= 0$
minimizes $\nJ\mcdot k$ (note that $\tilde p_\perp^2\leq 0$).

The cross section for the $\taum$ distribution is given by \eq{taucsfinal}, with
$\qB = \qBm\equiv xP$ and $\qJ=\qJm$, where $\qJm$ is the vector $\qJ$ in
\eq{truetau} that minimizes $\taum$. We will write $\qJm$ in terms of the vector
$\qJB = q+xP$ that was used to define the 1-jettiness $\tauB$.  Now, the vector
$\qJB$ has a direction $\nJB$ and magnitude $\wJB$, given by
\be
\label{qJB}
\qJB = \wJB \frac{\nJB}{2} = P_T e^Y \frac{n_z}{2} +
  P_T e^{-Y}\frac{\bn_z}{2} + P_T\hat n_T^J\,,
\ee
expressed in the CM frame, where $p_T,Y$ are given by \eq{pTxyrelations}. With
respect to $\nJB$ and $n_P$, the collinear fields in the jet function matrix
elements still have nonzero transverse labels. Now, for each $\tilde p_J^\perp$,
we rotate $\nJB$ to a vector $\nJm$ so that the transverse label with respect to
$\nJm$ is zero. This requires that the total label momenta in the two coordinate
systems be equal:
\be
\label{nJnJmrelation}
\wJB\frac{\nJB}{2} + \tilde p_J^\perp = \wJm\frac{\nJm}{2} \,,
\ee 
so $\nJm$ differs from $\nJB$ at most by a quantity of $\cO(\lambda)$. 
Now we express $q$ in $\nJm,n_P$ coordinates. In \eq{taucsfinal}, the transverse label on the collinear fields in the jet function is $q_\perp +  p_\perp$. The $\nJm,n_P$ coordinate system is defined as that which makes this quantity is zero, so $q_\perp =- p_\perp$. By using $q^2= -xys$ and $n_P\mcdot q = y\sqrt{s}$, $q$ is expressed as
\be
q = y\sqrt{s} \frac{\nJm}{\nJm\mcdot n_P}  - \Bigl(x\sqrt{s} + \frac{ p_\perp^2}{y\sqrt{s}} \Bigr) \frac{n_P}{2} -  p_\perp\,.
\ee
Then \eq{taucsfinal} takes the form
\be
\label{taumcs}
\begin{split}
\frac{d\sigma}{dx\,dQ^2\,d\taum} &= 
\int d^2\vect{p}_\perp \, \frac{d\sigma_{0}^\m}{dx\,dQ^2}\int dt_J dt_B dk_S \:
\delta\left(\taum - \frac{t_J}{Q^2} - \frac{t_B}{Q^2} 
           - \frac{k_S}{Q}\right) 
          J_q(t_J, \mu)   \Shemi(k_S,\mu)   \\
&\quad \times   \bigl[ H_{q}(\qJm,\qBm,Q^2,\mu) \cB_q(t_B, x,\vect{p}_\perp^2,\mu)   +   H_{\bar q}(\qJm,\qBm,Q^2,\mu) \cB_{\bar q}(t_B, x, \vect{p}_\perp^2,\mu)\bigr] \,,
\end{split}
\ee
where we used \tab{QJQB} to substitute for $s_{J,B}$ and $Q_R$.

The generalized beam functions appearing here explicitly depend on $p_\perp$.
The vector $\nJm$ appearing in $q_J^a$ implicitly depends on $p_\perp$. Now,
$\nJm$ differs from $\nJB$ (which is independent of $p_\perp$) by a quantity of
order $\lambda$. Here we can expand the hard and soft functions and the Born
cross section around $\nJm = \nJB + \cO(\lambda)$ and drop the power corrections
in $\lambda$. This makes everything in \eq{taumcs} independent of $p_\perp$
except for the generalized beam function. The integral over $p_\perp$ then turns
the generalized beam function into the ordinary beam function \eq{ordinarybeam}.
Thus the final factorization theorem for the $\taum$ cross section is
\begin{align} \label{taumcsfinal}
\frac{d\sigma}{dx\,dQ^2\,d\taum} &=   
\frac{d\sigma_{0}^\B}{dx\,dQ^2}   \int dt_J dt_B dk_S 
\: \delta\left(\taum - \frac{t_J}{Q^2} - \frac{t_B}{Q^2} 
                - \frac{k_S}{Q}\right)  J_q(t_J , \mu) \Shemi(k_S,\mu)
 \nn   \\
&\quad\times  \Bigl[ H_q^\B(y,Q^2,\mu) B_q(t_B, x,\mu) 
 + H^\B_{\bar q}(y,Q^2,\mu) B_{\bar q}(t_B, x,\mu)\Bigr] 
                  \,,
\end{align}
where the Born cross section is given by \eq{Bornxy} and $H^\B_{q,\bar
  q}(y,Q^2)$ is given by \eq{Hcoefficient} with $r$ in \eq{rB}.  The hard and
soft functions in \eq{taumcsfinal} are the same as those in \eq{tauBcs} for
$\tauB$. 

\eq{taumcsfinal} differs from \eq{tauBcs} in that the jet and beam functions are
no longer convolved together in the transverse momentum $p_\perp$.  The
1-jettiness $\taum$ is proportional to the invariant mass of the jet, while
$\tauB$ measures the projection $\qJB\cdot p_J$ onto the fixed axis $\qJB =
q+xP$. The emission of ISR with transverse momentum $p_\perp$, causing a shift in the jet momentum by the same amount due to momentum conservation, will not change the mass of the jet, but it will change the projection of the jet momentum onto the $\qJB$ axis. Thus $\taum$ involves no convolution over $p_\perp$, while $\tauB$ does.

\subsubsection{1-jettiness $\tauCM$}

For the 1-jettiness $\tauCM$, the directions $\nJ$ and $\nB$ are along the electron and proton directions, respectively:
\be
\nJ = n_e\,,\quad \nB = n_P\,, 
\ee
where $n_{e,P}$ are the light-cone directions of $\qJ=k = \omega_e n_e/2$ and
$\qB=P = \omega_P n_P/2$.
In the CM frame, $n_{e,P}$ are back-to-back, $n_e = n_z$ and $n_P = \bn_z$.
In this frame, $q$ is given by
\be
q = y\sqrt{s} \frac{n_z}{2} - xy\sqrt{s}\frac{\bn_z}{2} + q_\perp\,,
\ee
where $q_\perp = Q\sqrt{1-y}\,\hat n_\perp$.  Let us consider for a moment the
power counting of the argument of the jet function in \eq{taucsfinal} with
$\tauCM$. The requirement that $q_\perp\sim Q\lambda$ requires that $1-y\sim
\lambda^2$. This is ensured by measuring $\tauCM$ to be $\cO(\lambda^2)$. The
argument of the jet function (call it $m_J^2$) in the factorization theorem
\eq{taucsfinal} for $\tauCM$ is
\be
\label{mJCM}
\begin{split}
m_J^2 &= t_J - (1-y) Q^2 - 2Q\sqrt{1-y}\, \vect{\hat n}_\perp\mcdot \vect{p}_\perp - \vect{p}_\perp^2\,.
\end{split}
\ee
Now, the jet function will be proportional to a theta function $\theta(m_J^2)$,
requiring $m_J^2>0$. Measuring $\tauCM$ to be of order $\lambda^2$ and therefore
forcing $t_J$ to be of order $Q^2\lambda^2$ then enforces that
$1-y\sim\lambda^2$. Then, we can set $y=1$ to leading order everywhere in
\eq{taucsfinal} except in the argument of the jet function. In terms of $x$,
using the relation $xys=Q^2$, requiring $y\lesssim 1$ is equivalent to requiring
$x\gtrsim Q^2/s$, which sets a lower bound on $x$.

The normalization constants $s_{J,B}, Q_R$ in \eq{taucsfinal} are given for
$\tauCM$ in \tab{QJQB}.  The Born-level cross section and the coefficient
$r(\qJ,\qB)$ in the hard coefficients reduce to
\be
\label{BornCMrCM}
\frac{d\sigma_{0}^\CM}{dx\,dQ^2} = \frac{2\pi \alpha_{em}^2}{Q^4}\,, \qquad r(k,P)= 1\,,
\ee
where we see the Born cross section is now \eq{Bornxy} in the limit $y\to 1$.
This happens because the expression \eq{Born} is evaluated with $\qJ=k$, which
is the actual jet direction only near $y\to 1$. The hard coefficient is now
independent of $x,y$ and depends only on $Q^2$: $H_{q,\bar q}^\CM(Q^2,\mu) =
H_{q,\bar q}(k,P,Q^2,\mu)$.  From \eq{taucsfinal} the final factorization
theorem for the $\tauCM$ cross section is then given by
\begin{align} \label{tauCMcs}
\frac{d\sigma}{dx\,dQ^2\,d\tauCM} 
 &=  \frac{d\sigma_{0}^\CM}{dx\,dQ^2} \int d^2\vect{p}_\perp 
\int dt_J dt_B dk_S \:
\delta\left(\tauCM - \frac{t_J}{Q^2} - \frac{t_B}{xQ^2}- \frac{k_S}{\sqrt{x}\,Q}\right) 
  J_q(t_J - (\vect{q}_\perp + \vect{p}_\perp)^2, \mu) \, \Shemi(k_S,\mu) 
 \nn \\
& \quad \times \Bigl[ H_{q}^\CM(Q^2,\mu) \cB_q(t_B, x,\vect{p}_\perp^2,\mu)   
 +   H_{\bar q}^\CM(Q^2,\mu) \cB_{\bar q}(t_B, x, \vect{p}_\perp^2,\mu)\Bigr] 
 \,.
\end{align}
This is like the $\tauB$ cross section \eq{tauBcs} in that the jet and beam functions are convolved in the transverse momentum $\vect{p}_\perp$ of ISR, but in this case the momentum transfer $q$ itself has a nonzero transverse momentum with respect to the light-cone directions $n_{e,P}$. This will make the evaluation of the $\vect{p}_\perp$ 
integral considerably more involved than in the $\tauB$ cross section \eq{tauBcs}.

\end{widetext}

\section{Fixed-Order Predictions at $\cO(\as)$}
\label{sec:NLO}

In this Section we evaluate to $\cO(\as)$ the predictions of the factorization
theorems for the cross sections differential in the different versions of
1-jettiness in \eq{tauBcs} for $\tauB$, and \eq{taumcsfinal} for $\taum$ and
\eq{tauCMcs} for $\tauCM$. These formulas correctly predict the singular terms
at small $\taun$ in the fixed-order differential cross section, although they
have to be resummed to all orders in $\as$ to accurately predict the behavior at
small $\taun$. We will do this in the next Section. Also, for the predictions to
be correct for large $\taun$, they would have to be matched onto $\cO(\as)$ and $\cO(\as^2)$ fixed-order full QCD calculations, an analysis we defer to future work. 
Nevertheless we can estimate the size of these matching corrections by comparing our
predictions integrated up to large $\taun$ to the known total QCD cross section
$\sigma(x,Q^2)$ at $\cO(\as)$, which we will do in \sec{results}.

\subsection{Hard Function}

At $\cO(\as)$, the matching coefficients $C_{Vfq},C_{Afq}$ for the vector and axial currents \eq{JQCD} that appear in the hard coefficient in \eq{Hcoefficient} are equal and diagonal in flavor, and were calculated in \cite{Bauer:2003di,Manohar:2003vb}:
\begin{align}
\label{CNLO}
&C_{Vfq}(q^2) = C_{Afq}(q^2) = \delta_{fq}C(q^2) \,, \\
&C(q^2) = 1 + \frac{\as(\mu)C_F}{4\pi}\left( - \ln^2\frac{\mu^2}{-q^2} - 3 \ln\frac{\mu^2}{-q^2} - 8 + \frac{\pi^2}{6}\right).\nn
\end{align}
For DIS recall $q^2 = -Q^2$.  Then, the hard coefficients $H_{q,\bar q}$ in the cross section \eq{taucsfinal} are given to $\cO(\as)$ by
\be
\label{hardNLO}
H_{q,\bar q}(\qJ,\qB,Q^2,\mu)  = H(Q^2,\mu) L_{q,\bar q}(\qJ,\qB,Q^2) \,,
\ee
where we have defined the universal SCET 2-quark hard coefficient,
\begin{align}
\label{scalarH}
&H(Q^2,\mu) \equiv \abs{C(q^2,\mu)}^2 \\
&\quad = 1 + \frac{\as(\mu)C_F}{2\pi} \left(- \ln^2\frac{\mu^2}{Q^2} - 3\ln\frac{\mu^2}{Q^2} - 8 + \frac{\pi^2}{6}\right)\,, \nn
\end{align}
and the factor containing the components of the leptonic tensor \eq{leptontensor},
\be
\label{scalarL}
L_{q,\bar q}(\qJ,\qB,Q^2) = L_{gqq}^{VV} + L_{gqq}^{AA} \mp r(\qJ,\qB)(L_{\epsilon qq}^{VA} + L_{\epsilon qq}^{AV})\,,
\ee
where $r(\qJ,\qB)$ was defined in \eq{r}.

\subsubsection{$\taumB$ cross sections}

For the $\taumB$ cross sections \eqs{tauBcs}{taumcsfinal}, $\qB = xP$ and $\qJ=q+xP$, so that $r(\qJ,\qB)$ is given by \eq{rB}. Then the leptonic factor $L_{q,\bar q}$ in \eq{scalarL} becomes
\begin{align}
\label{Lqm}
&L_{q,\bar q}^\m(\qJ,\qB,Q^2) \\
&\quad = L_{gqq}^{VV} + L_{gqq}^{AA} \mp \frac{y(2-y)}{(1-y)^2+1}(L_{\epsilon qq}^{VA} + L_{\epsilon qq}^{AV}) \nn \\
&\quad = Q_q^2 - \frac{2Q_q v_q v_e}{1+m_Z^2/Q^2}   + \frac{(v_q^2+a_q^2)(v_e^2+a_e^2)}{(1+m_Z^2/Q^2)^2} \nn \\
&\qquad \mp \frac{2y(2-y)}{(1-y)^2 + 1}\frac{a_qa_e [Q_q(1+m_Z^2/Q^2) - 2v_qv_e]}{(1+m_Z^2/Q^2)^2} \,. \nn
\end{align}

\subsubsection{$\tauCM$ cross section}

For the $\tauCM$ cross section \eq{tauCMcs}, $\qJ=k$ and $\qB = P$, the electron and proton momenta, respectively. Then $r(k,P) = 1$ in \eq{r}, and the leptonic factor $L_{q,\bar q}$ in \eq{scalarL} becomes 
\begin{align}
\label{LqCM}
L_{q,\bar q}^\CM(k,P,Q^2) &= L_{gqq}^{VV} + L_{gqq}^{AA} \mp (L_{\epsilon qq}^{VA} + L_{\epsilon qq}^{AV}) \\
&= Q_q^2 - \frac{2Q_q(v_q v_e \pm a_q a_e)}{1+m_Z^2/Q^2}  \nn \\
&\quad   + \frac{(v_q^2+a_q^2)(v_e^2+a_e^2) \pm 4v_qa_q v_ea_e}{(1+m_Z^2/Q^2)^2}\,. \nn
\end{align}

\subsection{Soft Function}
\label{ssec:soft}

The soft function $\Shemi(k_S,\mu)$ that appears in the cross sections \eqss{tauBcs}{taumcsfinal}{tauCMcs} is given by \eqs{softrescaled}{Shemiprojection}. 
For $e^+e^-\to$ dijets, $\Shemi^\text{dijet}$ is known at $\cO(\as)$ \cite{Fleming:2007xt} and $\cO(\as^2)$ \cite{Hornig:2011iu,Kelley:2011ng,Monni:2011gb}. At 1-loop order the dijet soft function is the same for DIS. Beginning at 2-loop order, the finite part of the soft function for DIS could possibly differ due to switching incoming and outgoing Wilson lines, but the anomalous dimensions and thus the logs are the same.

To $\cO(\as)$, the soft function \eq{softrescaled} takes the form:
\be 
\label{softNLO}
\begin{split}
&\Shemi(k_s^J,k_s^B,\mu) = \delta(k_s^J)\delta(k_s^B)  \\
&+ S^{(1)}(k_s^J,\mu)\delta(k_s^B)  + \delta(k_s^J)S^{(1)}(k_s^B,\mu)\,,
\end{split}
\ee
where
\be
\label{Sren1}
\begin{split}
S^{(1)}(k_s,\mu) = \frac{\as C_F}{4\pi}  \biggl\{ \frac{\pi^2}{6}\delta(k_s)   - \frac{8}{\mu} \left[ \frac{\theta(k_s)\ln(k_s/\mu)}{k_s/\mu}\right]_+ \biggr\}\,,
\end{split}
\ee
and the projection \eq{Shemiprojection} is then given to $\cO(\as)$ by 
\be
\Shemi(k_S,\mu) = \delta(k_S) + 2S^{(1)}(k_S,\mu)\,.
\ee
It has previously been observed that the sizes $R_{J,B}$ of the regions $\cH_{J,B}$ to which soft radiation is confined enter the arguments of the logs in the soft function \cite{Ellis:2009wj,Ellis:2010rw,Jouttenus:2011wh}, which is due to changing the effective scale at which the soft modes live \cite{Bauer:2011uc}.

\begin{widetext}

\subsection{Jet Function}

The jet function \eq{jetfunction} is given to $\cO(\as)$ by \cite{Bauer:2003pi,Becher:2006qw}
\begin{align}
\label{jetNLO}
J_{q,\bar q}(t,\mu) = \delta(t) + \frac{\as(\mu)C_F}{4\pi} \biggl\{ (7-\pi^2) \delta(t)    - \frac{3}{\mu^2} \left[\frac{\theta(t)}{t/\mu^2}\right]_+   + \frac{4}{\mu^2}\left[\frac{\theta(t)\ln(t/\mu^2)}{t/\mu^2}\right]_+\biggr\}\,.
\end{align}
It is in fact known to two-loop order \cite{Becher:2006qw} and its anomalous dimension to three loops \cite{Becher:2006mr}.

\subsection{Beam Functions}

\subsubsection{Generalized Beam Functions}

The generalized beam functions in \eq{genbeamdef} can be matched onto ordinary
PDFs, defined in SCET as \cite{Bauer:2002nz}:
\begin{subequations}
\label{PDFs}
\begin{align}
f_q(\omega'/P^-,\mu) &= \theta(\omega') \bra{P_n(P^-)}\bar\chi_n(0) \frac{\bnslash}{2}[\delta(\omega'-\bn\cdot\cP)\chi_n(0)]\ket{P_n(P^-)}\,, \\
f_{\bar q}(\omega'/P^-,\mu) &= \theta(\omega') \bra{P_n(P^-)}\tr  \frac{\bnslash}{2}\chi_n(0)[\delta(\omega'-\bn\cdot\cP)\bar\chi_n(0)]\ket{P_n(P^-)} \, .
\end{align}
\end{subequations}
The matching result is \cite{Mantry:2009qz,Jain:2011iu}:
\be
\label{genbeamNLO}
\cB_i(t,x,\vect{k}_\perp^2,\mu) = \sum_j \int_x^1 \frac{d\xi}{\xi}\cI_{ij}\Bigl(t,\frac{x}{\xi},\vect{k}_\perp^2,\mu\Bigr)f_j(\xi,\mu)\Bigl[1 + \cO\Bigl(\frac{\Lqcd^2}{t},\frac{\Lqcd^2}{\vect{k}_\perp^2}\Bigr)\Bigr]\,,
\ee
where $i,j = q,\bar q,g$. This expansion is valid for perturbative beam radiation satisfying $t,\vect{k}_\perp^2\gg \Lqcd^2$. At tree level, $\cI_{ij}^{(0)}(t,z,\vect{k}_\perp^2,\mu) =  (1/\pi)\delta_{ij}\delta(t)\delta(1-z)\delta(\vect{k}_\perp^2)$, leading to $\cB_i^{(0)}(t,x,\vect{k}_\perp^2,\mu) = (1/\pi)\delta(t)\delta(\vect{k}_\perp^2)f_i(x,\mu)$.

To $\cO(\as)$, the nonzero matching coefficients in the generalized quark beam
function were computed in~\cite{Mantry:2009qz,Jain:2011iu}, and we use the
results from~\cite{Jain:2011iu}:
\begin{subequations}
\label{genbeamcoeffs}
\begin{align}
\cI_{qq}(t,z,\vect{k}_\perp^2,\mu) &= \frac{1}{\pi}\delta(t)\delta(1-z)\delta(\vect{k}_\perp^2)  + \frac{\as(\mu)C_F}{2\pi^2} \theta(z) \Biggl\{ \frac{2}{\mu^2}\left[\frac{\theta(t) \ln(t/\mu^2)}{t/\mu^2}\right]_+\delta(1-z)\delta(\vect{k}_\perp^2) \\
&\quad+ \frac{1}{\mu^2} \left[ \frac{\theta(t)}{t/\mu^2}\right]_+\left[ P_{qq}(z) - \frac{3}{2}\delta(1-z) \right] \delta\Bigl(\vect{k}_\perp^2 - \frac{(1-z)t}{z}\Bigr) \nn \\
&\quad   + \delta(t) \delta(\vect{k}_\perp^2) \Biggr[ \left[\frac{\theta(1-z)\ln(1-z)}{1-z}\right]_+ (1+z^2) - \frac{\pi^2}{6}\delta(1-z) + \theta(1-z) \Bigl( 1-z - \frac{1+z^2}{1-z}\ln z\Bigr) \Biggr] \Biggr\}\,, \nn \\ 
\cI_{qg}(t,z,\vect{k}_\perp^2,\mu) &= \frac{\as(\mu)T_F}{2\pi^2} \theta(z) \biggl\{ \frac{1}{\mu^2} \left[\frac{\theta(t)}{t/\mu^2}\right]_+ \!\!\!\! P_{qg}(z) \delta\Bigl(\vect{k}_\perp^2 \minus \frac{(1 \minus z)t}{z}\Bigr) + \delta(t) \delta(\vect{k}_\perp^2) \biggl[ P_{qg}(z)\ln\frac{1\minus z}{z} + 2\theta(1\minus z)z(1\minus z)\biggr] \biggr\}\,,
\end{align}
\end{subequations}
where $P_{qq,qg}$ are the $q\to qg$ and $g\to q\bar q$ splitting functions,
\begin{subequations}
\label{splittingfunctions}
\begin{align}
P_{qq}(z) &= \left[\frac{\theta(1-z)}{1-z}\right]_+ (1+z^2) + \frac{3}{2}\delta(1-z) = \left[ \theta(1-z)\frac{1+z^2}{1-z}\right]_+ \,,\\
P_{qg}(z) &= \theta(1-z) [(1-z)^2 + z^2]\,.
\end{align}
\end{subequations}
They appear in the anomalous dimensions of the PDFs, which to all orders obey
\be
\label{DGLAP}
\mu\frac{d}{d\mu}f_i(\xi,\mu) = \sum_j \int \frac{d\xi'}{\xi'} \gamma_{ij}^f\Bigl( \frac{\xi}{\xi'},\mu\Bigr)f_j(\xi',\mu) \,.
\ee
At $\cO(\as)$ the anomalous dimensions for the quark PDF are
\be
\gamma_{qq}^f(z,\mu) = \frac{\as(\mu)C_F}{\pi}\theta(z) P_{qq}(z) \,, 
 \quad\qquad
 \gamma_{qg}^f(z,\mu) = \frac{\as(\mu)T_F}{\pi}\theta(z) P_{qg}(z)\,.
\ee

\subsubsection{Ordinary Beam Functions}

The ordinary beam functions \eq{ordinarybeam} satisfy the matching condition
\cite{Fleming:2006cd,Stewart:2009yx,Stewart:2010qs}:
\be
\label{ordinarybeammatching}
B_i(t,x,\mu) = \sum_j \int_x^1 \frac{d\xi}{\xi}\cI_{ij}\Bigl(t,\frac{x}{\xi},\mu\Bigr)f_j(\xi,\mu)\Bigl[1 + \cO\Bigl(\frac{\Lqcd^2}{t}\Bigr)\Bigr]\,,
\ee
where at tree level $\cI_{ij}^{(0)}(t,z,\mu) = \delta_{ij}\delta(t)\delta(1-z)$,
leading to $B_i^{(0)}(t,x,\mu) = \delta(t)f_i(x,\mu)$.  To $\cO(\as)$, the
matching coefficients in the quark beam function are given by integrating
\eq{genbeamcoeffs} over $\vect{k}_\perp$~\cite{Stewart:2009yx,Stewart:2010qs}:
\begin{subequations}
\label{beamcoeffs}
\begin{align}
\cI_{qq}(t,z,\mu) &= \delta(t)\delta(1-z)  + \frac{\as(\mu)C_F}{2\pi} \theta(z) \Biggl\{ \frac{2}{\mu^2}\left[\frac{\theta(t) \ln(t/\mu^2)}{t/\mu^2}\right]_+\delta(1-z) + \frac{1}{\mu^2} \left[ \frac{\theta(t)}{t/\mu^2}\right]_+\left[ P_{qq}(z) - \frac{3}{2}\delta(1-z) \right] \nn \\
&\quad   + \delta(t) \Biggr[ \left[\frac{\theta(1-z)\ln(1-z)}{1-z}\right]_+ (1+z^2) - \frac{\pi^2}{6}\delta(1-z) + \theta(1-z) \Bigl( 1-z - \frac{1+z^2}{1-z}\ln z\Bigr) \Biggr] \Biggr\}\,,  \\ 
\cI_{qg}(t,z,\mu) &= \frac{\as(\mu)T_F}{2\pi} \theta(z) \biggl\{ \frac{1}{\mu^2} \left[\frac{\theta(t)}{t/\mu^2}\right]_+ P_{qg}(z) + \delta(t) \biggl[ P_{qg}(z)\ln\frac{1-z}{z} + 2\theta(1-z)z(1-z)\biggr] \biggr\}\,.
\end{align}
\end{subequations}

\subsection{Dijet Cross Section}

We can now form the SCET predictions for the $\taun$ cross section \eq{taucsfinal} to $\cO(\as)$ by plugging in the $\cO(\as)$ expressions for the hard function given by \eqss{hardNLO}{scalarH}{scalarL}, the soft function given by \eqs{softNLO}{Sren1}, the jet function given by \eq{jetNLO}, and the generalized beam function given by \eqs{genbeamNLO}{genbeamcoeffs}.
It is convenient to express the result in terms of the \emph{cumulant} $\taun$ distribution, defined by
\be
\label{cumulant}
\sigmac(x,Q^2,\taun) = \frac{1}{\sigma_0}\int_0^{\taun} d\taun' \frac{d\sigma}{dx\,dQ^2\,d\taun'}\,,
\ee
where $\sigma_0$ is the Born cross section defined in \eq{Born}.
We will give here the results for the $\taumB$ cumulants at $\cO(\as)$. The more complicated results for the $\tauCM$ and generic $\taun$ cumulants are given in \appx{NLO}.

\subsubsection{$\taum$ cross section}

Plugging in the $\cO(\as)$ results for the hard function given by \eqss{hardNLO}{scalarH}{Lqm}, the soft function given by \eqs{softNLO}{Sren1} with $s_{J,B}$ and $Q_R$ given in \tab{QJQB}, the jet function given by \eq{jetNLO}, and the ordinary beam function given by \eq{ordinarybeammatching} into the $\taum$ cross section \eq{taumcsfinal}, we obtain for the $\taum$ cumulant given by \eq{cumulant} in the CM frame:
\begin{align}
\label{NLOtaumcs}
\sigmac(x,Q^2,\taum) & = \theta( \taum )  \int_x^1 \frac{dz}{z} \bigl[ L_q^\m(x,Q^2)f_q(x/z,\mu) + L_{\bar q}^\m(x,Q^2) f_{\bar q}(x/z,\mu)\bigr]   \\
&\quad\times  \Biggl\{\delta(1-z)\biggl[ 1 - \frac{\as C_F}{4\pi}\Bigl(9 +\frac{2\pi^2}{3} + 6 \ln \taum + 4\ln^2\taum \Bigr) \biggr] \nn \\
&\qquad + \frac{\as C_F}{2\pi} \biggl[ \cL_1(1-z) (1+z^2) + \theta(1-z) \biggl( 1 - z - \frac{1+z^2}{1-z}\ln z\biggr) +  \ln\Bigl(\frac{Q^2\taum}{\mu^2} \Bigr)   P_{qq}(z)\biggr] \Biggr\} \nn \\
&\qquad  + \frac{\as T_F}{2\pi} \theta(\taum)   \bigl[ L_q^\m(x,Q^2) \plus L_{\bar q}^\m(x,Q^2)\bigr] \!  \int_x^1 \frac{dz}{z} f_g(x/z,\mu)\biggl\{ \ln\Bigl(\frac{Q^2\taum}{\mu^2}\frac{1\minus z}{z}\Bigr)  P_{qg}(z) +  2\theta(1 - z)z(1 - z)\biggr\} . \nn
\end{align}
The factorization scale $\mu$ still appears on the right-hand side of the equation, though the cross section is in fact independent of $\mu$. The $\mu$-dependence in the PDFs on the first line is cancelled by the $\mu$-dependence in the logs multiplying the splitting functions on the third and final lines to $\cO(\as)$. The residual $\mu$-dependence is $\cO(\as^2)$ and would be cancelled by the higher-order corrections.

\subsubsection{$\tauB$ cross section}

The $\tauB$ cross section is nearly identical to the $\taum$ cross section except for the presence of the $\vect{p}_\perp$-dependent generalized beam function in \eq{tauBcs} instead of the ordinary beam function. The effect of the nontrivial $\vect{p}_\perp$-dependent terms in the generalized beam function \eq{genbeamNLO} is simply to multiply the arguments of the $\mu$-dependent logs in \eq{NLOtaumcs} by $z$, giving the simple relation
\be
\label{NLOtauBcs}
\begin{split}
\sigmac(x,Q^2,\tauB) = \sigmac(x,Q^2,\taum)\Big\vert_{\taum\to\tauB} + \theta(\tauB) \frac{\as}{2\pi} \int_x^1 \frac{dz}{z} \ln z \Bigl\{&C_F\bigl[ L_q^\m(Q^2)f_q(x/z,\mu) + L_{\bar q}^\m(Q^2) f_{\bar q}(x/z,\mu)\bigr]  P_{qq}(z) \\
+ & T_F \bigl[ L_q^\m(Q^2) + L_{\bar q}^\m(Q^2)\bigr]  P_{qg}(z)f_g(x/z,\mu)\Bigr\}\,,
\end{split}
\ee

In \appx{NLO} we give the $\cO(\as)$ $\tauCM$ cross section. In the next section we resum the large logarithms of $\taumBCM$ that appear in these fixed-order expansions to all orders in $\as$ to NNLL accuracy.

\end{widetext}

\section{Resummed Predictions for $\taun$ Cross Sections}
\label{sec:NNLL}

The fixed-order predictions for the $\taun$ cross sections presented in the
previous Section contain logarithms of $\taun$ which grow large in the limit
$\taun\to 0$ and must be resummed to all orders in $\as$ to yield accurate
predictions for small $\taun$.  In this Section we use the factorization theorem
\eq{taucsfinal} for the $\taun$ 1-jettiness cross section and its specialized
cases \eqss{tauBcs}{taumcsfinal}{tauCMcs} for $\tauB,\taum,\tauCM$ to predict
the cross sections differential in these variables to next-to-next-to-leading
logarithmic (NNLL) accuracy, estimate the perturbative uncertainty by
appropriate scale variations, and discuss power corrections due to
hadronization, including their universality and impact in the tail and peak
regions.

\subsection{Perturbative Resummation to NNLL}

The hard, jet, beam, and soft functions in \eq{taucsfinal} obey renormalization group (RG) evolution equations whose solutions allow us to resum large logarithms of ratios of the separated hard, jet, beam, and soft scales. These solutions allow us to express any of these functions $G=\{H,J,B,S\}$ at one scale $\mu$ which contains logs of $\mu$ over some scale $Q_G$ in terms of the function evaluated at a different scale $\mu_G \sim Q_G$ where the logs are small.

The hard function $H(Q^2,\mu)$ obeys the RG equation
\be
\label{hardRGE}
\mu\frac{d}{d\mu} H(Q^2,\mu) = \gamma_H(\mu) H(Q^2,\mu)\,,
\ee
where the anomalous dimension $\gamma_H$ has the form
\be
\gamma_H(\mu) = \Gamma_H[\as(\mu)]\ln\frac{Q^2}{\mu^2} + \gamma_H[\as(\mu)]\,,
\ee
with a \emph{cusp} piece $\Gamma_H[\as] = 2\Gamma^q_\cusp$ and a \emph{non-cusp} piece $\gamma_H[\as]$ (which is conventionally denoted by the same symbol as the total anomalous dimension). Their expansions in $\as$ are given below in \eq{anomdimexpansion} and \eqs{Gacuspexp}{gaHexp}. 
Similarly the jet and beam functions which are both functions of a dimension-2 variable $t$ obey RG equations of the form
\be
\label{BJRGE}
\mu\frac{d}{d\mu} G(t,\mu) = \int dt' \gamma_G(t-t',\mu) G(t',\mu)\,,
\ee
where the anomalous dimension $\gamma_G$ takes the form
\be
\gamma_G(t,\mu) = \Gamma_G[\as(\mu)] \frac{1}{\mu^2} \left[\frac{\theta(t/\mu^2)}{t/\mu^2}\right]_+ + \gamma_G[\as(\mu)]\delta(t)\,,
\ee
where here $G=\{J,B\}$, and the plus distribution is defined in \appx{L}. The cusp pieces $\Gamma_{J,B}
= -2\Gamma^q_\cusp$ and non-cusp pieces $\gamma_{J,B}$ of the jet and beam
anomalous dimensions are given in \eqs{Gacuspexp}{gaBexp}. The beam function
also depends on $x$ and the generalized beam function also depends on
$\vect{p}_\perp^2$, but they do not change the structure of the RG equation
\eq{BJRGE}.

Finally, the soft function in \eq{softrescaled} obeys the RG equation
\be
\label{softRGE}
\begin{split}
&\mu\frac{d}{d\mu}\Shemi(k_J,k_B,\mu) = \int dk_J' dk_B' \\
&\qquad \times \gamma_S(k_J-k_J',k_B-k_B',\mu) \Shemi(k_J',k_B',\mu)\,,
\end{split}
\ee 
where the anomalous dimension factorizes into the form
\be
\gamma_S(k_J,k_B,\mu)  = \gamma_S(k_J,\mu)\delta(k_B) + \gamma_S(k_B,\mu)\delta(k_J),
\ee
which is required by $\mu$-independence of the total cross section \eq{taucsfinal}  \cite{Hoang:2007vb}. Each piece $\gamma_S(k,\mu)$ takes the form
\be
\gamma_S(k,\mu) = 2\Gamma_S[\as(\mu)]\frac{1}{\mu } \left[\frac{\theta(k/\mu )}{k/\mu }\right]_+ + \gamma_S[\as(\mu)]\delta(k)\,,
\ee
where $\Gamma_S = \Gamma_\cusp$, and the non-cusp piece is given by $\gamma_S = -\gamma_H/2 - \gamma_J$. 

The cusp and non-cusp pieces of the anomalous dimensions of all the functions above all have perturbative expansions in $\as$:
\be
\label{anomdimexpansion}
\Gamma_G[\as] = \sum_{n=0}^\infty \Gamma_G^n\Bigl(\frac{\as}{4\pi}\Bigr)^{\! n+1} , \ \gamma_G[\as] = \sum_{n=0}^\infty \gamma_G^n\Bigl(\frac{\as}{4\pi}\Bigr)^{\! n+1}\,,
\ee
which defines the coefficients $\Gamma_G^n,\gamma_G^n$. Furthermore, the cusp
pieces of the anomalous dimension are proportional to the same \emph{cusp
  anomalous dimension} $\Gamma^q_\cusp[\as]$, whose perturbative expansion along
with the non-cusp anomalous dimensions are given in \appx{rge}. The explicit
solutions to the RG equations for the hard, jet, beam, and soft functions
individually are given in \appx{rge}.

The solutions of the RG equations \eqss{hardRGE}{BJRGE}{softRGE} allow us to
express the hard, jet, beam, and soft functions at any scale $\mu$ in terms of
their values at different scales $\mu_{H,J,B,S}$ where logarithms of $\mu_G/Q_G$
in their perturbative expansions are small. There are a different conventional
ways in the literature to express the resummed cross section in terms of the
solutions for hard, jet, beam, and soft functions to the RG equations. One
method \cite{Ligeti:2008ac,Abbate:2010xh} performs the exact inverse transform
back from Fourier space, and carries out analytically the convolution of all the
evolution factors and the fixed-order functions for the $\taun$ factorization
theorem \eq{taucsfinal} in momentum space. In this section we use this method
and formalism, relegating some of the required formulas to \appx{RGconstants}.
We give an alternative equivalent form of the resummed cross sections in
\appx{Laplace}, using a method \cite{Becher:2006mr,Becher:2006nr} that first
Laplace transforms the cross section and writes certain corrections as
derivative operators before transforming back to momentum space. This avoids taking
explicit convolutions of the evolution factors and the fixed-order functions. If
one carries out these derivatives analytically then the final results from the two
formalisms are identical.

In this section we give just the final results for the RG improved cross
sections for the 1-jettinesses $\taumBCM$ using the formalism of
\cite{Ligeti:2008ac,Abbate:2010xh}. We will express the results in terms of the
\emph{cumulant} $\taumBCM$ distributions:
\be
\label{sigmac}
\sigmac(x,Q^2,\taun) = \frac{1}{\sigma_0}\int_0^{\taun} d\taun' 
  \frac{d\sigma}{dx\,dQ^2\,d\taun'}\,,
\ee
where we note that $\sigmac$ is dimensionless due to the division by $\sigma_0$.
The differential cross section can be obtained by taking the derivative of
$\sigmac(x,Q^2,\taun)$ with respect to $\taun$. Care must be exercised in this
procedure because $\sigmac$ also depends on $\taun$ dependent jet/beam and soft
scales in the factorization theorem \eq{taucsfinal},
$\sigmac(x,Q^2,\taun,\mu_i(\tilde\tau_1))$. The appropriate procedure is to use,
for $\epsilon\to 0$,
\begin{align}
  \frac{d\hat\sigma}{d\tau_1} &=
  \frac{\sigma_c(x,Q^2,\tau_1\plus\epsilon,\mu_i(\tau_1))
  -\sigma_c(x,Q^2,\tau_1\minus\epsilon,\mu_i(\tau_1))}{2\epsilon},
\end{align}
where $d\hat\sigma/d\tau_1=(1/\sigma_0)d\sigma/d\tau_1$. See
Ref.~\cite{Abbate:2010xh} for further discussion of this point.

\subsubsection{$\taumB$ cross sections}

The cross section in \eq{taucsfinal} is expressed as a convolution of jet, beam,
and soft functions in momentum space. To resum the large logs, each function is
RG evolved from a scale where the logs are small, an operation which is in the
form of a convolution of an RG evolution kernel and the fixed order function as
in \eqs{Brun_full}{softRGEsolution}. The evolution kernels $U_{J,B,S}$ in
\eqs{Brun_full}{SBrun} are plus distributions, and each fixed order function can
also be written as a sum of plus distributions as in \appx{SnJn}. Thus, the
resummed cross section contains numerous convolutions of plus distributions
$\cL^\eta$, $\cL_n$, which we can compute by repeatedly applying the plus
distribution convolution identity in \eq{LL}. The cross section then gets
written as a resummation factor times sums of products of coefficients called
$V$ in \appx{Vcoeff} and $J_n$, $I_n^{qq,qg}$, and $S_n$ in \appx{SnJn}.

The resummed $\taum$ and $\tauB$ cross sections in \eqs{tauBcs}{taumcsfinal},
obtained from RG evolution of the hard, jet, beam, and soft functions, are given
by:
 \begin{widetext}
\begin{subequations}
\label{resummedtaumB}
\begin{align}
\label{resummedtaumB1}
\sigmac(x,Q^2,\tau) &= 
\frac{ e^{\cK-\gamma_E\Omega}}{\Gamma(1+\Omega)}
\left(\frac{Q}{\mu_H}\right)^{\eta_H(\mu_H,\mu)}
\left(\frac{\tau\,Q^2}{\mu_{B}^2}\right)^{\eta_{B}(\mu_B,\mu)}
\left(\frac{\tau\,Q^2}{\mu_{J}^2}\right)^{\eta_{J}(\mu_J,\mu)}
\left(\frac{\tau\,Q}{\mu_{S}}\right)^{2\eta_{S}(\mu_S,\mu)}
\nn \\&
\quad\times \bigg[ \sum_j  L_q^\m(x,Q^2)\, \int_x^1 \, \frac{dz}{z} \,
  f_j(x/z,\mu_B)\,\left[ W_{qj}(z,\tau)+ \Delta W_{qj}(z)\right] 
  +(q\leftrightarrow \bar{q}) 
  \bigg]
\,,  \\[10pt]
\label{W}
W_{qj}(z,\tau) &=
H(Q^2,\mu_H) \, 
\sum_{ \substack{ n_1,n_2, \\ n_3=-1 } } ^{1}
 J_{n_1}\Big[\alpha_s(\mu_J),\frac{\tau Q^2}{\mu_J^2}\Big]\,
 I^{qj}_{n_2}\Big[\as(\mu_B), z, \frac{\tau Q^2}{\mu_B^2} \Big]
\,\,
S_{n_3}\Big[\alpha_s(\mu_{S}),\frac{\tau Q}{\mu_{S}} \Big] 
\nn\\&\quad\times
\sum_{\ell_1=-1}^{n_1+n_2+1}\sum_{\ell_2=-1}^{\ell_1+n_3+1}
V_{\ell_1}^{n_1 n_2} V_{\ell_2}^{\ell_1 n_3}  \,
 V^{\ell_2}_{-1}(\Omega) \,
\,, \\[5pt]
\label{DW}
\Delta W_{qj}(z) &
=\left\{
\begin{array}{ll}
   0 &\text{for}\quad \taum
\\
\frac{\as(\mu_B) }{2\pi} \left[ \delta_{jq} C_F P_{qq}(z) +\delta_{jg} T_F P_{qg}(z)\right]\, \ln z 
  &\text{for}\quad \tauB
\end{array}
\right.
\end{align}
\end{subequations}
Here $j$ sums over quark flavors and gluons, and the $+(q\leftrightarrow \bar q)$
includes the term where the virtual gauge boson couples to an
antiquark.  In \eq{resummedtaumB1} the exponent is a resummation factor that
resums the large logs and the terms $W_{qj}$ and $\Delta W_{qj}$ are fixed-order
factors which do not contain large logs.  The evolution kernels
$\mathcal{K}$ and $\Omega$ are given by
\begin{subequations}
\begin{align} \label{Kappa}
\cK &\equiv \mathcal{K}(\mu_H,\mu_J,\mu_B,\mu_S,\mu) = K_H(\mu_H,\mu) + K_J(\mu_J,\mu) + K_B(\mu_B,\mu) +2 K_S(\mu_S,\mu) \\
\label{Omega}
\Omega &  \equiv \Omega(\mu_J,\mu_B,\mu_S,\mu) = \eta_J(\mu_J,\mu) + \eta_B(\mu_B,\mu) + 2\eta_S(\mu_S,\mu)\,,
\end{align}
\end{subequations}
where the individual evolution kernels $K_H$, $K_J=K_B$, $K_S$, $\eta_J=\eta_B$,
and $\eta_S$ are given below in \eqss{Hrun}{Brun_full}{SBrun}.  Note that $\cK$
and $\Omega$ are indpendent of $\mu$ because the $\mu$ dependence cancels
between the various $K_i$ and $\eta_i$ factors in the sums.  Their expressions
to NNLL accuracy are given in \eq{Keta} \appx{rge}.  The coefficients $J_{n}$,
$I^{qq}_{n}$, $I^{qg}_{n}$, $S_{n}$ in \eq{W} are given in \eq{Fn}.  The
constants $V_k^{mn}$ and $V_k^n(\Omega)$ are given in \appx{Vcoeff}.

Note that in the resummed cross section \eq{resummedtaumB} the coefficients
$J_n$, $I_n^{qj}$, and $S_n$ are functions of logarithms of their last argument
as shown in \eq{Fn} and the hard function also depends on the logarithm
$\ln(Q^2/\mu_H^2)$. The logs in these fixed-order factors can be minimized by
choosing the canonical scales
\be \label{canonicalscales}
 \mu_H = Q\,,\quad \mu_J = \mu_B = Q\sqrt{\taumB}\,,\quad \mu_S = Q\taumB\,.
\ee
Large logs of ratios of these scales to the arbitrary factorization scale $\mu$
are then resummed to all orders in $\as$ in the evolution kernels $\cK$ and
$\Omega$.  The choices in \eq{canonicalscales} are appropriate in the tail
region of the distribution where $\taun$ is not too close to zero and not too
large so that the logs of $\taun$ are still large enough to dominate non-log
terms and need to be resummed. Near $\taun\sim 0$ and $\taun\sim 1$, we will
need to make more sophisticated choices for the scales, which we will discuss in
\ssec{profiles}.

\subsubsection{$\tauCM$ cross section}

The resummed $\tauCM$ cross section obtained from RG evolution of the hard, jet, beam, and soft functions in \eq{tauCMcs} is given by

\begin{align}\label{resummedtauCM}
\sigma^\CM_\cumulant (x,Q^2,\tau) &= 
\frac{e^{K-\gamma_E \Omega}}{\Gamma(1+\Omega)} \!
\left(\frac{Q}{\mu_B}  \right)^{\!\eta_H(\mu_H,\mu)} \!
\left(\frac{(\tau\minus1 \plus y)\, x Q^2}{\mu_{ B}^2} \right)^{\!\eta_{B}(\mu_H,\mu)} \!
\left(\frac{(\tau\minus 1 \plus y)\,  Q^2}{\mu_{ J}^2} \right)^{\!\eta_{J}(\mu_H,\mu)} 
\left(\frac{(\tau\minus 1 \plus y)\,\sqrt{x}Q}{\mu_{S}} \right)^{\!2\eta_{S}(\mu_H,\mu)} \!
\nn\\ & \quad\times
L_q^\CM(x,Q^2) \int_x^1 \frac{dz}{z} f_j(x/z)  \,\left[ W^\CM_{qj} (z,\tau-1+y)+\Delta W^\CM_{qj} (z,\tau-1+y)\right]
+(q\leftrightarrow \bar{q})\,,
\end{align}
where $W^\CM_{qj}$ and $\Delta W^\CM_{qj}$ are the fixed-order terms from jet, beam, and soft functions:
\begin{subequations}
\begin{align}
\label{WCM}
W_{qj}^\CM(z,\tau) &=
H(Q^2,\mu_H) \, 
\sum_{ \substack{ n_1,n_2, \\ n_3=-1 } } ^{1}
 J_{n_1}\left[\alpha_s(\mu_J),\frac{\tau Q^2}{\mu_J^2}\right]
 I^{qj}_{n_2}\left[\as(\mu_B), z, \frac{\tau x Q^2}{\mu_B^2} \right]
\,\,
S_{n_3}\left[\alpha_s(\mu_{S}),\frac{\tau\sqrt{x} Q}{\mu_{S}} \right] 
\\&\hspace{50pt}\times
\sum_{\ell_1=-1}^{n_1+n_2+1}\sum_{\ell_2=-1}^{\ell_1+n_3+1}
V_{\ell_1}^{n_1 n_2} V_{\ell_2}^{\ell_1 n_3} \,
V^{\ell_2}_{-1}(\Omega) \,
\,,  \nn \\ \label{DWCM}
\Delta W_{qj}^\CM(z,\tau)&= 
\frac{\as(\mu_B) }{2\pi} \big[ \delta_{jq} C_F P_{qq}(z) +\delta_{jg} T_F P_{qg}(z)\big]
\Bigg[
 \theta\left( \tau \right)\, \left\{ \ln\left[\frac{\tau (1-X)}{(1-y)X} \right]
        -H(-1-\Omega) \right\}
 \\ &\quad\quad\quad
+\theta\left(\frac{\tau}{(1-y)X}+1 \right)\,
        \left\{
        \frac{1}{\Omega}        \left(\frac{|\tau|}{(1-y)X}\right)^{-\Omega }\,
         {}_2F_1\left[-\Omega,-\Omega;1-\Omega;-\frac{\tau}{(1-y)X}\right]
- \theta\left(-\tau \right)
 \frac{\pi}{\sin(\pi \Omega)}
\right\}
\Bigg]
\,, \nn
\end{align}
\end{subequations}
where $X \equiv x(1-z)/(x+z-xz)$.  Note that the $\tau$ in $W_{qj}^\CM(z,\tau)$
and $\Delta W_{qj}^\CM(z,\tau)$ gets shifted by $1-y$ in \eq{resummedtauCM}.
$H(n)$ is the harmonic number and $_2F_1(a,b;c;z)$ is the hypergeometric
function.  The additional more complicated terms in $\Delta W_{qj}^\CM$ are due
to the nontrivial $\vect{p}_\perp$ integral in \eq{tauCMcs} which convolves the
terms in the generalized beam function with nontrivial $\vect{p}_\perp^2$
dependence with the dependence of the jet function on
$(\vect{q}_\perp+\vect{p}_\perp)^2$, with $\vect{q}_\perp\neq0$ when $y<1$. Note
that the term on the last line of \eq{DWCM} contributes  below $\tauCM=1-y$ when plugged into \eq{resummedtauCM}, but
that the size of the correction in this region is very small.

The second arguments of $J_n$, $I^{qj}_n$, and $S_n$ in \eq{WCM} show that the
canonical scales should be chosen to minimize the logs of the agruments, which
are the fixed-order terms in the hard, jet, beam, and soft functions.  
\be
\label{canonicalscalesCM}
\mu_H = Q\,,\quad \mu_J = Q\sqrt{\tauCM - 1+y} \,, \quad \mu_B = Q\sqrt{x(\tauCM - 1+y)} \,,\quad \mu_S = \sqrt{x}Q(\tauCM-1+y)\,.
\ee
Here the whole cross section is shifted to the right by an amount $1-y$ due to
the nonzero $\vect{q}_\perp$ and choice of axes for $\tauCM$. Unlike $\taumB$,
the jet and beam scales are separated by a factor $\sqrt{x}$ due to the
different normalization of the $\qB$ reference vector in the definition of
$\tauCM$. For $\tauCM$, $\qBCM = P$ while for $\taumB$, $\qBmB = xP$. The soft
scale is also rescaled by $\sqrt{x}$.  We will discuss below a more
sophisticated choice of scales than \eq{canonicalscalesCM} that give rise to
proper behavior in the limits $\tauCM \to 1-y$ and $\tauCM\sim 1$.

\subsubsection{Logarithms included in our LL, NLL, and NNLL results}

It is worth briefly discussing the logarithmic accuracy of our resummed results.
Although this discussion is standard in the literature, sometimes the same
notation is used for different levels of resummed precision, so it is worth
being specific about our notation.
The order in $\as$ to which the anomalous dimensions, running coupling, and
fixed-order hard, jet, beam, and soft functions are known determines the
accuracy to which the logarithms of $\tau$ in cross section are
resummed. It is most straightforward to count the number of logs thus resummed in the Laplace transform of the cross section (equivalently we could consider the Fourier transform to position space), 
\be
\tilde \sigma(x,Q^2,\nu) = \int_0^\infty d\tau \, e^{-\nu\tau} \frac{d\sigma}{dx\,dQ^2\,d\tau}\,.
\ee
The fixed-order expansion of $\tilde\sigma(x,Q^2,\nu)$ takes the form,
\begin{align}
\label{fixedorderexpansion}
\tilde\sigma(x,Q^2,\nu) =\    1 \ 
 + \ \frac{\as}{4\pi}(c_{12}L^2 & + c_{11}L + c_{10} + \tilde d_1(\nu))  \\
+ \left(\frac{\as}{4\pi}\right)^2(c_{24}L^4 & + c_{23}L^3 
 + c_{22} L^2 + c_{21}L + c_{20} + \tilde d_2(\nu)) 
  \nn  \\
+ \left(\frac{\as}{4\pi}\right)^3(c_{36}L^6 &+ c_{35}L^5 
 + c_{34} L^4 + c_{33} L^3 + c_{32} L^2 + c_{31} L 
 + c_{30} + \tilde d_3(\nu)) + \cdots
  \nn \,,
\end{align}
where $L\equiv \log\nu$. The largest log at each order in $\as$ is $\as^n L^{2n}$. Our results in \eqs{resummedtaumB}{resummedtauCM}, once Laplace transformed, reorganize and resum the logarithms into the form:
\begin{align} \label{resummedexpansion}
\tilde\sigma(x,Q^2,\nu) =\   \exp\ \biggl[ \ \frac{\as}{4\pi}(C_{12}L^2 &+ 
   C_{11}L + C_{10})   \\
  +  \left(\frac{\as}{4\pi}\right)^2(C_{23}L^3 &+ C_{22} L^2 + 
  C_{21}L + C_{20})  \nn \\
  +  \left(\frac{\as}{4\pi}\right)^3(C_{34}L^4 &+ C_{33} L^3 + 
  C_{32}L^2 + C_{31}L + C_{30}) + \cdots\biggr] + \tilde d(x,Q^2,\nu)
 \,, \nn
\end{align}
\begin{table}[t]
$$
\begin{array}{|c|c|c|c|c|}
\hline
 & \Gamma[\as] & \gamma[\as] & \beta[\as] & \{H,J,B,S\}[\as] \\ \hline
 \text{LL} & \as & 1 & \as & 1 \\ \hline
  \text{NLL} & \as^2 & \as & \as^2 & 1 \\ \hline
    \text{NNLL} & \as^3 & \as^2 & \as^3 & \as \\ \hline
\end{array}
\qquad
\begin{array}{|c|c|c|c|c|}
\hline
 & \Gamma[\as] & \gamma[\as] & \beta[\as] & \{H,J,B,S\}[\as] \\ \hline
 \text{LL} & \as & 1 & \as & 1 \\ \hline
  \text{NLL}' & \as^2 & \as & \as^2 & \as \\ \hline
    \text{NNLL}' & \as^3 & \as^2 & \as^3 & \as^2 \\ \hline
\end{array}
$$
\caption{Orders of logarithmic accuracy and required order of cusp ($\Gamma$) and non-cusp ($\gamma$) anomalous dimensions, beta function $\beta$, and fixed-order hard, jet, beam, and soft matching coefficients $H,J,B,S$. The ``primed'' counting includes the fixed-order coefficients to one higher order in $\as$. 
\label{tab:logaccuracy}}
\end{table}
\end{widetext}
where the largest log at each order in the exponent is $\as^n L^{n+1}$.
The coefficients $c_{nm}$, $C_{nm}$, and $d_n(\nu)$ are functions of $x$ and
$Q^2$.  The function $\tilde d(x,Q^2,\nu)$ contains terms  $\tilde d_n(\nu)$ and is a
non-singular function of $\nu$ that vanishes as $\nu\to\infty$ ($\tau\to 0$). Transforming \eqs{fixedorderexpansion}{resummedexpansion} back to momentum space using
\be
\frac{d\sigma}{dx\,dQ^2\,d\tau} = \int_{\gamma-i\infty}^{\gamma+i\infty} \frac{d\nu}{2\pi i}e^{\nu\tau}\tilde\sigma(x,Q^2,\nu)\,,
\ee
where $\gamma$ lies to the right of all singularities of the integrand in the complex plane,
defines the accuracy to which logs of $\tau$ in the cross section and its cumulant $\sigmac(x,Q^2,\tau)$ are resummed.

 Our main
results in \eqs{resummedtaumB}{resummedtauCM} resum singular logarithmic
terms $\alpha_s^n \ln^m\tau$, but not the terms in the non-singular
$d(x,Q^2,\tau)$ (inverse transform of $\tilde d$).  The $d(x,Q^2,\tau)$ must either be calculated by comparing a
full QCD perturbation theory calculation with the resummed result and
determining the difference order by order in $\alpha_s$, or by determining the
next-to-singular infinite towers of logarithmic terms in $d(x,Q^2,\tau)$ by
carrying out a factorization and resummation analysis in SCET at subleading
power.

Fixed-order perturbation theory sums the series in \eq{fixedorderexpansion}
row-by-row, order-by-order in $\as$. When the logs are large this expansion is
not well behaved. Resummed perturbation theory instead sums the exponent in
\eq{resummedexpansion} column-by-column, in a modified power expansion that
counts $\ln\tau\sim 1/\as$ when the logs are large. Everything in the first
column of \eq{resummedexpansion} is $\cO(1/\as)$ [leading log (LL)], the second
$\cO(1)$ [next-to-leading-log (NLL)], the third $\cO(\as)$ (NNLL), etc. in this
counting. Each order of logarithmic accuracy is achieved by calculating the cusp
and non-cusp anomalous dimensions, running coupling, and fixed-order hard, jet,
beam, and soft functions to the orders
given in \tab{logaccuracy}. Another common counting used in the literature
(eg.~\cite{Fleming:2007xt,Ligeti:2008ac,Abbate:2010xh}) is the primed counting
which accounts for the fixed-order matching coefficients $H,J,B,S$ at one higher order than in the unprimed
counting. This primed counting is particularly useful when one also requires
predictions for transition regions where the size of the logarithmic and
non-logarthmic $d_n(\tau)$ terms are comparable. Since in this paper we have not
considered the nonsingular terms we adopt the unprimed counting (LL, NLL, NNLL)
throughout.

\subsection{Comparison to NLL DIS Thrust $\tau_Q$}
\label{ssec:tauQ}

As discussed above in \sec{versions} there are several versions of
DIS thrust discussed in the literature.  Here we consider the version of thrust
called $\tau_Q$ in \cite{Antonelli:1999kx}, to which we can directly compare our
results for $\tauB$, since as shown above in \eq{tauBtauQ} of
\sec{BeamRegion} they are one and the same.  We will see that at NLL
accuracy the result in \cite{Antonelli:1999kx} for $\tau_Q$ is equivalent to our
result in \eq{resummedtaumB} for $\tauB$ for the particular scale choices
\eq{canonicalscales} in the SCET cross section.  The NLL resummed cross section
given in \cite{Antonelli:1999kx} in the $\MSbar$ scheme is, in our notation,
\begin{align}
\label{Antonelli}
&\sigmac(\tau_Q) = \theta(\tau_Q)\biggl\{\sum_{q} e_q^2 \biggr[ f_q(x,\sqrt{\tau} Q) \\
&\qquad\qquad\qquad  + \frac{\as(Q)}{2\pi} \int_x^1 \frac{dz}{z} C_{1q}(z)f_q(x/z,Q)\biggr] \nn \\
& + \biggl(\sum_{q} e_q^2\biggr)\frac{\as(Q)}{2\pi} \!\! \int_x^1\!
\frac{dz}{z}C_{1g}(z) f_g(x/z,Q)\biggr\} e^{-g_1\ln\tau + g_2}\,, \nn
\end{align}
where
\begin{align}
C_{1q}(z) &= C_F\biggl[ 2\cL_1(1-z) - (1+z) \ln(1-z) \nn \\
&\quad + 1 - z  -
\Bigl(\frac{\pi^2}{3} + \frac{9}{2}\Bigr) \delta(1-z)\biggr] \,, \\
C_{1g}(z) &= T_F [ P_{qg}(z) \ln(1-z) + 2z(1-z)]\,, \nn
\end{align}
and complete expressions for the resummation constants $g_{1,2}$ can be found in
Ref.~\cite{Antonelli:1999kx}. They have fixed-order expansions in
$\alpha_s=\alpha_s(Q)$ given by
\begin{align} \label{gcoeffs}
g_1 \ln\tau &= G_{12} \frac{\as}{2\pi} \ln^2\tau -
G_{23}\Bigl(\frac{\as}{2\pi}\Bigr)^2 \ln^3\tau + \cdots \\
 g_2 &= -G_{11} \frac{\as}{2\pi} \ln\tau + G_{22}\Bigl(\frac{\as}{2\pi}\Bigr)^2
\ln^2\tau + \cdots \,, \nn
\end{align}
with the coefficients
\begin{align} \label{Gcoeffs}
G_{12} &= -2C_F\,,\quad  G_{11} = 3C_F\,, \quad  
 G_{23} = 2\pi\beta_0 G_{12}\,,  \\
 G_{22}  &= -\frac{4}{3}\pi^2 C_F^2 + \Bigl( \frac{\pi^2}{3}
- \frac{169}{36}\Bigr) C_A C_F + \frac{11}{18} C_F n_f\,.  \nn
\end{align}
Note that the cross section \eq{Antonelli} includes only the photon contribution
for the intermediate gauge boson mediating the scattering, so for the comparison
we specialize our results to this case.

By comparing to
the resummed cross section in \eq{resummedtaumB}, we find that the result of
\cite{Antonelli:1999kx} given in \eq{Antonelli} is equivalent to the SCET photon
induced cross section at NLL order with the following fixed choices for the
scales in the evolution factors:
\begin{align}
\mu_H &= Q\,, 
&\mu & = \mu_{J}=\mu_{B} = Q\sqrt{\taun}\,,
& \mu_S & = Q\taun\,.
\end{align}
Thus the two results agree at NLL order.

We note that in the fixed-order coefficient in \eq{Antonelli}, the choice
$\mu=Q\sqrt{\tau}$ has been made in the tree level term, but the $\cO(\as)$
terms have been evaluated at $\mu= Q$. In the SCET result \eq{resummedtaumB} (or \eq{appx:resummedtaumB})
pieces of the $\alpha_s(\mu)$ terms are evaluated at $\mu=Q$,
$\mu=Q\sqrt{\tau}$, or $\mu=Q\tau$ according to whether they come from the hard,
beam/jet, or soft functions, while the PDFs are evaluated at $\mu=Q\sqrt{\tau}$.
The difference between the SCET result and \eq{resummedtaumB} is NNLL, since the
error is $\sim \as^2 \ln\tau$ in the fixed-order coefficient.  In our counting
taking the correct scales for $\alpha_s(\mu)$ is required for NLL$'$ accuracy,
since this provides the appropriate boundary conditions for the full NNLL
result. Thus the result in \eq{Antonelli} with $\alpha_s(Q)$ is at an
intermediate level of accuracy between NLL and NLL$'$. 

Note that the SCET expression \eq{resummedtaumB} still shows the full dependence
on the individual scales $\mu_{H,B,J,S}$ instead of the single scale $\mu$, and
the dependence on each of these scales cancels out to the order in resummed
perturbation theory that we are working. The remaining scale dependence  thus provides a useful way
to estimate the theoretical uncertainty due to the truncation of higher order terms in resummed perturbation theory.

\subsection{Scale Profile Functions}
\label{ssec:profiles}

In general there are three relevant regions with different power
counting
\begin{align} \label{eq:regions}
  &  \text{peak region:} & & \tau_1  \sim  2\Lambda_{\rm QCD}/Q_R \ll 1\nn\,, \\
  &  \text{tail region:} & & 2\Lambda_{\rm QCD}/Q_R  \ll \tau_1 \ll 1 \,, \\
  & \text{far-tail region:} & & \tau_1 \sim 1 \,. \nn
\end{align}
For the peak and tail regions of the distribution we have $\tau_1\ll 1$ and we
must sum the large logarithms.  In the tail region the results in
\eqs{canonicalscales}{canonicalscalesCM} above are the canonical scales for
$\mu_{H,J,B,S}$ for which the logs in the fixed-order hard, jet, beam, and soft
functions are minimized.  Evolution from these scales to another scale $\mu$
resums the logs of the ratios $\mu/\mu_{H,J,B,S}$ to all orders in $\as$.  In
the peak region for small $\taun$, the scale $\mu_{S}\sim Q\taun$ goes towards
the nonperturbative region.  The validity of our resummation analysis relies on
there being a perturbative expansion for the soft function anomalous dimensions
at the scale $\mu_S$, $\Gamma_S[\alpha_s(\mu_S)]$ and
$\gamma_S[\alpha_s(\mu_S)]$.  Therefore in the SCET approach it is mandatory
that we stop the renormalization group evolution at a scale $\mu_S\sim 1\,{\rm
  GeV}$ that can still be considered perturbative. This requires the scales to
deviate from the canonical form.  Finally, for larger values of $\taun$ the
logs are no longer large, and the nonsingular terms in the fixed-order expansion
become equally important. In this large $\taun$ region we should turn off the
resummation, which will revert the results to a fixed-order expansion in
$\as$. Again this forces the scales to deviate from the canonical ones.

To achieve these properties we use \emph{profile functions} to describe the
functional dependence of the scale $\mu_{S,B,J}$ on $\taun$.  First we will
consider the profile functions for the $\taumB$ cross sections and then for
$\tauCM$.

\subsubsection{$\taumB$ profile functions}
\label{sssec:profiles}

For the $\taumB$ cross sections, the canonical scales are given in
\eq{canonicalscales}, $\mu_S\sim\taun$, $\mu_{B,J}\sim \sqrt{\taun}Q$,
$\mu_H\sim Q$. The perturbative resummation of large logs of ratios of these
scales is valid when $\Lqcd\ll\mu_S \ll\mu_{B,J} \ll \mu_H$, which is the tail
region.  We will define boundaries , $t_1<\taun<t_2$ for the region of $\taun$
where this condition is satisfied, and use scales that are within a factor of
$2$ of the canonical ones.  Beyond this region, when $\taun>t_2$, $\taun$ is of
${\cal O}(1)$, and the logs are the same order as the nonsingular terms in the
fixed-order expansion.  In this region, the scales must be taken to be of the
same order, $\mu_S\simeq \mu_{B,J}\simeq \mu_H\sim Q$, which turns off the
resummation in \eq{resummedtaumB}.  Finally for $\taun<t_1$, the soft scale
approaches $\Lqcd$ and nonperturbative corrections become important.  In this
region we freeze the soft scale $\mu_S$ used in the perturabtive cross section
to a value above $\Lqcd$: $\mu_S\sim 1$--$2~\mathrm{GeV}$.  The hard scale
is $\mu_H\sim Q$ and beam and jet scale are determined by hard and soft scales
as $\sqrt{\mu_H \mu_S}\sim \mu_{J,B}$.

Profile functions for scales that satisfy the above criteria have been used for
other cross sections in \cite{Abbate:2010xh,Ligeti:2008ac,Berger:2010xi}.  Here,
we adopt the profile functions in \cite{Berger:2010xi}.  The hard, beam, jet,
and soft scales we use are given by
\begin{align} \label{profiles}
\mu_H &=\mu
\,, \\
\mu_{B,J}(\taun) &=\biggl[1+e_{B,J}\theta(t_3\minus\taun)\Bigl(1\minus\frac{\taun}{t_3}\Bigr)^2 \biggr] \! \sqrt{\mu\, \mu_\mathrm{run}(\taun\,,\mu)}
\,, \nn \\
\mu_{S}(\taun) &=\biggl[1+e_S\theta(t_3\minus\taun)\Bigl(1\minus
\frac{\taun}{t_3}\Bigr)^2 \biggr] \mu_\mathrm{run}(\taun\,,\mu)
 \,, \nn
\end{align}
where $e_{B,J,S}$ are parameters used to vary the jet, beam, and soft scales to estimate theoretical uncertainty of the perturbative predictions.
$t_3$ is the point above which all scales are set equal, $\mu_H=\mu_{B,J}=\mu_{S}=\mu$.
The common function $\mu_\mathrm{run}(\taun,\mu)$ is given by
\begin{align}
\label{murun}
& \mu_\mathrm{run}(\taun,\mu) =
\begin{cases}
\mu_0 + a\taun^2/t_1 & \taun \leq t_1
\,,\\
2a\, \taun + b & t_1 \leq \taun \leq t_2
\,,\\
\mu - a (\taun-t_3)^2/(t_3 - t_2) & t_2 \leq \taun \leq t_3
\,,\\
\mu & \taun > t_3
\,,\end{cases}
\nn \\
&
a= \frac{\mu_0-\mu}{t_1-t_2-t_3}
\,, \qquad
b = \frac{\mu t_1 - \mu_0 (t_2 + t_3)}{t_1-t_2-t_3}\,,
\end{align}
The function $\mu_\mathrm{run}(\taun,\mu)$ quadratically approaches  $\mu_0$ 
below $t_1$   and $\mu$ above $t_2$, and it is linearly increasing from $t_1$ to $t_2$.
The continuity of $\mu_\mathrm{run}(\taun,\mu)$ and its derivative at $t_1$ and $t_2$ 
determines $a$ and $b$.

\begin{figure}[t]{
     \includegraphics[width=\columnwidth]{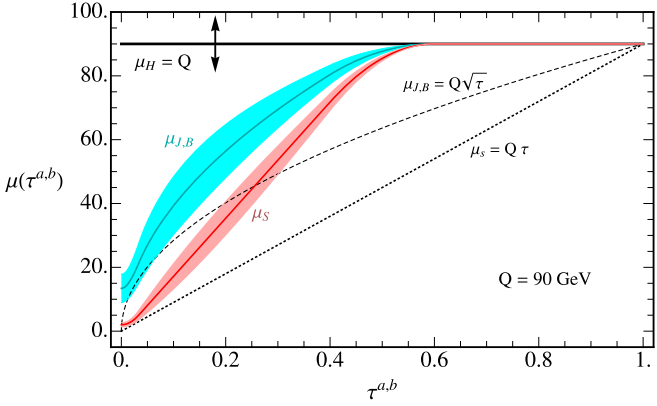}   
     \vspace{-2em}
    { \caption[1]{$\taumB$ profile functions for the scales $\mu_H$, $\mu_{B,J}(\taun)$, $\mu_S(\taun)$ with $Q=90$ GeV used in the resummed factorized cross section \eq{resummedtaumB}. The double arrow and the colored bands illustrate the scale variations in \eq{scalevariations} used to obtain theoretical uncertainty estimates.}
  \label{fig:profile}} }
  \vspace{-1em}
\end{figure}

The default values of parameters we will use for what we will consider the ``central values'' of the $\taumB$  cross sections are: 
\be
\label{profileparam}
\begin{split}
\mu&=Q\,,\quad e_{B,J}=e_S=0\,,\quad \mu_0=2~\mathrm{GeV}\,,\quad 
\\
t_1&=\frac{3~\mathrm{GeV}}{Q}\,,\quad t_2=0.4\,,\quad t_3=0.6\,.
\end{split}
\ee
To estimate theoretical uncertainty due to missing higher order terms in
fixed-order and resummed perturbation theory, we vary the parameters $\mu$,
$e_{B,J}$, and $e_S$ from their default values by ${\cal O}(1)$ factors in order
to vary corresponding scales $\mu_H$, $\mu_{B,J}$, and $\mu_S$ by ${\cal O}(1)$
factors, respectively. We separately vary the parameters one by one and keep the
others at their default values.  The total number of variations we perform
around the central values are as follows:
\begin{subequations}
\label{scalevariations}
\begin{align}
1) \quad \mu &= 2^{\pm 1} Q\,,\quad e_{B,J} = 0\,,\quad e_S = 0 \\
2) \quad \mu &= Q \,, \quad e_{B,J} = \pm \frac{1}{3},\pm\frac{1}{6}\,,\quad e_S = 0 \\
3) \quad \mu &= Q\,, \quad e_{B,J} = 0\,,\quad e_S = \pm\frac{1}{3},\pm\frac{1}{6}\,.
\end{align}
\end{subequations}
Variation 1 moves all the scales in \fig{profile} together up and down by
factors of 2, and corresponds to the scale variation used to estimate the fixed
order theoretical uncertainty in perturbation theory. Variations 2 and 3 are
additional variations we are able to perform because of having independent
$\mu_{J,B}$ and $\mu_S$ scales in the resummed cross section \eq{resummedtaumB}
and give an estimate of the uncertainty at each order in
logarithmic accuracy in resummed perturbation theory that can not be achieved by
varying the single scale $\mu$. Variation 1 alone underestimates the total
uncertainty. 

The size of the cross section at a given value of $\taun$ may not vary
monotonically with $e_{J,B},e_S$, and ideally we would vary them continuously
within some finite band to find the maximum uncertainty. The four values we test
for $e_{J,B},e_S$ in \eq{scalevariations} are a discrete approximation to such a
procedure that remains computationally tractable. We take the largest and
smallest values of the cross section among these points and use them to define
the width of the uncertainty band from $e_{J,B}$ or $e_S$ variation.

To make a conservative estimate of the total uncertainty, we sum in quadrature
the uncertainties we get from variations 1, 2, and 3 individually.  We find that
the total size of the bands provided by \eq{scalevariations} are reasonable
estimates of the theoretical uncertainty when we compare the cross sections at
different orders of logarithmic accuracy.

\begin{figure}[t!]{
     \includegraphics[width=\columnwidth]{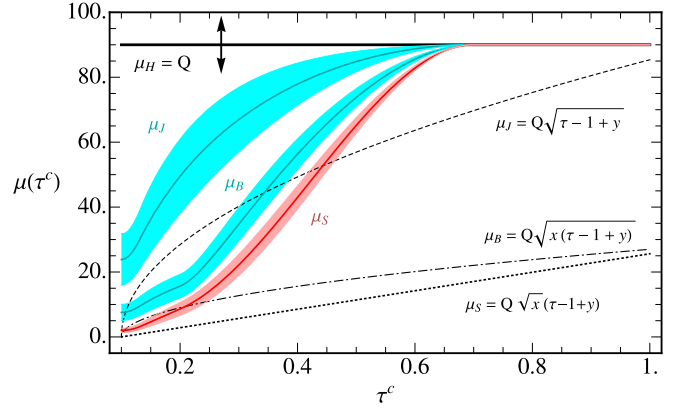}   
     \vspace{-2em}
    { \caption[1]{$\tauCM$ profile functions for the scales $\mu^\CM_H$, $\mu^\CM_{B,J}(\taun)$, $\mu^\CM_S(\taun)$ with $x=0.1$, $y=0.9$, and $Q=90$ GeV, along with the simple canonical scales \eq{canonicalscalesCM}. The double arrow and colored bands illustrate the scale variations in \eq{scalevariations}.}
  \label{fig:profileCM}} }
  \vspace{-1em}
\end{figure}

\fig{profile} shows profile functions for $\mu_H$, $\mu_{B,J}(\taun)$,
$\mu_S(\taun)$ with $Q=90$ GeV.  The solid lines are the central values of the
scales with default values in \eq{profileparam}, the double-headed arrow implies
variation 1 and the bands represent variations 2 and 3 in \eq{scalevariations}.
The dashed and dotted lines are the canonical scales in \eq{canonicalscales}.

\subsubsection{$\tauCM$ profile functions}
\label{sssec:profileCM}

For $\tauCM$, the canonical scales in \eq{canonicalscalesCM} are 
\begin{align}
\mu_H&\sim Q\,, &
\mu_S&\sim Q\sqrt{x}(\taun-1+y)
\,,\nn\\
\mu_{B}&\sim Q\sqrt{x(\taun-1+y)}
\,,&
\mu_{J}&\sim Q\sqrt{\taun-1+y}
\,,
\end{align}
where they satisfy the relation $\mu_{B,J}^2=x^{\pm1/2} \mu_H \mu_S$
 Compared to the canonical scales for
$\taumB$ in \eq{canonicalscales}, there are two differences in the canonical
scales for $\tauCM$.  First, $\taun$ is replaced by $\taun-(1-y)$ because the
transverse momentum of jet is nonzero, which is $(1-y)Q^2$ at tree level and the
projection onto $q_J^\CM$ differs from the projection onto jet axis by
$(1-y)Q^2$.  This requires that canonical scales in \eq{canonicalscales} and
profile in \eq{profiles} are shifted by $1-y$.  Second, the soft scale and beam
scale are multiplied by $\sqrt{x}$ because of rescaling of the beam axis from
$xP$ for $\taumB$ to $P$ for $\tauCM$.

In this paper we consider the case $\sqrt{x}\sim {\cal O}(1)$ and this factor
changes the scales by ${\cal O}(1)$, which is the size of perturbative
uncertainties from varying $\mu$, $e_{B,J}$ and $e_S$.  This means that
multiplying $\mu_{B}$ and $\mu_S$ by $\sqrt{x}$ in \eq{profiles} should not make
a difference within the perturbative uncertainty.  So, we could use the profile
in \eq{profiles} but shifted by $1-y$.  On the other hand, by modifying the
profile the canonical relations among the scales $\mu_{H,J,B,S}$ for $\tauCM$
\eq{canonicalscalesCM} can be maintained and we can account for the extra
factors of $\sqrt{x}$.  Therefore, for these profiles we define
$\mu^\CM_{H,J,B,S}$ as
\begin{align}
\label{profilesCM}
&\mu_H^\CM =\mu\,,\quad  \mu^\CM_{B,J,S}(\taun)=\mu_{B,J,S}(x,\taun-1+y)
\,,\\[5pt] &
 \mu_{J}(x,\taun) = \! \Bigl[1\plus e_{J}\theta(t_3 \minus \taun)\Bigl(1 \minus \frac{\taun}{t_3}\Bigr)^{\! 2} \Bigr] \!\sqrt{\!\mu \,\mu_\mathrm{run}^\CM(x,\taun,\mu,0)}
, \nn \\ &
\mu_{B}(x,\taun) = \!\Bigl[1\plus e_{B}\theta(t_3\minus \taun)\Bigl(1 \minus \frac{\taun}{t_3}\Bigr)^{\! 2} \Bigr] \!\sqrt{\!\mu \,\mu_\mathrm{run}^\CM(x,\taun,\mu,1)}
, \nn \\ &
\mu_{S}(x,\taun) = \! \Bigl[1\plus e_{S}\theta(t_3\minus\taun)\Bigl(1-\frac{\taun}{t_3}\Bigr)^2 \Bigr] 
\mu_\mathrm{run}^\CM(x,\taun,\mu,1/2).\nn
\end{align}
The $\mu_\mathrm{run}^\CM$ used here depend on $x$ and index $0,1,2$ that is different for $\mu_{J}$,
$\mu_{B}$, $\mu_{S}$.  We want $\mu_\mathrm{run}^\CM(x\,,\taun\,,\mu\,,n)\sim x^{n}
\taun \mu$ with $n=0,1/2,1$ so that the canonical scaling for $\mu_{J,S,B}$ in
\eq{canonicalscalesCM} is respected in the small $\taun$ region.  In the large
$\taun$ limit, $\mu_\mathrm{run}^\CM(x\,,\taun\,,\mu\,,n)$ should go to $\mu$, so
that $\mu_S$ and $\mu_{B,J}$ both go to $\mu$.

As in \eq{murun} $\mu_\mathrm{run}^\CM$ should run linearly between $t_1$ and $t_2$.
However, the slope of $\mu_\mathrm{run}^\CM$ in \eq{profilesCM} should be different
for the three cases $n=0,1/2,1$.  Therefore, we cannot use \eq{murun} to define $\mu_\mathrm{run}^\CM$  because all parameters in $\mu_\mathrm{run}$ are fixed by matching
boundary conditions and the slope is fixed.  Instead, by replacing the quadratic
polynomial in \eq{murun} by a cubic polynomial one can introduce a free
parameter and this parameter can be chosen such that
$\mu_\mathrm{run}^\CM(x\,,\taun\,,\mu\,,n)\sim x^{n} \taun \mu$ between $t_1$ and
$t_2$.  We define $\mu_\mathrm{run}^\CM$ as
\begin{align}
\label{murunCM}
& \mu_\mathrm{run}^\CM(x\,,\taun\,,\mu\,,n) =
\begin{cases}
x^{n-\frac{1}{2}} \mu_0 + a(n)\,\taun^2/t_1 & \taun \leq t_1
\,,\\
2a(n)\, \taun + b(n) & t_1 \leq \taun \leq t_2
\,,\\
\mu_{\text{cubic}}(x\,,\taun\,,\mu\,,n)& t_2 \leq \taun \leq t_3
\,,\\
\mu & \taun > t_3
\,,\end{cases}
\nn \\[5pt]
& 
\mu_{\text{cubic}}(x\,,\taun\,,\mu\,,n) =
\mu \minus  c(n) \left(\frac{\taun\minus t_3}{t_3 \minus  t_2}\right)^{\!2} \minus  d(n) \left(\frac{\taun\minus t_3}{t_3\minus  t_2}\right)^{\!3} 
 \nn \\&
b(n) = x^{n-1/2}\mu_0-a(n) t_1\,,
\nn \\
&
c(n)=3(\mu-x^{n-1/2}\mu_0)-a(n)(2t_3+4t_2-3t_1)
\,,\nn \\
&
d(n)=2(\mu-x^{n-1/2}\mu_0)-2a(n)(t_3+t_2-t_1) 
\,.  \end{align}
Here the parameters $b(n),~c(n),~d(n)$ are determined by continuity of
$\mu_\mathrm{run}$ and its derivative at $t_1,~t_2,~t_3$. The slope $a(n)$ is a
free parameter which is chosen to satisfy $a(n)\sim x^{n} \mu$ to achieve
canonical scaling of jet, beam, and soft scales:
\be\label{an}
a(n)=x^n \,\frac{\mu-x^{-1/2}\mu_0}{t_3+t_2-t_1}.
\ee
Note that in $x\to1$ limit, \eq{murunCM} reduces to \eq{murun} and profiles for
$\tauCM$ in \eq{profilesCM} reduce to the profiles in \eq{profiles} for $\taum$
and $\tauB$.

We choose the same default parameters and scale variations as for $\taumB$
in \eqs{profileparam}{scalevariations} except for $t_2$:
\be\label{tau2CM}
t_2=0.1
\,.
\ee
Because of the different definition of the profiles for $\tauCM$ this value of
$t_2$ must be smaller than the value for the $\taumB$ profiles. This occurs
because $\mu_\mathrm{run}$ in \eq{murunCM} changes faster than that the
$\mu_\mathrm{run}$ in \eq{murun} between $t_2$ and $t_3$. As can be seen from
\fig{profileCM} the final profiles for $\mu_S$ have similar shapes.

\fig{profileCM} shows $\tauCM$ profile functions for $\mu^\CM_H$,
$\mu^\CM_{B,J}(\taun)$, $\mu^\CM_S(\taun)$ defined in \eq{profilesCM}
with $x=0.1$, $y=0.9$, and $Q=90$ GeV.  The solid lines are the
central values of the scales with default values in \eq{tau2CM} for
$t_2$ and in \eq{profileparam} for all other parameters.  The
double-headed arrow represents variation 1 and the uncertainty bands
are variations 2 and 3 in \eq{scalevariations}.  The dashed, dotted,
and dotted-dashed are the canonical scales in \eq{canonicalscalesCM}.

\subsection{Nonperturbative Soft Function}
\label{ssec:NPfunction}

The hemisphere soft function defined in \eq{Shemiprojection} describes soft
radiation between jets at the nonperturbative scale $\Lqcd$ as well as at
perturbative scales above $\Lqcd$.  The results given in
\eqs{softNLO}{softRGEsolution} are valid in the perturbative region.  In the
$\overline{\rm MS}$ scheme the soft function valid at both scales is given by a
convolution between a purely perturbative function $\Shemi^\mathrm{pert}$ and a
nonperturbative model function $F$~\cite{Hoang:2007vb}:
\be
\label{NPsoftfunction}
\Shemi(k,\mu) = \int dk'\, \Shemi^{\text{pert}}(k-k',\mu)\, F(k')\,.
\ee
The function $F(k)$ contains information about physics at the nonperturbative
scale and has support for $k\sim \Lambda_{\rm QCD}$, falling off exponentially
outside this region.  Inserting \eq{NPsoftfunction} into the factorization
formula in \eq{taucsfinal} one obtains the convolved form for the cross section:
\be
\label{NPcrosssection}
\begin{split}
\frac{d\sigma(\taun)}{d\taun}=\int dk\,\frac{d\sigma^\mathrm{pert}}{d\taun} \left(\taun- \frac{k}{Q_R}\right)\,F(k)\,,
\end{split}
\ee
where $d\sigma^\mathrm{pert}/d\taun$ is the cross section calculated by using
only the perturbative soft function and $Q_R$ is given by \eq{QR}.
\eq{NPcrosssection} correctly describes both the peak region $Q_R\tau_1
\sim \Lambda_{\rm QCD}$ where the entire function $F(k)$ is required, as well
as the tail region $Q_R\tau_1 \gg  \Lambda_{\rm QCD}$ where only its first moment
is required since we can expand in $\Lambda_{\rm QCD}/(Q_R\tau_1)$.

For the peak region, various ways to parametrize models for $F(k)$ have been
proposed \cite{Korchemsky:2000kp,Hoang:2007vb,Ligeti:2008ac}. We will adopt one
proposed in \cite{Ligeti:2008ac} that expands $F$ systematically in an infinite
set of basis functions:
\be
\label{Fbasis}
F(k)=\frac{1}{\lambda}\left[\sum_{n=0}^{N} c_n f_n\left(\frac{k}{\lambda} \right) \right]^2\,,
\ee
where in principle we can choose any complete basis of functions $f_n$.  We
adopt the same basis that has already been used in
\cite{Ligeti:2008ac,Abbate:2010xh}, and exhibits fast convergence of the
expansion.  The normalization condition $\int dk \, F(k) =1$ gives the
constraint $\sum_{i} c_i^2=1$.  The characteristic scale $\lambda$ of size
${\cal O}(\Lqcd)$ is an additional parameter if the sum is truncated at finite
$N$, as we will do in practice.  

In the tail region where $Q_R \taun\gg \Lqcd$, \eq{NPcrosssection} is
consistent with the power correction from an operator product expansion (OPE),
\begin{align}
\label{OPEcrosssection}
\frac{d\sigma(\taun)}{d\taun}&= 
 \biggl\{\frac{d\sigma^\mathrm{pert}(\taun)}{d\taun}
 -\frac{2\Omega_1^{a,b,c}}{Q_R}\,\frac{d^2\sigma^\mathrm{pert}(\taun)}{d\taun^2}\biggr\}
 \\
 &\quad \times \biggl[ 1+ \cO\left( \frac{\as\Lqcd}{Q\taun}\right) + \cO\left(\frac{\Lqcd^2}{Q^2\taun^2 }\right) + \cdots \biggr]\,. \nn
\end{align}
To lowest order in $\Lambda_{\rm QCD}/(Q\tau_1)$ this result agrees with a simple shift
$\tau_1\to \tau_1-2\Omega_1/Q_R$.  Here the coefficient of the power correction
$2\Omega_1^{a,b,c}$ is a nonperturbative matrix element and it corresponds to the
first moment of the nonperturbative function $\int dk\, k\, F(k)$ which could in
principle differ for each of $\taumBCM$. The first set of power corrections
indicated on the second line of \eq{OPEcrosssection} comes from perturbative
corrections to the leading power correction \cite{Mateu:2012nk}, and the second
set involves purely nonperturbative corrections at subleading order. In the next
section we will consider the question of universality of the $\Omega_1^{a}$,
$\Omega_1^b$, $\Omega_1^c$ parameters for the observables $\taum$, $\tauB$,
$\tauCM$.

In the peak region the parameters $c_i$ and $\lambda$ should be determined by
fitting to experimental data. Since data is not yet available, our only purpose
here will be to get an idea of the impact of the nonperturbative shape function.
We take the simplest function $F(k)$ with $N=0$.  Then, $c_0=1$ by normalization
and $\lambda$ is the only parameter. To get the right first moment, we require
$\lambda=2\, \Omega_1$.  We use $\Omega_1 = 0.35$~GeV, which is determined from
measurements of $e^+ e^-\to \mathrm{dijets}$ \cite{Abbate:2010xh}. However,
$\Omega_1$ in DIS is not necessarily the same as in $e^+e^-$ collisions, and we
merely consider this to be an illustrative but reasonable value.

\subsection{Universality Classes for $\Omega_1$ Parameters Defined with Different 
Directions}
\label{ssec:universality}

The various versions of 1-jettiness $\taumBCM$ or the generic version
\eq{tau1def} depend on different choices of the axes $q_J=\omega_J n_J/2$ and
$q_B=\omega_B n_B/2$. In this section we will show that the $1$-jettiness power
correction parameter is universal under changes to the axes used in its
definition, by exploiting properties of operators~\cite{Lee:2006fn,Lee:2006nr}
and including hadron mass effects~\cite{Salam:2001bd,Mateu:2012nk}.

If we use different axes for the decomposition of four-momenta then they can all
be written in a form similar to the event shapes given in \cite{Mateu:2012nk}:
\be
\label{tau1mY}
\taun = \frac{1}{2Q_R} \sum_i m_i^\perp f(r_i,{\cal Y}_{JB}^i)\,,
\ee
where $Q_R$ is defined in \eq{QR}, $i$ sums over hadrons, and $m_i^\perp, r_i,
\cY_{JB}^i$ are defined with respect to the vector $q_{J,B}$ by:
\be
\label{mrYdefs}
m^\perp \equiv \sqrt{\vect{p}_\perp^2 + m^2}\,, \ r_i \equiv \frac{p_\perp}{m^\perp}\,,\quad \cY_{JB} \equiv \frac{1}{2}\ln \frac{q_B\mcdot p}{q_J\mcdot p}\,,
\ee
where $m$ is the mass of the hadron whose momentum is $p^\mu$.  For
the 1-jettinesses $\taun$ given in \eq{tau1def} we have
\be \label{ftau1}
f(r,\cY) = e^{-\abs{\cY}}\,.
\ee
For each different $\taun$, i.e. each choice of $q_{J,B}$, the definition of
$m^\perp$ and $\cY_{JB}$ change since they are computed with different
coordinates. The $Q_R$ also depends on $q_J\cdot q_B$, as given in \eq{QR}.

Following the logic in \cite{Mateu:2012nk} for massive hadrons and
\cite{Lee:2006fn,Lee:2006nr} for massless particles, the leading power
correction in the expansion \eq{OPEcrosssection} of distributions in event
shapes of the form \eq{tau1mY} is always described by the nonperturbative matrix
element
\begin{align}
\label{rYintegral}
2\Omega_1^{a,b,c} &= \int_0^1 \! dr\int_0^\infty \!\! d\cY_{JB} f(r,\cY_{JB}) \\
&\quad\times\bra{0} Y_{n_B}^\dag Y_{n_J} \hat\cE_T^{a,b,c}(r,\cY_{JB})
 Y_{n_J}^\dag Y_{n_B}\ket{0}\,.\nn 
\end{align}
Here $\cE_T$ is a ``transverse velocity operator'' defined as in
\cite{Mateu:2012nk}, but now using the axes given by $q_J$ and $q_B$, 
\be
\hat\cE_T(r,\cY_{JB})\ket{X}
  = \sum_{i\in X} m_i^\perp \delta(r-r_i) \delta(\cY_{JB} - \cY_{JB}^i) \ket{X}\,.
\ee
It measures the total transverse mass of particles flowing in a slice of
velocity and rapidity around $r$ and $\cY_{JB}$. 
Now consider making an RPI-III transformation \cite{Manohar:2002fd} in the matrix element in \eq{rYintegral} which takes
$n_J\to n_J/\zeta$ and $n_B\to \zeta n_B$. This transformation leaves the vacuum
and the Wilson lines $Y_{n_J}$ and $Y_{n_B}$ invariant, but shifts
$\cE_T(r,\cY_{JB})$ to $\cE_T(r,\cY_{JB} + \cY')$ where $\cY'=\ln\zeta$. This is
the analog of the boost argument for back-to-back $n_J$ and $n_B$ in
Ref.~\cite{Lee:2006fn,Lee:2006nr}. Thus, the matrix element inside the integral
in \eq{rYintegral} is independent of $\cY_{JB}$, and we can integrate over
$\cY_{JB}$ to obtain the power correction $\Omega_1^{a,b,c}$ for $\taumBCM$, using
the $f$ given in \eq{ftau1}:
\be
\label{Omega1def}
 \biggl[\int_{-\infty}^\infty d\cY_{JB}\: f(r,\cY_{JB})\biggr] \Omega_1^{JB}(r,\mu) 
 = 2\, \Omega_1^{JB}(r,\mu)\,,
\ee
where the renormalized matrix element is
\be
\label{Omega1r}
\Omega_1^{JB}(r,\mu) = \bra{0} Y_{n_B}^\dag Y_{n_J} 
 \hat\cE_T(r,0) Y_{n_J}^\dag Y_{n_B}\ket{0}
 \,.
\ee
This matrix element still depends on the choices of axes through $n_{B,J}$. By
rescaling $n_J$ and $n_B$ as in \eq{softrescaled} we find it is independent of
$n_J\cdot n_B$. It still depends on these axes through the parameter $r$, since
the transverse momenta $p_\perp$ inside $r$ depends on the choice of these axes.
 
However, in the tail region the $\Omega_1^{JB}(r)$ always appears inside an
integral. At LL order we have the resummed coefficient $C_1^{\rm LL}(k,r,\mu)$
from~\cite{Mateu:2012nk} for any $\tau_1$ and the shape function OPE is
\begin{align} \label{Fope}
  F(k) &= \delta(k) +\! \int_0^1\!\! dr\, C_1^{\rm LL}(k,r,\mu) 2 \Omega_1^{JB}(r,\mu) 
 + {\cal O}\Big(\frac{\Lambda_{\rm QCD}^2}{k^3}\Big) 
  \nn\\ 
  & =  \delta(k) +\! \int_0^1\!\! dr\, C_1^{\rm LL}(k,r,\mu) 2 \Omega_1(r,\mu) 
+ {\cal O}\Big(\frac{\Lambda_{\rm QCD}^2}{k^3}\Big) 
   \,,
\end{align}
where in the second line we removed the ${}^{JB}$ superscript on $\Omega_1$ by using the fact that the only axis dependence occurs
through the parameter $r$ which is now just a dummy variable. It would be
interesting to consider the universality beyond LL order for this Wilson
coefficient.

Thus we see that at least to LL order there is a universal power correction
$\Omega_1 (r)$ for all three versions of $1$-jettiness, $\taumBCM$. Taking the
tree level result $C_1^{\rm LL}(k,r,\mu)\to -\delta'(k)$ yields
\eq{OPEcrosssection} and leads to the identification
\begin{align}
 \Omega_1^{a,b,c}= \int_0^1\! dr\ \Omega_1(r) \,.
\end{align}
Eq.~(\ref{Fope}) also implies universality of the shift parameter appearing in
\eq{OPEcrosssection} for $\taumBCM$:
\begin{align}
 \Omega_1^a=\Omega_1^b=\Omega_1^c \,.
\end{align}

\section{Results}
\label{sec:results}

In this section we present our numerical results for the three versions of DIS
1-jettiness: $\taum$, $\tauB$, and $\tauCM$.  We plot the cross sections
accurate for small $\taun$ resummed from LL to NNLL accuracy, and also the
singular terms at fixed order $\cO(\as)$ (NLO) for comparison. (We estimate the
size of the small missing non-singular terms by comparing to the known
$\cO(\as)$ cross section integrated over all $\taun$.)  We start by describing
the $\taum$ spectrum in detail, and then compare the features of the $\tauB$ and
$\tauCM$ cross sections relative to the results for $\taum$.  We choose
$s=(300~\text{GeV})^2$ as in the H1 and ZEUS experiments.  For the PDFs, we use
the MSTW2008 \cite{Martin:2009iq} set at NLO and include five quark and
antiquark flavors excluding top.  To be consistent with the $\alpha_s$ used in
the NLO PDFs we use the 2-loop beta function for running $\alpha_s$ and
$\alpha_s(m_Z)=0.1202$.

We present results for the cumulant cross section $\sigmac(\taun)$ defined in
\eq{sigmac} and the dimensionless distribution 
\be
\label{dsigma}
\frac{d\hat\sigma}{d\taun}= \frac{1}{\sigma_0} \frac{d\sigma}{d\taun}
 = \frac{d}{d\taun} \sigmac(\taun) \,.
\ee
Note that both the cumulant $\sigmac(\taun)$ and the differential
distribution $d\hat\sigma/d\taun$ are differential in $x$ and $Q^2$. However, for notational
simplicity we made their $x$ and $Q^2$ dependences implicit in this section.


%

\subsection{$\taum$ cross section}

In this subsection, we present results for the cumulant cross section
$\sigmac(\taun)$ and differential cross section $d\hat\sigma/ d\taun$ for the
``aligned'' 1-jettiness $\taun=\taum$.

\begin{figure}[t!]{
      \includegraphics[width=\columnwidth]{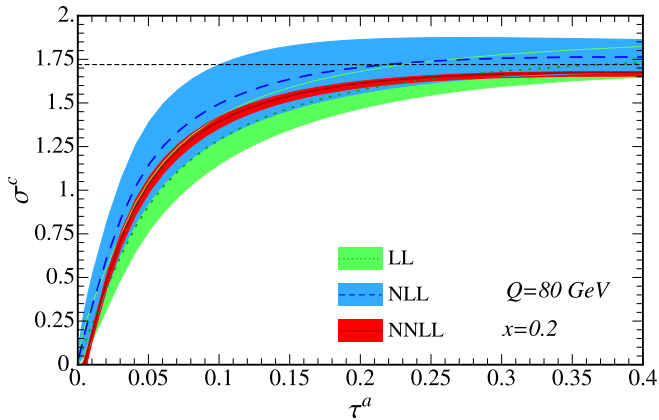}
      \vspace{-2em}
       { \caption[1]{Cumulant cross section in $\taum$ at $Q=80\GeV$ and $x=0.2$. Colored bands show theoretical uncertainties around central values (lines) to LL (dotted line, green band), NLL (dashed line, blue band), and NNLL (solid line, red band) accuracy and the horizontal dashed line is the total cross section at fixed $x,Q^2$.}
  \label{fig:taumcumulant}} }
    \vspace{-1.5em}
\end{figure}

\fig{taumcumulant} shows the $\taum$ cumulant cross section, defined by
\eq{sigmac}, at $Q=80\GeV$ and $x=0.2$. In order to illustrate perturbative
convergence the results resummed to LL, NLL, and NNLL accuracy are shown.  The
bands indicate perturbative uncertainties by varying the scales $\mu_{H,B,J,S}$
given by ``profile functions'' as described in \sssec{profiles}, and there is
excellent order-by-order convergence, and beautiful precision at NNLL order.
The cumulant cross section increases monotonically from the small $\taum$
region and begins to saturate near for large $\taum$ where the integral
defining this cumulant becomes that for the total cross section.  There is a
small gap between the total cross section at ${\cal O}(\alpha_s)$ (dashed
horizontal line) and our NNLL cumulant at large $\taum$, reflecting the small
size of nonsingular terms not taken into account in this paper. Note however
that these terms are important at the level of precision of our cumulant cross
section, and hence they will be considered in the future.

We can characterize the $d\hat\sigma/d\taum$ cross section by three distinct
physical regions: the peak region ($\taum\sim 2\Lqcd/Q$), the tail region
($2\Lqcd/Q\ll \taum \ll 1$), and the far-tail region ($\taum\sim{\cal O}(1)$).
We will do this with four plots.  We first show the purely perturbative cross
section to study convergence and the impact of resummation compared to fixed
order results. Next we show the impact of nonperturbative effects, which in the
tail region produce a simple shift in the distribution, and have a significant
impact on the shape of the spectrum in the peak region.  We also illustrate the
dependence of the cross section on $x$ and $Q^2$ at fixed $\taum$.

\begin{figure}[t!]{
    \includegraphics[width=\columnwidth]{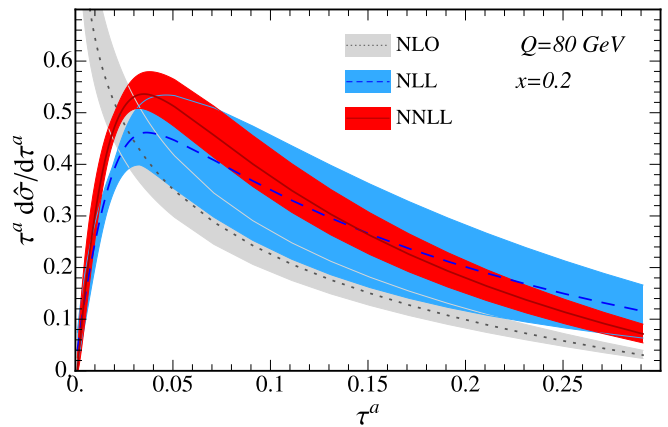} \vspace{-4ex}
    { \caption[1]{Weighted differential cross section in $\taum$ at $Q=80\GeV$
        and $x=0.2$.  Colored bands show theoretical uncertainties around
        central values (lines) at fixed order $\alpha_s$ (dotted line, gray
        band) and resummed to NLL (dashed line, blue band) and NNLL (solid line,
        red band) accuracy.}
  \label{fig:taumdistribution}} }
  \vspace{-1em}
\end{figure}

\fig{taumdistribution} shows the weighted differential cross section $\taum\,
d\sigma/ d\taum$ at $Q=80\GeV$ and $x=0.2$.  The results are weighted by $\taum$
for better visibility because the differential cross section falls very rapidly
with $\taum$.  In the tail region, the overlap in resummed results shows a good
perturbative convergence from NLL to NNLL. The large deviation between NLO and
NNLL shows the large effect of resummation and the underestimated uncertainty of
a pure fixed-order result.  In the peak region, NLO result blows up as
$(\ln\taun)/\taun$, while the NLL and NNLL results converge into a peak due to
resummation of the large logs to all orders in $\as$. Again the uncertainty
bands overlap fairly well. In the far-tail region for larger $\taun$, the
resummation effect becomes small and the size of the deviation is reduced.  Near
the far-tail region ($\taun\sim 0.3$), the NNLL curve begins to depart from the NLL
band.  In this region the nonlogarithmic $\alpha_s^2$ term and nonsingular terms
neglected in our NNLL result may begin to be significant.

\begin{figure}[t!]{
    \includegraphics[width=\columnwidth]{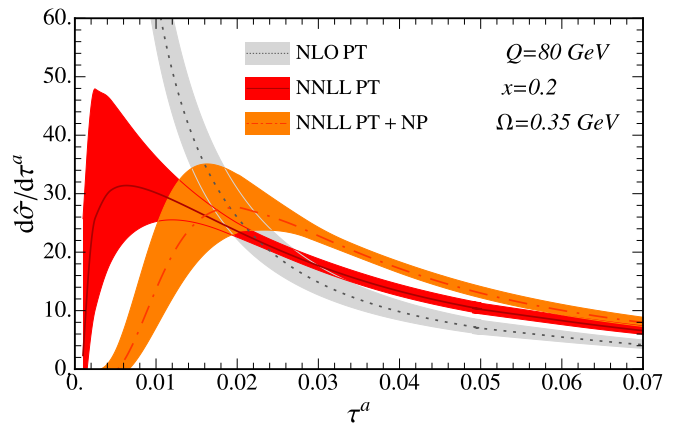}
    \vspace{-4ex} { \caption[1]{Differential cross section in $\taum$ at
        $Q=80\GeV$ and $x=0.2$ in the peak region, NNLL with nonperturbative
        shape function taken into account (NNLL PT+NP, dashed, orange), and
        without NP shape function at fixed-order $\alpha_s$ (NLO PT, dotted,
        gray) and resummed (NNLL PT, solid, red).}
  \label{fig:taumNP}} }
\end{figure}

\fig{taumNP} shows the differential cross section $d\sigma/ d\taun$ at
$Q=80\GeV$ and $x=0.2$ in the peak region at fixed order and NNLL resummed
accuracy.  Note that it is not scaled by $\taun$ as in \fig{taumdistribution}.
In this plot, the NNLL result convolved with a nonperturbative shape function
(NNLL PT + NP) is shown in comparison with purely perturbative fixed-order NLO
and resummed NNLL results (NLO PT and NNLL PT).  As discussed in
\ssec{NPfunction} we use the simplest shape function with one basis function
$N=0$ in \eq{Fbasis} with a reasonable choice $\Omega_1=0.35\GeV$ for the value
of the first moment just to illustrate the impact of the nonperturbative
effects.  For practical analysis, a shape function with more basis functions
should be used and the parameters $c_i,\lambda$ in the model function
\eq{Fbasis} should be determined from experimental data.  In the endpoint
region, there is significant change from NLO and NNLL due to the resummation of
large perturbative logs, and there is another large change from perturbative
NNLL to the result convolved with the shape function due to nonperturbative
effects.  As we move into the tail region, the size of nonperturbative
correction reduces to ${\cal O} (\Lqcd/\taun Q)$ and the correction simplifies
to the power correction in \eq{OPEcrosssection}.

\begin{figure}[t!]{
\vspace{-1ex}
      \includegraphics[width=\columnwidth]{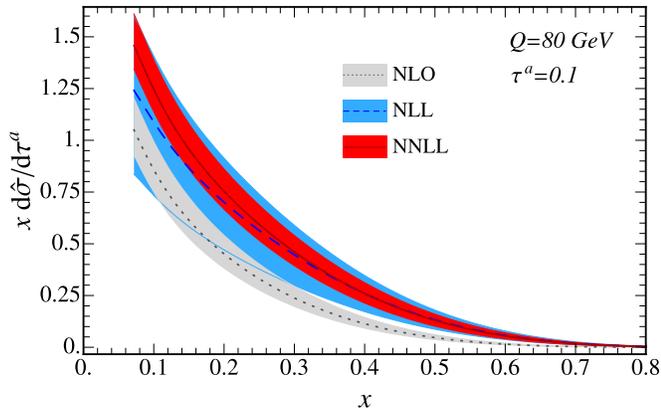}
      \vspace{-2em}
       {\caption[1]{$x$ dependence of $\taum$ differential cross section at $Q=80\GeV$ and $\taum=0.1$. Colored bands show theoretical uncertainties around central values at fixed-order $\alpha_s$ (dotted, gray) and resummed to NLL (dashed, blue) and NNLL accuracy (solid, red).}
       \vspace{-1em}
  \label{fig:taumx}} }
\end{figure}

\fig{taumx} shows the weighted differential cross section $x\, d\sigma/
(dx\,dQ^2\, d\taun)$ as a function of $x$ at $Q=80\GeV$ and $\taum=0.1$.  Note
that the lower bound $x \ge Q^2/s$ is set by the relation $xys=Q^2$ in \eq{y}
and the constraint $y\leq 1$.  The $x$ dependence comes from the quark and
anti-quark beam functions and the decreasing curves with increasing $x$ are
characteristic patterns of PDFs contained in the beam function.  With decreasing
$x$, NLO and NNLL curves rise faster than NLL curve because they contain the
gluon PDF, which rises faster than the quark PDF, and whereas the NLL result
only contains the tree-level beam function which is just the quark PDF.

\fig{taumQ} shows the $Q$ dependence of the differential cross section at
$x=0.2$ and $\taum=0.1$.  Overall, $Q$ dependence is mild.  In the naive parton
model the cross section is insensitive to $Q$ because of the approximate scaling
law in the Bj\"{o}rken limit where $Q,s\to \infty$ with $x$ fixed.  This scaling
is broken by logarithms of $Q$ in QCD.  It is also broken by the $Z$ boson mass
with the factors $1/(1+m_Z^2/Q^2)$ in \eq{Lqm}.  As shown in the plot, well
below $m_Z= 91.2\GeV$ the curves vary gently in $Q$ and near and above $m_Z$
they increase due to the factor $Q^2/(Q^2+m_Z^2)$.


\subsection{$\tauB$ cross section}

\begin{figure}[t!]{
      \includegraphics[width=.95\columnwidth]{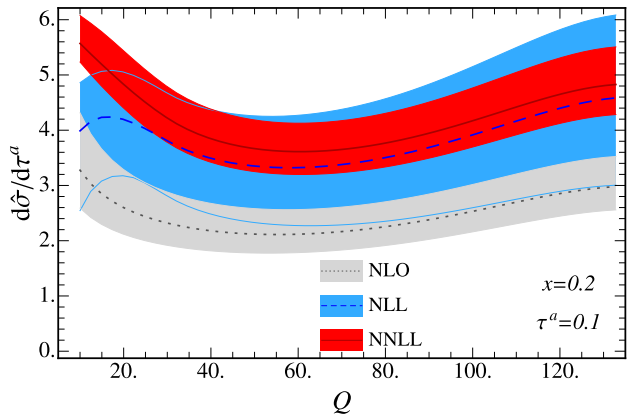}
\vspace{-1.5ex}
       { \caption[1]{$Q$ dependence of $\taum$ differential cross section at $x=0.2$ and $\taum=0.1$, with theoretical uncertainties at fixed-order $\alpha_s$ (dotted, gray) and resummed to NLL (dashed, blue) and NNLL accuracy (solid, red)}
   \vspace{-1em} \label{fig:taumQ}} }
\end{figure}

\begin{figure}[t!]{
      \includegraphics[width=\columnwidth]{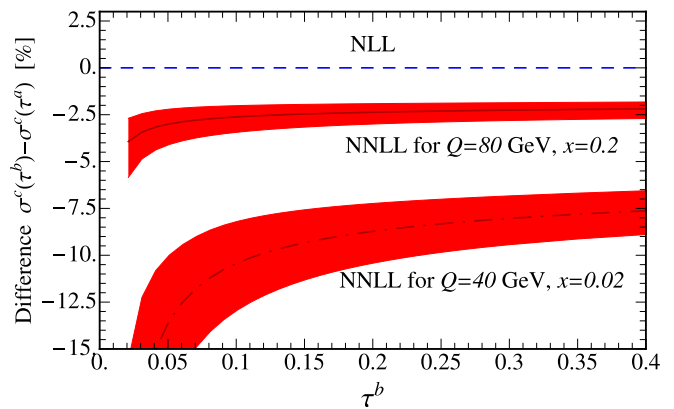}
      \vspace{-2em}
       { \caption[1]{Difference between $\tauB$ and $\taum$ cumulant cross sections at $Q=80\GeV$ and $x=0.2$ and at $Q=40\GeV$ and $x=0.02$. The difference at NLL is zero for both parameter sets.}
  \label{fig:tauB}} }
\end{figure}

The $\tauB$ cumulant cross section is different from $\taum$ by a single term at
NLO in \eq{resummedtaumB}.  The term contains $\ln z$ where $z$ is integrated
over from $x$ to 1, and so the term becomes larger for smaller $x$.  \fig{tauB}
shows their percent difference at NLL and NNLL for two sets of $(Q,x)$:
$(80,0.2)$ and $(40,0.02)$.  The difference at NLL is zero because at LO fixed
order $\taum$ and $\tauB$ cross section are identical and the NLL logs are the
same.  At NNLL for $x=0.2$ the size of difference is small, a few percent.  The
difference at the value $x=0.02$ is larger than that for $x=0.2$, becoming now a
10-15\% effect.  This difference is roughly constant in $Q$ because of the mild
$Q$ dependence in \fig{taumQ}.

\subsection{$\tauCM$ cross section}

\begin{figure}[t!]{
      \includegraphics[width=.95\columnwidth]{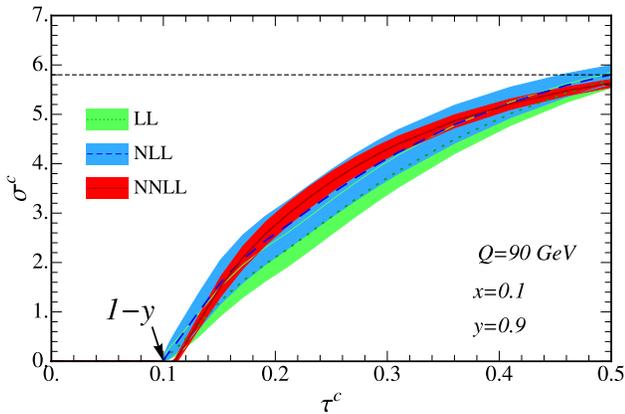}
      \vspace{-1em}
       { \caption[1]{$\tauCM$ cumulant cross section at $Q=90\GeV$ and $x=0.1$, giving $y=0.9$.
Colored bands show theoretical uncertainties around central values for resummed results to LL (dotted, green), NLL (dashed, blue), and NNLL (solid, red) accuracy. The horizontal line is the total cross section. The arrow at $1-y$ indicates the threshold in $\tauCM$ spectrum.}
\vspace{-1em}
  \label{fig:tauCM}} }
\end{figure}

The 1-jettiness $\tauCM$ is designed to measure a jet close to the $z$ axis
(incoming electron direction), and the factorization theorem for $\tauCM$ in
\eq{tauCMcs} is valid for a jet with small transverse momentum
$q_\perp^2=(1-y)Q^2$.  So, the parameters $Q$ and $x$ should be chosen such that
$1-y \ll 1$ in other words, $Q^2/(xs)\approx 1$.  The parameters in
\fig{taumcumulant} cannot be used because $y\approx 0.36$ for $Q=80\GeV$ and
$x=0.2$.  For $\tauCM$ in \figs{tauCM}{tauCMtaum} we choose $Q=90\GeV$ and
$x=0.1$ for which $y=0.9$.  Note that the profile functions for $\tauCM$ given
in \eq{profilesCM} are also different from those for $\taumB$.

\fig{tauCM} shows the cumulant $\tauCM$ cross section resummed to LL, NLL, and
NNLL accuracy.  The most notable feature in the $\tauCM$ spectrum is the
threshold $\theta(\tauCM -1+y)$ indicated by an arrow in the plot.  The
threshold is exactly respected in LL and NLL results and is effectively true at
NNLL because, although \eq{DWCM} contains terms violating this threshold at
$\cO(\as)$, their size is numerically small $(\,\sim 0.1\%)$. In the region
near this threshold nonperturbative corrections are quite important, and the
purely perturbative cross section actually has a small negative dip (almost
invisible in the plot).

\begin{figure}[t!]{
      \includegraphics[width=.98\columnwidth]{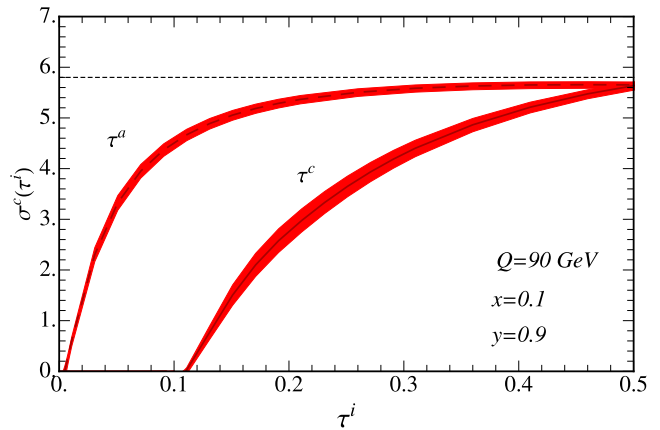}
      \vspace{-1em}
       { \caption[1]{$\tauCM$ cumulant cross section in comparison to $\taum$ result at $Q=90\GeV$ and $x=0.1$ which gives $y=0.9$. The horizontal dashed line is the total cross section at this $x,Q^2$.}
  \label{fig:tauCMtaum}} }
  \vspace{0em}
\end{figure}

\fig{tauCMtaum} shows $\tauCM$ in comparison with the $\taum$ cumulant cross
section at NNLL.  In addition to the threshold discussed in \fig{tauCM}, the
$\tauCM$ curve increases more slowly than the $\taum$ curve does.  This is
because the normalization of the $\tauCM$ axes in \eq{CMvectors} are different
from those for $\taum$.  The beam axis $\qB$ for $\tauCM$ is larger than for
$\taum$ by a factor of $1/x$ while the jet axis $\qJ$ is approximately the same in
the limit $y\to 1$.  This increases the projection of the particle momentum
$\qB\cdot p_i$ by the factor of $1/x$ in 1-jettiness \eq{tau1def}, but $\tauCM$
is not increased by quite the same factor because fewer particles are grouped into the
${\cal H}_B$ region due to the minimum in \eq{tau1def}.  Still, in \fig{tauCMtaum}
for the same value of the cross section the departure of $\tauCM$ from its
threshold is larger than that of $\taum$ due to this factor.

%

\section{Conclusions}
\label{sec:conclusions}

We have predicted 1-jettiness ($\taun$) cross sections in DIS to NNLL accuracy
in resummed perturbation theory, accurate for small $\taun$ where hadrons
 in the final state are collimated into two jets, including one from
ISR. We used three different versions of 1-jettiness, $\taumBCM$, which group
final-state hadrons into ``beam'' and ``jet'' regions differently and have
different sensitivity to the transverse momentum of ISR relative to the proton
direction.

Each $\taun$ is similar to thrust, measuring how closely final-state hadrons are
collimated along ``beam'' and ``jet'' reference axes, but with important
variations. $\taum$ measures the small light-cone momentum along two axes
aligned with the proton direction and the actual jet direction, and averages
over the transverse momentum of ISR in the calculation of the cross section.
$\tauB$ projects onto fixed axes such that the beam and jet regions are
back-to-back hemispheres in the Breit frame. The fixed jet axis is not quite
equal to the physical jet axis in the final state, causing $\tauB$ to be
sensitive to the transverse momentum $p_\perp$ of ISR and requiring a
convolution over $p_\perp$ in the jet and beam functions in the $\tauB$
factorization theorem. Finally $\tauCM$ groups final-state hadrons into
back-to-back hemispheres in the CM frame, projecting momenta onto the initial
proton and electron directions, and also requires a convolution over the
transverse momenta of the ISR and final-state jets. Furthermore, the case of
small $\tauCM$ also requires the DIS variable $y$ to be near 1.

We proved factorization theorems for all three versions of $\taun$ using the
tools of SCET, carefully accounting for the differing dependences on the
transverse momentum of ISR. These differences lead to the appearance of the
ordinary beam function in the $\taum$ factorization theorem and the generalized
$k_\perp$-dependent beam function in the $\tauB$ and $\tauCM$ factorization
theorems. We were able to relate the soft function appearing in any of these
factorization theorems in any reference frame to the ordinary DIS hemisphere
soft function by suitable rescaling of the arguments, using boost invariance.

The relevant hard, jet, beam, and soft functions and their anomalous dimensions are known to sufficiently high order that we could immediately achieve NNLL resummed accuracy in our predictions for the $\taumBCM$ cross sections (using the factorization theorems we derived). We gave predictions for the differential and
cumulant $\taun$ cross sections, illustrating the differences among $\taumBCM$
due to the different dependences on the transverse momentum of ISR. We presented
numerical predictions at $x$ and $Q^2$ values explored at the HERA collider, but
our analytical predictions can easily be applied to a much wider range of
kinematics relevant at other experiments, such as at JLab \cite{Dudek:2012vr}
and the future EIC \cite{Accardi:2012hwp} and LHeC \cite{AbelleiraFernandez:2012cc}.

The resummed predictions we presented are accurate for small values of $\taun$
where final-state hadrons are well collimated into two jets. For large $\taun$
our predictions have to be matched onto fixed-order predictions of non-singular
terms in $\taun$ from full QCD. We leave the performance of this matching at
$\cO(\as)$ and beyond to future work. However, we compared our cumulant $\taun$
cross sections for large $\taun$ to the known total cross section at fixed $x$
and $Q^2$, and found that the cumulative effect of these corrections on
the whole cross section is roughly at the several percent level for the
kinematics we considered.

To achieve higher perturbative accuracy in the overall $\taun$ distributions we
require both singular and the above-mentioned non-singular corrections to higher
order. Here we achieved NNLL resummed accuracy, but without non-singular
matching corrections needed to achieve NNLL+NLO accuracy. To go to NNLL$'$+NNLO
accuracy, we need the fixed-order hard, jet, beam, and soft functions in SCET
and non-singular terms in full QCD to $\cO(\as^2)$. These are already known for
the hard and jet functions. The soft function (known for $e^+e^-$ to
$\cO(\as^2)$ but not yet for DIS) and beam function (including both $t$ and
$\vect{p}_\perp$ dependence for $\tauBCM$) are not yet known. Once they are, we
could actually achieve N$^3$LL accuracy immediately since the necessary
anomalous dimensions are all known to sufficiently high order. In extractions of
$\as$ from $e^+e^-$ event shapes, it was found that adding another order of
accuracy in the fixed-order SCET and full QCD calculations (i.e. adding a $'$)
reduces theoretical uncertainty in the final value for $\as$ by about a factor
of 2.5 at a time, with a precision of order 1--2\% possible using N$^3$LL or
N$^3$LL$'$ results \cite{Abbate:2010xh}.  We may anticipate similar future
precision in extracting $\as$ from DIS event shapes.

We showed how to incorporate nonperturbative hadronization corrections into our
predictions by inclusion of a shape function that is convolved together with the
perturbative soft function. The first moment of the shape function gives the
parameter $\Omega_1$ which describes the shift to the distribution in the tail
region. We demonstrated that this parameter is universal for our three event
shapes $\taumBCM$ and for any values of $x,Q^2$, and so it can be extracted from
one set of data to predict others.  We also made a simple illustration of the
effects of a shape function numerically on the cross section. We leave a more
extensive study of nonperturbative effects and extractions of the model
parameters from data to future work. We note that extraction of $\as$ from DIS
data (along the lines of \cite{Abbate:2010xh} for $e^+e^-$) using the above
rigorous factorization theorem based treatment of the power correction
$\Omega_1$ has yet to be performed.

The extension of our results to $N$-jettiness $\tau_N$ in DIS with $N>1$ is
straightforward, at least if we define $\tau_N$ similarly to the 1-jettiness
$\taum$ that we defined in \eq{truetau}. That is,
\be \label{truetauN}
 \tau_N^a = \frac{2}{Q^2}\sum_{i\in X}
 \min \{ \qBm\cdot p_i, q_1^a\cdot p_i,\dots, q_N^a\cdot p_i\}\,,
\ee
where $q_B^a = xP$ and $q_i^a$ is the jet axis of the $i$th non-ISR jet in the
final state as given by a jet algorithm or by minimization of the sum
\eq{truetauN} over the directions of $q_1^a,\dots,q_N^a$. As long as these jet
reference axes are aligned with the physical jet axes, the transverse momentum
$k_\perp$ of ISR will not affect the value of $\tau_N^a$ at leading order in
$\lambda$. The factorization theorem will then look like \eq{taumcsfinal}, with
suitable generalizations of the hard and soft functions and additional jet
functions (cf. \cite{Stewart:2010tn}):
\begin{align} \label{tauNfactorization}
&\frac{d\sigma}{dx\,dQ^2\,d\tau_N^a} 
 = \frac{d\sigma_0}{dx\,dQ^2} \int dt_J^1\cdots dt_J^N\, dt_B\,dk_S \\
 &\quad\times
 \delta\Big( \tau_N^a  - \frac{t_B+t_J^1+\cdots + t_J^N }{Q^2}- \frac{k_S}{Q} \Big)
  \nn\\
&\quad \times \sum_{i,\kappa} B_i(t_B, x,\mu) 
  J_{\kappa_1}(t_J^1,\mu)\cdots J_{\kappa_N}(t_J^N,\mu)  \nn \\
&\quad \times \tr \widehat H_{i\to\kappa} (\{q_m\}, L,\mu) \widehat S_N^{i\to
  \kappa} (k_S, \{ q_m \},\mu)\,, \nn
\end{align}
where $\widehat H_{i\to\kappa}(\{q_m\},L,\mu)$ contains the underlying hard
interaction $i(q_B) e(k) \to e(k') \kappa_1(q_1)\cdots \kappa_N(q_N)$, where
$i,\kappa_j$ denote parton types, $L$ denotes the dependence on the leptonic
states $e(k),e(k')$ and the exchanged virtual boson, and the sum over $i,\kappa$
is over all relevant partonic channels. The hard and soft functions $\widehat
H,\widehat S$ are matrices in color space, and the trace is over these colors.
$B_i$ is the ordinary beam function for the initial-state parton of flavor $i$.
Since \eq{truetauN} uses reference axes $q_{j}^a$ that are aligned with the
physical jet axes, the arguments $t_J^{j}$ of the jet functions are the
invariant masses of the jets and are not shifted by any transverse momentum
$k_\perp$ of ISR. Thus only the ordinary beam function $B_i$ appears in
\eq{tauNfactorization}, $k_\perp$ having been averaged over. We leave the
explicit evaluation of \eq{tauNfactorization} for $N$-jettiness cross sections
in DIS with $N>1$ to future work.

Our results bring to the arena of DIS the power of SCET that has already vastly
improved the precision of theoretical predictions of event shapes in $e^+e^-$
collisions and $pp$ collisions. The factorization theorems derived here point
the way to methods to improve the precision of parton distributions, hadron
structure, and the strong coupling $\as$ that we can extract from existing and
future experiments. With further advances in our calculations to greater
perturbative accuracy and improved modeling of the nonperturbative effects, the
frontiers of the study of the strong interaction using jets in DIS can be pushed
to higher precision.

\vspace{0.1cm}
{\bf Note Added:} While this paper was being finalized, Ref.~\cite{Kang:2013wca}
appeared which also considers the event shape we call $\tau_1^a$ at NNLL order.
A complete derivation of the factorization theorem was not presented there,
where the focus is instead the use of 1-jettiness to probe nuclear PDFs and
power corrections from dynamical effects in the nuclear medium.

\begin{acknowledgments}
  
  CL is grateful to the MIT Center for Theoretical Physics for hospitality
  during the course of this work. The work of DK and IS is supported by the
  Office of Nuclear Physics of the U.S. Department of Energy under Contract
  DE-FG02-94ER40818, 
  and the work of CL by DOE Contract DE-AC52-06NA25396 and by the LDRD office at
  Los Alamos.

\end{acknowledgments}

\appendix

\section{Generalized Rapidity Gap $\Delta Y$}
\label{app:observables}

The 1-jettiness $\taun$ in \eq{tauQJQB} is just one possible combination of jet
and beam momenta that we can choose to measure in DIS. It is quite
straightforward to keep $\nJ\mcdot p_J, \nB\mcdot p_B$ as independent
observables in the factorization theorem \eq{Wfinal}, and then to form other
observables by taking different combinations of $\nJ\cdot p_J,\nB\cdot p_B$. In
this Appendix we consider one of these possibilities---the \emph{generalized
  rapidity gap} $\Delta Y$ between the beam jet and the other final-state jet.

The rapidity of a particle with momentum $p$ with respect to the  $z$-axis is given by
\be
Y_{\nz\bnz}(p)=\frac12\ln \frac{\bnz\mcdot p}{\nz\mcdot p}.
\label{Ynbn}
\ee
If $p$ is $\nz$-collinear, the rapidity $Y_{\nz\,\bnz}$ is large and positive,
while it is large and negative if $p$ is $\bnz$-collinear.  Two jets produced in
DIS are not, in general, back-to back, and the reference vectors that measure
jets are not always aligned along one ($z$) axis, as \fig{tau} illustrates.  The
rapidity in \eq{Ynbn} can be generalized by replacing $n_{z,\bar z}$ with $\nB$
and $\nJ$ as follows:
\be
\label{YnBnJ}
Y_{\nJ \nB}(p)=\frac12\ln \frac{\nB\mcdot p}{\nJ\mcdot p},
\ee
where $Y_{\nJ \nB}$ is large and positive for the $\nJ$-collinear jet
and is large and negative for the $\nB$-collinear jet.
The generalized rapidity difference between two jets of momenta $p_J$ and $p_B$ is given by
\be
\label{DYdef}
\Delta Y\equiv Y_{\nJ \nB}(p_J)-Y_{\nJ \nB}(p_B) 
=\frac12\ln \frac{\nB\mcdot p_J}{\nJ\mcdot p_J} 
            \frac{\nJ\mcdot p_B}{\nB\mcdot p_B}.
\ee
The $\nBJ$ in \eq{DYdef} can be replaced by $\qBJ$ because the energy factors
$\omega_{J,B}/2$ in the numerator and denominator cancel. By using \eq{tauBJ}
$\qBJ\mcdot p_{B,J}$ can be expressed in terms of $\tau_{B,J}$. So, \eq{DYdef}
can be rewritten as
\be
\label{DYtauBJ}
\Delta Y = \frac12\ln \frac{4\, \qJ\mcdot p_B\, \qB\mcdot p_J}{\tau_J \,\tau_B Q^4}\,,
\ee
where the products $2\, \qJ\mcdot p_B$ and $2\, \qB\mcdot p_J$ are $\cO (Q^2)$
and $\Delta Y$ is $\cO \left[\ln (1/\sqrt{\tau_J \tau_B}) \right] \sim \cO\left[
  \ln(1/\lambda^2)\right]$.  \eq{DYtauBJ} can be specified for DIS by using
$\qJ\mcdot p_B\approx \qJ\mcdot (P+q)$ and $\qB\mcdot p_J\approx \qB\mcdot q$
where we use momentum conservation $P+q=p_B+p_J$ and suppress $p_B^2$ and
$p_J^2$.  As we have three versions of $\taun$, there are three versions of
$\Delta Y$:
\be 
\Delta Y^{\m,\B}=\frac12 \ln\frac{1-x}{x\tau_J\tau_B}\,, \quad 
 \Delta Y^\CM= \frac12 \ln\frac{1-x}{x^2\tau_J\tau_B}\,.
\ee

\section{Tensors and contractions}

The symmetric and asymmetric tensors transverse to both $\nB$ and $\nJ$ are defined by
\begin{subequations}
\label{geperp}
\begin{align}
\label{gperp}
g_\perp^{\mu\nu}&
=  g^{\mu\nu} - \frac{\nJ^\mu \nB^\nu + \nJ^\nu \nB^\mu}{\nJ\mcdot \nB}
\,,\\&
= g^{\mu\nu} - \frac{\nJ^\mu \bnJ^\nu + \nJ^\nu \bnJ^\mu}{2} 
= g^{\mu\nu} - \frac{\nB^\mu \bnB^\nu + \nB^\nu \bnB^\mu}{2} \, , \nn\\
\label{eperp}
\epsilon^\perp_{\mu\nu} 
&= \frac{1}{\nJ\mcdot \nB}\epsilon_{\mu\nu\alpha\beta}\nJ^\alpha \nB^\beta
\,,\\&
= \frac{1}{2}\epsilon_{\mu\nu\alpha\beta}\nJ^\alpha \bnJ^\beta
= \frac{1}{2}\epsilon_{\mu\nu\alpha\beta}\bnB^\alpha \nB^\beta\,, \nn
\end{align}
\end{subequations}
where $\bnB$ and $\bnJ$ are conjugate to $\nB$ and $\nJ$ as defined in \eq{nBbarnJbar} .

In order to calculate the contraction of the lepton tensor $L^{\mu\nu}$ 
with the hard function $H^{II'}_{q \bar q\,\mu\nu}$ as in \eq{leptonhard}, we must compute two tensor contractions: $g^T_{\mu\nu} g_\perp^{\mu\nu}$ and $\epsilon^T_{\mu\nu} \epsilon_\perp^{\mu\nu}$, where $g^T,\epsilon^T$ are defined in \eq{metrictensor} and $g_\perp,\epsilon_\perp$ in \eq{geperp}.
These contractions are given by
\begin{align}
\label{gg}
g^T_{\mu\nu} g_\perp^{\mu\nu} 
&= \Bigl(g_{\mu\nu} - 2\frac{k_\mu k'_\nu \plus k_\nu k'_\mu}{Q^2}\Bigr)
\Bigl(g^{\mu\nu} - \frac{\nJ^\mu \nB^\nu \plus \nB^\nu \nJ^\mu}{\nJ\mcdot \nB}\Bigr) \,,\nn \\
&= \frac{4}{\nJ\mcdot \nB \, Q^2} (\nJ \mcdot k \, \nB \mcdot k' + \nJ \mcdot k' \nB \mcdot k).
\end{align}
and
\begin{align}
\label{ee}
\epsilon^T_{\mu\nu} \epsilon_\perp^{\mu\nu} 
&= \frac{2}{\nJ\mcdot \nB \, Q^2} \epsilon_{\alpha\beta\mu\nu}\epsilon_{\gamma\delta}^{\mu\nu} k^\alpha {k'}^\beta \nJ^\gamma \nB^\delta \,,\nn\\
&= \frac{4}{\nJ\mcdot \nB \, Q^2} (\nJ \mcdot k' \, \nB \mcdot k - \nJ \mcdot k \, \nB \mcdot k') \,.
\end{align}
The ratio \eq{gg} over \eq{ee} is the coefficient $r(\qJ,\qB)$ defined in \eq{r}.

\section{Plus distribution}
\label{app:L}
The standard plus distribution for some function $q(x)$ is given by
\begin{align} \label{eq:genplus}
\bigl[q(x)\bigr]_+ & = \lim_{\e\to 0}\, \frac{\df}{\df x} \bigl[ \theta(x - \e)\, Q(x) \bigr]
\,,\nn\\& 
= \lim_{\e\to 0}\,  \bigl[ \theta(x - \e)\, q(x) + \delta(x-\e)\, Q(x)
\bigr]\,,
\end{align}
where
\be Q(x)= \int_{1}^{x}\,dx' q(x') \,. \ee
Integrating against a test function $f(x)$, we have
\begin{align}
\label{plusonf}
& \int_{-\infty}^{x_\mathrm{max}} dx  [\theta(x)\, q(x)]_+\, f(x)
\,,\nn\\
&= \int_{0}^{x_\mathrm{max}}dx\,  q(x)\,[f(x) - f(0)] 
+ f(0)\, Q(x_\mathrm{max})\,,
\end{align}
for $x_{\text{max}}> 0$.

For the special cases $q(x) = 1/x^{1-a}$ with $a>-1$
and $q(x) = \ln^n x/x$ with integer $n \geq 0$, we define:
\begin{align}
\label{eq:cLa_def}
\cL^a(x) &= \biggl[\frac{\theta(x)}{x^{1-a}} \biggr]_+
\,,\\
\label{eq:cLn_def}
\cL_n(x) &= \biggl[\frac{\theta(x)\ln^n x}{x}\biggr]_+
\,,\quad\quad n\geq 0
\,.\end{align}
For convenience we also define
\begin{equation} \label{Lm1}
\cL_{-1}(x) \equiv  \delta(x)
\,.\end{equation}
The plus function $\cL_n$ obeys the rescaling relation,
\be \label{cLn_rescale}
\lambda\, \cL_n(\lambda x)
= \sum_{k = 0}^n \binom{n}{k} \ln^k\!\lambda\, \cL_{n-k}(x)
  + \frac{\ln^{n+1}\!\lambda}{n+1}\, \delta(x),
\ee
where $\lambda>0$.

\section{Renormalization Group Evolution}
\label{app:rge}

In this appendix we collect results relevant for resummation of the DIS 1-jettiness cross section \eq{taucsfinal}
and its special cases \eqss{taumcs}{tauBcs}{tauCMcs} for $\taumBCM$.

The RGE and anomalous dimension for the hard Wilson coefficient $C$ in \eq{CNLO} for the two-quark operator are~\cite{Manohar:2003vb, Bauer:2003di}
\begin{align} \label{C_RGE}
\mu \frac{\df}{\df\mu} C(q^2, \mu) &= \gamma_C^q(q^2, \mu)\, C(q^2, \mu)
\,,
\nn\\
\gamma_C^q(q^2, \mu) &=
\Gamma_\cusp^q(\alpha_s) \ln\frac{- q^2}{\mu^2} + \gamma_C^q(\alpha_s)
\,.\end{align}
The anomalous dimension for the hard function $H$ in \eq{hardNLO} is given by
\begin{align} \label{H_RGE}
\mu \frac{\df}{\df\mu} H(Q^2, \mu) &= \gamma_H(Q^2, \mu)\, H(Q^2, \mu)
\,,
\nn\\
\gamma_H(Q^2, \mu) &=
2\Gamma_\cusp^q(\alpha_s) \ln\frac{Q^2}{\mu^2} + \gamma_H(\alpha_s)\,,
\end{align}
where $\gamma_H = 2\gamma_C^q$.
The expansions in $\as$ of $\Gamma_\cusp^q(\alpha_s)$ and $\gamma_C^q(\alpha_s)$ are given below in \eqs{Gacuspexp}{gaHexp}. 

The solution of the RGE in \eq{C_RGE} yields for the RG evolved hard function:
\begin{align} \label{Hrun}
H(Q^2, \mu) &= H(Q^2, \mu_0)\, U_H(Q^2, \mu_0, \mu)
\,, \nn\\
U_H(Q^2, \mu_0, \mu)
&= e^{K_H(\mu_0, \mu)} \Bigl(\frac{Q }{\mu_0}\Bigr)^{\eta_H(\mu_0, \mu)} 
\,, \nn\\
K_H(\mu_0,\mu) &= -4K_{\Gamma^q}(\mu_0,\mu) + K_{\gamma_H}(\mu_0,\mu)
\,, \nn\\
\eta_H(\mu_0,\mu) &= 4\eta_{\Gamma^q}(\mu_0,\mu)
\,,\end{align}
where the functions $K_{\Gamma^q}(\mu_0, \mu)$, $\eta_{\Gamma^q}(\mu_0, \mu)$ and $K_\gamma$ are given below in \eqs{Keta_def}{Keta}.

The quark beam function RGE is given by
\begin{gather}
\label{beamRGE}
\mu \frac{\df}{\df \mu} B_q(t, x, \mu) = \int\! \df t'\, \gamma_B^q(t-t',\mu)\, B_q(t', x, \mu)
\,,\\
\gamma_B^q(t, \mu)
= -2 \Gamma^q_{\cusp}(\alpha_s)\,\frac{1}{\mu^2}\cL_0\Bigl(\frac{t}{\mu^2}\Bigr) + \gamma_B^q(\alpha_s)\,\delta(t)
\,, \nn
\end{gather}
and its solution is~\cite{Balzereit:1998yf, Neubert:2004dd, Fleming:2007xt, Ligeti:2008ac}
\begin{align} \label{Brun_full}
B_q(t,x,\mu) & =  \int\! \df t'\, B_q(t - t',x,\mu_0)\, U_{B_q}(t',\mu_0, \mu)
\,, \nn\\ 
U_{B_q}(t, \mu_0, \mu) &= \frac{e^{K_{B_q} -\gamma_E\, \eta_{B_q}}}{\Gamma(1 + \eta_{B_q})}\,
\biggl[\frac{\eta_{B_q}}{\mu_0^2} \cL^{\eta_{B_q}} \Bigl( \frac{t}{\mu_0^2} \Bigr) + \delta(t) \biggr]
\,,  \nn\\
K_{B_q}(\mu_0,\mu) &= 4 K_{\Gamma^q}(\mu_0,\mu) + K_{\gamma_{B}^q}(\mu_0,\mu)
\,, \nn\\
\eta_{B_q}(\mu_0,\mu) &= -2\eta_{\Gamma^q}(\mu_0,\mu)
\,.\end{align}
The solution of the RGE for $B_q$ given by \eq{Brun_full} can be derived from the form of the solution \eq{Hrun} for the hard function by first Laplace transforming the beam function:
\be
\widetilde B_q(\nu,x,\mu) = \int_0^\infty dt \, e^{-\nu t} B_q(t,x,\mu) \,,
\ee
which obeys the RGE
\be
\label{beamLaplaceRGE}
\mu\frac{d}{d\mu} \widetilde B_q(\nu,x,\mu) = \tilde \gamma_{B_q}(\nu,\mu) \widetilde B_q(\nu,x,\mu)\,,
\ee
with the Laplace transformed anomalous dimension,
\be
\tilde\gamma_{B}^q(\nu,\mu) = 2\Gamma_{\text{cusp}}^q(\as) \ln(\mu^2\nu e^{\gamma_E}) + \gamma_{B}^q(\as)\,.
\ee
The evolution of $\widetilde B_q$ in \eq{beamLaplaceRGE} is multiplicative, of the same form as the hard function RGE \eq{H_RGE}, and therefore its solution is just like the hard function \eq{Hrun}, given by 
\be
\label{beamLaplacesolution}
\widetilde B_q(\nu,x,\mu) = \widetilde B_q(\nu,x,\mu_0) \widetilde U_{B_q}(\nu,\mu_0,\mu)\,,
\ee
where
\be
\label{beamevolution}
\widetilde U_{B_q}(\nu,\mu_0,\mu) = e^{K_{B_q}(\mu_0,\mu)} (\mu_0^2 \nu e^{\gamma_E})^{-\eta_{B_q}(\mu_0,\mu)} \,,
\ee
with $K_{B_q},\eta_{B_q}$ given by the same expressions as in \eq{Brun_full}. The inverse Laplace transform of the solution \eq{beamLaplacesolution} gives the momentum space solution for $B_q(t,x,\mu)$ in \eq{Brun_full}. 

The jet function obeys the same RGE as the beam function. They are defined by matrix elements of the same operator. The solution for the Laplace transformed jet function $\widetilde J_q(\nu,\mu)$ is given by the same form, \eqs{beamLaplacesolution}{beamevolution} with $B\to J$, and for the momentum-space jet function $J_q(t,\mu)$ by the same form \eq{Brun_full}, with $B\to J$.

The hemisphere soft function in \eq{softNLO} obeys the RGE
\be
\label{softRGE2}
\begin{split}
&\mu\frac{\df}{\df\mu}\Shemi(k_J,k_B,\mu) = \int dk_J' dk_B' \\
&\qquad  \times \gamma_S(k_J-k_J',k_B-k_B',\mu)\Shemi(k_J',k_B',\mu)
\,,\end{split}
\ee
where the dependence of the anomalous dimension on the two variables separates \cite{Hoang:2007vb}:
\be
\gamma_S(k_J,k_B,\mu) = \gamma_S(k_J,\mu)\delta(k_B) 
+ \gamma_S(k_B,\mu)\delta(k_J)\,,
\ee
with each piece of the anomalous dimension taking the form
\be
\label{softanomdim}
\gamma_S(k,\mu) = 2\Gamma_{\cusp}^q(\as) \frac{1}{\mu}\cL_0\Bigl(\frac{k}{\mu}\Bigr) + \gamma_S(\as)\delta(k)\,,
\ee
where $\gamma_S = -\gamma_C^q - \gamma_B^q$. 
The solution to the soft RGE \eq{softRGE2} is given by
\be
\label{softRGEsolution}
\begin{split}
 &\Shemi(k_J,k_B,\mu) = \int dk_J' dk_B' \Shemi(k_J',k_B',\mu_0)\\
 &\quad\qquad \times U_S(k_J-k_J',\mu_0,\mu)U_S(k_B-k_B',\mu_0,\mu)  
   \end{split}
\ee
where
\begin{align} \label{SBrun}
U_S(k, \mu_0, \mu) & = \frac{e^{K_S -\gamma_E\, \eta_S}}{\Gamma(1 + \eta_S)}\,
\biggl[\frac{\eta_S}{\mu_0} \cL^{\eta_S} \Big( \frac{k}{\mu_0} \Big) + \delta(k) \biggr]
\,,  \nn\\
K_S(\mu_0,\mu) &= -2K_{\Gamma^q}(\mu_0,\mu) + K_{\gamma_S}(\mu_0,\mu)
\,, \nn\\
\eta_S(\mu_0,\mu)& = 2\eta_{\Gamma^q}(\mu_0,\mu)
\,.\end{align}
This solution can be derived as for the beam and jet functions above by first taking the Laplace transform:
\be
\begin{split}
&\tShemi(\nu_J,\nu_B,\mu) 
\\&\quad
= \int_0^\infty \!\!dk_J  \int_0^\infty \!\! dk_B\,e^{-\nu_J k_J-\nu_B k_B}\Shemi(k_J,k_B,\mu)\,,
\end{split}
\ee
which obeys the RGE
\begin{align}
\label{softLaplaceRGE}
\mu\frac{d}{d\mu} \tShemi(\nu_J,\nu_B,\mu) &= \tShemi(\nu_J,\nu_B,\mu) \\
&\quad \times[\tilde\gamma_S (\nu_J,\mu) + \tilde \gamma_S(\nu_B,\mu)]  \,, \nn
\end{align}
where each part of the anomalous dimension takes the form
\be
\tilde\gamma_S(\nu,\mu) = -2\Gamma_{\text{cusp}}^q \ln(\mu \nu e^{\gamma_E}) + \gamma_S(\as)\,.
\ee
Solving the soft RGE \eq{softLaplaceRGE}, we obtain
\be
\label{softLaplacesolution}
\begin{split}
\tShemi(\nu_J,\nu_B,\mu) &= \tShemi(\nu_J,\nu_B,\mu_0) \\
&\quad \times \widetilde U_S(\nu_J,\mu_0,\mu)\widetilde U_S(\nu_B,\mu_0,\mu)\,,
\end{split}
\ee
where each soft evolution factor takes the form
\be
\label{softLaplaceevolution}
\widetilde U_S(\nu,\mu_0,\mu) = e^{K_S(\mu_0,\mu)} (\mu_0 \nu e^{\gamma_E})^{-\eta_S(\mu_0,\mu)}\,,
\ee
where $K_S,\eta_S$ are given by \eq{SBrun}. Taking the inverse Laplace transform of \eq{softLaplacesolution} gives the solution to the RGE for the soft function in momentum space $\Shemi(k_J,k_B,\mu)$ given in \eqs{softRGEsolution}{SBrun}.

In the 1-jettiness cross sections in this paper, we always encounter the soft function \eq{softRGEsolution} projected onto a function of a single variable $k$, according to \eq{Shemiprojection}. It obeys the RGE
\be
\mu\frac{d}{d\mu}\Shemi(k,\mu) = \int dk' \, 2\gamma_S(k-k',\mu) \Shemi(k',\mu),
\ee
where $\gamma_S(k,\mu)$ is given by \eq{softanomdim}. In Laplace space,
\be
\mu\frac{d}{d\mu} \tShemi(\nu,\mu) = 2\tilde\gamma_S(\nu,\mu) \tShemi(\nu,\mu)\,.
\ee
The solutions to these RGEs are given by
\begin{subequations}
\begin{align}\label{Sprojectrun}
\Shemi(k,\mu) &= \! \int \! dk'  \Shemi(k',\mu_0) U_S^2(k \minus k',\mu_0,\mu) \\
\tShemi(\nu,\mu) &= \tShemi(\nu,\mu_0) \widetilde U_S(\nu,\mu_0,\mu)^2 \,,
\end{align}
\end{subequations}
where $U_S^2(k,\mu_0,\mu)$ is given by \eq{SBrun} with $K_S,\eta_S \to 2K_S,2\eta_S$, and $\widetilde U_S(\nu,\mu_0,\mu)$ is given by \eq{softLaplaceevolution}.

The functions $K_{\Gamma^q}(\mu_0, \mu)$, $\eta_{\Gamma^q}(\mu_0, \mu)$, $K_\gamma(\mu_0, \mu)$ in the above RGE solutions are defined as
\begin{align} \label{Keta_def}
K_{\Gamma^q}(\mu_0, \mu)
& = \int_{\alpha_s(\mu_0)}^{\alpha_s(\mu)}\!\frac{\df\alpha_s}{\beta(\alpha_s)}\,
\Gamma_\cusp^q(\alpha_s) \int_{\alpha_s(\mu_0)}^{\alpha_s} \frac{\df \alpha_s'}{\beta(\alpha_s')}
\,,\nn\\
\eta_{\Gamma^q}(\mu_0, \mu)
&= \int_{\alpha_s(\mu_0)}^{\alpha_s(\mu)}\!\frac{\df\alpha_s}{\beta(\alpha_s)}\, \Gamma_\cusp^q(\alpha_s)
\,, \nn\\
K_\gamma(\mu_0, \mu)
& = \int_{\alpha_s(\mu_0)}^{\alpha_s(\mu)}\!\frac{\df\alpha_s}{\beta(\alpha_s)}\, \gamma(\alpha_s)
\,.\end{align}
 
Expanding the beta function and anomalous dimensions in powers of $\alpha_s$,
\begin{gather}
\label{anomdimexpansions}
\beta(\alpha_s) =
- 2 \alpha_s \sum_{n=0}^\infty \beta_n\Bigl(\frac{\alpha_s}{4\pi}\Bigr)^{n+1}
\,, \\
\Gamma^q_\cusp(\alpha_s) = \sum_{n=0}^\infty \Gamma^q_n \Bigl(\frac{\alpha_s}{4\pi}\Bigr)^{n+1}
\,, \ \gamma(\alpha_s) = \sum_{n=0}^\infty \gamma_n \Bigl(\frac{\alpha_s}{4\pi}\Bigr)^{n+1}
\,, \nn
\end{gather}
their explicit expressions to NNLL accuracy are (suppressing the superscript $q$ on  $\Ga^q$),
\begin{widetext}
\begin{align} \label{Keta}
K_\Gamma(\mu_0, \mu) &= -\frac{\Gamma_0}{4\beta_0^2}\,
\biggl\{ \frac{4\pi}{\alpha_s(\mu_0)}\, \Bigl(1 - \frac{1}{r} - \ln r\Bigr)
   + \biggl(\frac{\Gamma_1 }{\Gamma_0 } - \frac{\beta_1}{\beta_0}\biggr) (1-r+\ln r)
   + \frac{\beta_1}{2\beta_0} \ln^2 r
\nn\\ & \hspace{10ex}
+ \frac{\alpha_s(\mu_0)}{4\pi}\, \biggl[
  \biggl(\frac{\beta_1^2}{\beta_0^2} - \frac{\beta_2}{\beta_0} \biggr) \Bigl(\frac{1 - r^2}{2} + \ln r\Bigr)
  + \biggl(\frac{\beta_1\Gamma_1 }{\beta_0 \Gamma_0 } - \frac{\beta_1^2}{\beta_0^2} \biggr) (1- r+ r\ln r)
  - \biggl(\frac{\Gamma_2 }{\Gamma_0} - \frac{\beta_1\Gamma_1}{\beta_0\Gamma_0} \biggr) \frac{(1- r)^2}{2}
     \biggr] \biggr\}
\,, \nn\\
\eta_\Gamma(\mu_0, \mu) &=
 - \frac{\Gamma_0}{2\beta_0}\, \biggl[ \ln r
 + \frac{\alpha_s(\mu_0)}{4\pi}\, \biggl(\frac{\Gamma_1 }{\Gamma_0 }
 - \frac{\beta_1}{\beta_0}\biggr)(r-1)
 + \frac{\alpha_s^2(\mu_0)}{16\pi^2} \biggl(
    \frac{\Gamma_2 }{\Gamma_0 } - \frac{\beta_1\Gamma_1 }{\beta_0 \Gamma_0 }
      + \frac{\beta_1^2}{\beta_0^2} -\frac{\beta_2}{\beta_0} \biggr) \frac{r^2-1}{2}
    \biggr]
\,, \nn\\
K_\gamma(\mu_0, \mu) &=
 - \frac{\gamma_0}{2\beta_0}\, \biggl[ \ln r
 + \frac{\alpha_s(\mu_0)}{4\pi}\, \biggl(\frac{\gamma_1 }{\gamma_0 }
 - \frac{\beta_1}{\beta_0}\biggr)(r-1) \biggr]
\,.\end{align}
Here, $r = \alpha_s(\mu)/\alpha_s(\mu_0)$ and the running coupling is given to three-loop order by the expression
\begin{align} \label{alphas}
\frac{1}{\alpha_s(\mu)} &= \frac{X}{\alpha_s(\mu_0)}
  +\frac{\beta_1}{4\pi\beta_0}  \ln X
  + \frac{\alpha_s(\mu_0)}{16\pi^2} \biggr[
  \frac{\beta_2}{\beta_0} \Bigl(1-\frac{1}{X}\Bigr)
  + \frac{\beta_1^2}{\beta_0^2} \Bigl( \frac{\ln X}{X} +\frac{1}{X} -1\Bigr) \biggl]
\,,\end{align}
where $X\equiv 1+\alpha_s(\mu_0)\beta_0 \ln(\mu/\mu_0)/(2\pi)$.
In our numerical analysis we use the full NNLL expressions for $K_{\Gamma,\gamma},\eta_\Gamma$ in \eq{Keta}, but to be consistent with the value of $\as(\mu)$ used in the NLO PDFs we only use the two-loop truncation of \eq{alphas}, dropping the $\beta_2$ and $\beta_1^2$ terms, to obtain numerical values for $\alpha_s(\mu)$. (The numerical difference between using the two-loop and three-loop $\alpha_s$ is numerically very small and well within our theory uncertainties.) Up to three loops, the coefficients of the beta function~\cite{Tarasov:1980au, Larin:1993tp} and cusp anomalous dimension~\cite{Korchemsky:1987wg, Moch:2004pa} in $\overline{\mathrm{MS}}$ are
\begin{align} \label{Gacuspexp}
\beta_0 &= \frac{11}{3}\,C_A -\frac{4}{3}\,T_F\,n_f
\,,\nn\\
\beta_1 &= \frac{34}{3}\,C_A^2  - \Bigl(\frac{20}{3}\,C_A\, + 4 C_F\Bigr)\, T_F\,n_f
\,, \nn\\
\beta_2 &=
\frac{2857}{54}\,C_A^3 + \Bigl(C_F^2 - \frac{205}{18}\,C_F C_A
 - \frac{1415}{54}\,C_A^2 \Bigr)\, 2T_F\,n_f
 + \Bigl(\frac{11}{9}\, C_F + \frac{79}{54}\, C_A \Bigr)\, 4T_F^2\,n_f^2
\,,\nn\\[2ex]
\Gamma^q_0 &= 4C_F
\,,\nn\\
\Gamma^q_1 &= 4C_F \Bigl[\Bigl( \frac{67}{9} -\frac{\pi^2}{3} \Bigr)\,C_A  -
   \frac{20}{9}\,T_F\, n_f \Bigr]
\,,\nn\\
\Gamma^q_2 &= 4C_F \Bigl[
\Bigl(\frac{245}{6} -\frac{134 \pi^2}{27} + \frac{11 \pi ^4}{45}
  + \frac{22 \zeta_3}{3}\Bigr)C_A^2 
  + \Bigl(- \frac{418}{27} + \frac{40 \pi^2}{27}  - \frac{56 \zeta_3}{3} \Bigr)C_A\, T_F\,n_f
\nn\\* & \hspace{8ex}
  + \Bigl(- \frac{55}{3} + 16 \zeta_3 \Bigr) C_F\, T_F\,n_f
  - \frac{16}{27}\,T_F^2\, n_f^2 \Bigr]
\,.\end{align}

The $\overline{\mathrm{MS}}$ non-cusp anomalous dimension $\gamma_H = 2\gamma_C^q$ for the hard function $H$ can be obtained~\cite{Idilbi:2006dg, Becher:2006mr} from the IR divergences of the on-shell massless quark form factor $C(q^2,\mu)$ which are known to three loops~\cite{Moch:2005id},
\begin{align} \label{gaHexp}
\gamma^q_{C\,0} &= -6 C_F
\,,\nn\\
\gamma^q_{C\,1}
&= - C_F 
\Bigl[
  \Bigl(\frac{82}{9} - 52 \zeta_3\Bigr) C_A
+ (3 - 4 \pi^2 + 48 \zeta_3) C_F
+ \Bigl(\frac{65}{9} + \pi^2 \Bigr) \beta_0 \Bigr]
\,,\nn\\
\gamma^q_{C\,2}
&= -2C_F \Bigl[
  \Bigl(\frac{66167}{324} - \frac{686 \pi^2}{81} - \frac{302 \pi^4}{135} - \frac{782 \zeta_3}{9} + \frac{44\pi^2 \zeta_3}{9} + 136 \zeta_5\Bigr) C_A^2
\nn\\ & 
\quad + \Bigl(\frac{151}{4} - \frac{205 \pi^2}{9} - \frac{247 \pi^4}{135} + \frac{844 \zeta_3}{3} + \frac{8 \pi^2 \zeta_3}{3} + 120 \zeta_5\Bigr) C_F C_A
\nn\\ & 
\quad + \Bigl(\frac{29}{2} + 3 \pi^2 + \frac{8\pi^4}{5} + 68 \zeta_3 - \frac{16\pi^2 \zeta_3}{3} - 240 \zeta_5\Bigr) C_F^2
+ \Bigl(-\frac{10781}{108} + \frac{446 \pi^2}{81} + \frac{449 \pi^4}{270} - \frac{1166 \zeta_3}{9} \Bigr) C_A \beta_0
\nn\\ & 
\quad + \Bigl(\frac{2953}{108} - \frac{13 \pi^2}{18} - \frac{7 \pi^4 }{27} + \frac{128 \zeta_3}{9}\Bigr)\beta_1
+ \Bigl(-\frac{2417}{324} + \frac{5 \pi^2}{6} + \frac{2 \zeta_3}{3}\Bigr)\beta_0^2
\Bigr]
\,.\end{align}
As shown in \cite{Stewart:2010qs}, the anomalous dimension for the beam function equals that of the jet function, $\gamma_B^q = \gamma_J^q$, 
so the non-cusp three-loop anomalous dimension for the jet and beam functions are both given by \cite{Becher:2006mr},
\begin{align}\label{gaBexp}
\gamma_{B\,0}^q = \gamma_{J\,0}^q &= 6 C_F
\,,\nn\\
\gamma_{B\,1}^q = \gamma_{J\,1}^q
&= C_F 
\Bigl[
  \Bigl(\frac{146}{9} - 80 \zeta_3\Bigr) C_A
+ (3 - 4 \pi^2 + 48 \zeta_3) C_F
+ \Bigl(\frac{121}{9} + \frac{2\pi^2}{3} \Bigr) \beta_0 \Bigr]
\,,\nn\\
\gamma_{B\,2}^q = \gamma_{J\,2}^q
&= 2 C_F\Bigl[
  \Bigl(\frac{52019}{162} - \frac{841\pi^2}{81} - \frac{82\pi^4}{27} -\frac{2056\zeta_3}{9} + \frac{88\pi^2 \zeta_3}{9} + 232 \zeta_5\Bigr)C_A^2
\nn\\ & 
\quad + \Bigl(\frac{151}{4} - \frac{205\pi^2}{9} - \frac{247\pi^4}{135} + \frac{844\zeta_3}{3} + \frac{8\pi^2 \zeta_3}{3} + 120 \zeta_5\Bigr) C_A C_F
\nn\\ &
\quad + \Bigl(\frac{29}{2} + 3 \pi^2 + \frac{8\pi^4}{5} + 68 \zeta_3 - \frac{16\pi^2 \zeta_3}{3} - 240 \zeta_5\Bigr) C_F^2
+ \Bigl(-\frac{7739}{54} + \frac{325}{81} \pi^2 + \frac{617 \pi^4}{270} - \frac{1276\zeta_3}{9} \Bigr) C_A\beta_0
\nn\\ &
\quad + \Bigl(-\frac{3457}{324} + \frac{5\pi^2}{9} + \frac{16 \zeta_3}{3} \Bigr) \beta_0^2
+ \Bigl(\frac{1166}{27} - \frac{8 \pi^2}{9} - \frac{41 \pi^4}{135} + \frac{52 \zeta_3}{9}\Bigr) \beta_1
\Bigr]
\,.\end{align}
\end{widetext}
The anomalous dimension for the soft function is obtained from $\gamma_S = -\gamma_C^q - \gamma_B^q$.
At NNLL, we only need the one- and two-loop coefficients of $\gamma_{H,B,J,S}$. The three-loop coefficients are given for completeness. They would be required at N$^3$LL, along with the four-loop beta function and cusp anomalous dimension, the latter of which has not yet been calculated. In addition, the full N$^3$LL result would also require the two-loop fixed-order corrections, which are known for the hard function, but not yet for the beam and soft functions.

\section{Coefficients in Momentum-Space Resummed Cross Section} 
\label{app:RGconstants}

The resummed cross sections for $\taumBCM$ in \sec{NNLL} are obtained by plugging the solutions to the RG equations for the hard function and for the momentum-space jet, beam, and soft functions given in \appx{rge} into the factorization theorems derived in \ssec{csresults}. Performing the convolutions in these factorization theorems of the jet, beam, and soft evolution kernels given in \appx{rge} and fixed-order functions requires computing the convolutions of plus functions with each other. The results of these convolutions produce the expressions given in \eqs{resummedtaumB}{resummedtauCM}, given in terms of coefficients $J_n,I_n,S_n$ of the logs in the fixed-order jet, beam, and soft functions and coefficients $V_{k}^{mn}$ and $V_{k}^n(a)$ that are the result of the convolutions of plus functions. In this Appendix we tabulate these coefficients. For more details see Refs.~\cite{Abbate:2010xh,Ligeti:2008ac}.

\subsection{Jet, Beam, and Soft Coefficients $J_n$, $I^{qq,qg}_n$, $S_n$}
\label{app:SnJn}
The fixed-order results at $\cO(\as)$ of soft, jet, and beam functions 
can be written as sum of plus distributions as
\be
\label{Fseries1}
G(t,\mu)=\frac{1}{\mu^{n_G}} \sum_{n=-1}^{1} G_n[\alpha_s(\mu)] \, \cL_n\left(\frac{t}{\mu^{n_G}}\right)
\,.\ee
where $G(t,\mu)$ represents the single-variable soft function $S(t,\mu)$ in
\eq{NPsoftfunction}, jet function $J(t,\mu)$ in \eq{jetNLO}, or the coefficient
$I^{qq,qg}(t,z,\mu)$ inside the beam function in \eq{beamcoeffs}.  The index
$n_F=1$ for the soft function and $n_F=2$ for the jet and beam function.  In the
case of the beam function, the $z$ dependence in $F(t,\mu)$ is implicit.
The coefficients $F_n$ in \eq{Fseries1} for the three functions are $S_n$, $J_n$, and $I_n^{qq,qg}$.
The soft coefficients at order $\as$ are given by
\be
\label{Sn}
\begin{split}
&S_{-1}(\as)=1+\frac{\as C_F}{4\pi} \frac{\pi^2}{3}\,,
\\
&S_0(\as)= 0\,, \quad
S_1(\as)= \frac{\as C_F}{4\pi} (-16)\,,
\end{split}
\ee
the jet coefficients by
\be
\begin{split}
\label{Jn}
&J_{-1}(\as)=1+\frac{\as C_F}{\pi} \left(\frac74 -\frac{\pi^2}{4}  \right)\,,
\\
&J_0(\as)= -\frac{\as C_F}{\pi} \frac{3}{4}\,, \quad
J_1(\as)= \frac{\as C_F}{\pi}\,,
\end{split}
\ee
and the beam function coefficients by
\begin{align}
\label{Iqqn}
I^{qq}_{-1}(\as,z)&=\cL_{-1}(1-z)+\frac{\as C_F}{2\pi} \bigg[ \cL_1(1-z) (1+z^2)
 \nn \\&
\minus\frac{\pi^2}{6} \cL_{-1}(1\minus z)+\theta(1 \minus z)\Bigl( 1-z-\frac{1\plus z}{1 \minus z}\ln z\Bigr)
\bigg]\,, \nn
\\
I^{qq}_0(\as,z)&=\frac{\as C_F}{2\pi} \theta (z) \left( P_{qq}(z) -\frac32 \cL_{-1}(1-z)\right)\,, \nn
\\
I^{qq}_1(\as,z)&= \frac{\as C_F}{2\pi}\, 2\cL_{-1}(1-z)\,,
\end{align}
and
\begin{align}
\label{Iqgn}
I^{qg}_{-1}(\as,z)&=\frac{\as T_F}{2\pi}\theta(z)\! \Bigl[P_{qg}(z)\ln\frac{1 \minus z}{z} \plus 2\theta(1\minus z) z(1 \minus z)   \!\Bigr] ,
\nn \\
I^{qg}_0(\as,z)&= \frac{\as T_F}{2\pi}\theta(z) P_{qg}(z)\,,
\end{align}
where the splitting functions $P_{qq,qg}(z)$ are given in \eq{splittingfunctions}.

The argument of the plus distributions $\cL_n$ in \eq{Fseries1} can rescaled by using the identity \eq{cLn_rescale}.
\eq{Fseries1} can be rewritten in terms of the rescaled distribution as
\be
\label{Fseries2}
G(t,\mu)=\frac{1}{\lambda\mu^{n_G}} \sum_{n=-1}^{1} G_n[\alpha_s(\mu),\lambda] \, \cL_n\left(\frac{\lambda^{-1} t}{\mu^{n_G}}\right)
\,,\ee
where the coefficents $G_n(\as,\lambda)$ in \eq{Fseries2} are
expressed in terms of the coefficients in \eq{Fseries1} by using the rescaling identity in \eq{cLn_rescale}
as
\begin{align}
\label{Fn}
G_{-1}(\as,\lambda) &= G_{-1}(\as) +\sum_{n=0}^{\infty} G_n(\as)\frac{\ln^{n+1} \lambda}{n+1}\,,
\nn\\
G_n(\as,\lambda) &=\sum_{k=0}^{\infty}\frac{(n+k)!}{n!\, k!} G_{n+k}(\as) \ln^k \lambda\,,
\end{align}
where $G_n = \left\{S_n,J_n,I_n^{qq,qg}\right\}$.
Explicit expressions for $S_n(\as,\lambda)$, $J_n(\as,\lambda)$, and $I^{qq,qg}_n(\as,\lambda)$ are obtained 
by inserting Eqs.~\eqref{Sn}, \eqref{Jn}, \eqref{Iqqn}, and \eqref{Iqgn} into \eq{Fn}.

\subsection{Results of convolving plus functions}
\label{app:Vcoeff}

Convolutions of plus distributions in the jet, beam, and soft evolution kernels and the fixed-order functions produce the functions $V_k^n(\Omega)$ and the coefficients $V_{k}^{mn}$ in the resummed cross sections \eqs{resummedtaumB}{resummedtauCM}.
There are three types of convolutions of plus distribtions $\cL_n$ and $\cL^a$.
and we write them in useful form as
\begin{align}
\label{LL}
&\int dy \cL_m (x-y) \cL_n (y) = \sum_{\ell=-1}^{m+n+1} V_\ell^{mn} \cL_\ell(x),
\\
&\int dy \left[ a \cL^a(x-y) +\delta(x-y)\right] \left[b \cL^b(y) +\delta(y)\right]
\nn\\
&\hspace{50pt}= \frac{\Gamma(1+a)\Gamma(1+b)}{\Gamma(1+a+b)}\,(a+b)\left[ \cL^{a+b} (x) +\delta(x)\right],
\nn\\
&\int dy \left[ a \cL^a(x-y) +\delta(x-y)\right] \cL_n(y)
=\sum_{k=-1}^{n+1}\, V^n_k(a) \,\cL^a_k(x)\,.\nn
\end{align}
The coefficients $V_k^n(a)$ and $V_k^{mn}$
are related to the Taylor series expansion of $V(a,b)$ around $a = 0$ and $a = b
= 0$, where $V(a,b)$ is defined by
\be \label{Vab}
V(a,b)=\frac{\Gamma(a) \Gamma(b)}{\Gamma(a+b)} -\frac{1}{a}-\frac{1}{b}\,,
\ee
which satisfies $V(0,0) = 0$.
The $V_k^n(a)$ for $n\ge 0$ are
\begin{align} \label{eq:Vkna_def}
V_k^n(a) &= \begin{cases}
   \displaystyle a\, \frac{\df^n}{\df b^n}\,\frac{V(a,b)}{a+b}\bigg\vert_{b = 0}\,,
  &\!\!\!\!\!\!\!\!\!\!   k=-1\,, \\[10pt]
   \displaystyle  a\, \binom{n}{k}   \frac{\df^{n-k}}{\df b^{n-k}}\, V(a,b)
   \bigg\vert_{b = 0} + \delta_{kn} \,, \qquad
  &\!\!\!\!\!\!\!\!\!\! 0\le k\le n \,,  \\[10pt]
  \displaystyle  \frac{a}{n+1} \,,
  &\!\!\!\!\!\!\!\!\!\! k=n+1  \,.
\end{cases}
\end{align}
The $V_k^{mn}$ are symmetric in
$m$ and $n$, and for $m,n\ge 0$ they are
\be
\label{eq:Vkmn_def}
V_k^{mn}=
 \begin{cases}
 \displaystyle \frac{\df^m}{\df a^m}\, \frac{\df^n}{\df b^n}\,\frac{V(a,b)}{a+b}\bigg\vert_{a = b = 0} \,,
   & k=-1\,, \\[15pt]
\displaystyle  \sum_{p = 0}^m \sum_{q = 0}^n \delta_{p+q,k}\,\binom{m}{p} \binom{n}{q}
\\\hspace{20pt} \displaystyle\times
\frac{\df^{m-p}}{\df a^{m-p}}\, \frac{\df^{n-q}}{\df b^{n-q}} \ V(a,b)
  \bigg\vert_{a = b = 0}\,, & 
\\
   &\hspace{-30pt} 0\le k \le m+n \,,\\[15pt]
 \displaystyle  \frac{1}{m+1} + \frac{1}{n+1}\,,
   &\hspace{-30pt}  k=m+n+1 \,.
\end{cases}
\ee
Using \eq{Lm1} we can extend these definitions to include the cases
$n = -1$ or $m = -1$.  The relevant coefficients are
\begin{equation}
\begin{split}
V_{-1}^{-1}(a) &= 1
\,,\qquad
V_0^{-1}(a) = a \\
V_{k \geq 1}^{-1}(a) &= 0
\,,\qquad
V^{-1,n}_k = V^{n,-1}_k = \delta_{nk}
\,.
\end{split}
\end{equation}

\section{Resummed cross section from Laplace transforms} \label{app:Laplace}

An alternative way \cite{Becher:2006mr,Becher:2006nr} to express the resummed cross sections in \sec{NNLL} is to utilize the Laplace-transformed jet, beam, and soft functions given in \appx{rge} and their RGE solutions. The method avoids taking explicit convolutions of plus functions in the evolution factors and in the fixed-order jet, beam, and soft functions.

Each of the RGE solutions for the jet, beam, and soft functions is given by a function of the form
\be
\label{Fsolution}
\widetilde G(\nu,\mu) = \widetilde G(\nu,\mu_0) e^{K_G(\mu_0,\mu)} [\mu_0(\nu e^{\gamma_E})^{1/j_G}]^{-j_G\eta_G(\mu_0,\mu)}\,.
\ee
For the jet and beam functions, $j_G=2$, while for the soft function $j_G = 1$. The fixed-order expansion of $\widetilde G(\nu,\mu_0) \equiv \widetilde G(L_G,\mu_0)$ can be considered to be a function of the log $L_G \equiv \ln Q_G/\mu_0$, where $Q_G = (\nu e^{\gamma_E})^{-1/j_G}$. To $\cO(\as^2)$,
\begin{align}
\label{fixedorder}
&\widetilde G(L_G,\mu_0) = 1 + \frac{\as(\mu_0)}{4\pi} \Bigl( -\Gamma_G^0 L_G^2 - \gamma_G^0 L_G + c_G^1\Bigr) \\
&\qquad + \Bigl(\frac{\as(\mu_0)}{4\pi}\Bigr)^2 \biggl[ \frac{1}{2}(\Gamma_G^0)^2 L_G^4  + \Gamma_G^0 \Bigl( \gamma_G^0 + \frac{2}{3}\beta_0\Bigr) L_G^3 \nn \\
&\qquad\qquad\qquad\quad + \Bigl(\frac{1}{2}(\gamma_G^0)^2 + \gamma_G^0\beta_0 - \Gamma_G^1 - c_G^1\Gamma_G^0\Bigr) L_G^2 \nn \\
&\qquad\qquad\qquad\quad   - (\gamma_G^1 + c_G^1\gamma_G^0 + 2c_G^1\beta_0) L_G+ c_G^2\biggr]\,. \nn
\end{align}
Each power of $L_G$ can be generated by taking derivatives with respect to $\eta_G$ in \eq{Fsolution}:
\be
\widetilde G(\nu,\mu) = e^{K_G(\mu_0,\mu)} \tilde g(\partial_{\eta_F},\mu_0) \big[ \mu_0 (\nu e^{\gamma_E})^{1/j_G}\bigr]^{-j_G\eta_G(\mu_0,\mu)}\,,
\ee
where $\tilde g(\partial_\eta,\mu_0)$ is the operator constructed by replacing each $L_G$ in \eq{fixedorder} with $\partial_\eta/j_G$:
\begin{align}
\label{foperator}
&\tilde g(\partial_\eta,\mu_0) = 1 + \frac{\as(\mu_0)}{4\pi} \Bigl( -\Gamma_G^0 \frac{\partial_\eta^2}{j_G^2} - \gamma_G^0 \frac{\partial_\eta}{j_G} + c_G^1\Bigr) \\ 
&\qquad + \Bigl(\frac{\as(\mu_0)}{4\pi}\Bigr)^2 \biggl[ \frac{1}{2}(\Gamma_G^0)^2 \frac{\partial_\eta^4}{j_G^4}  + \Gamma_G^0 \Bigl( \gamma_G^0 + \frac{2}{3}\beta_0\Bigr) \frac{\partial_\eta^3}{j_G^3} \nn \\
&\qquad\qquad\qquad\quad + \Bigl(\frac{1}{2}(\gamma_G^0)^2 + \gamma_G^0\beta_0 - \Gamma_G^1 - c_G^1\Gamma_G^0\Bigr) \frac{\partial_\eta^2}{j_G^2} \nn\\
&\qquad\qquad \qquad\quad - (\gamma_G^1 + c_G^1\gamma_G^0 + 2c_G^1\beta_0) \frac{\partial_\eta}{j_G} + c_G^2\biggr]\,. \nn
\end{align}
Now it is easy to take the inverse Laplace transform of \eq{Fsolution},
\begin{align}
G(t,\mu) &= \int_{c-i\infty}^{c+i\infty} \frac{d\nu}{2\pi i} e^{\nu t} \widetilde G(\nu,\mu)  \\
&= e^{K_G(\mu_0,\mu)} \tilde g(\partial_{\eta_F},\mu_0) \frac{e^{-\gamma_E\eta_G}}{\Gamma(\eta_G)} \frac{(t/\mu_0^{j_G})^{\eta_G}}{t}\,, \nn
\end{align} 
where $\eta_G \equiv \eta_G(\mu_0,\mu)$. The derivatives with respect to $\eta_G$ automatically generate the results of taking convolutions of the logs inside $G(t,\mu_0)$ with the evolution kernel $U_G(t,\mu_0,\mu)$ in RGE solutions like \eqs{Brun_full}{softRGEsolution}.

\begin{widetext}
\subsection{$\taumB$ cross sections}

Using the above formalism, we obtain for the Laplace transforms of the $\taumB$ differential cross sections $(1/\sigma_0)d\sigma/d\taumB$ in \eqs{tauBcs}{taumcs}, 
\be
\label{resummedLaplace}
\begin{split}
\tilde\sigma(x,Q^2,\nu^{\m,\B}) &= H(Q^2,\mu_H)\tilde j(\partial_{\eta_J},\mu_J) [L_q(x,Q^2)\tilde b_q^{\m,\B}(\partial_{\eta_B},x,\mu_B) + L_{\bar q}(x,Q^2) \tilde b_{\bar q}^{\m,\B}(\partial_{\eta_B},x,\mu_B)]\tilde s(\partial_{2\eta_S},\mu_S) \\
&\quad\times e^{K_H(\mu_H,\mu) + K_J(\mu_J,\mu) + K_B(\mu_B,\mu) + 2K_S(\mu_S,\mu)}\\
&\quad\times \Bigl(\frac{Q}{\mu_H}\Bigr)^{\eta_H(\mu_H,\mu)}\Bigl(\frac{Q^2}{\mu_J^2 e^{\gamma_E} \nu^{\m,\B}}\Bigr)^{\eta_J(\mu_J,\mu)} \Bigl(\frac{Q^2}{\mu_B^2 e^{\gamma_E} \nu^{\m,\B}}\Bigr)^{\eta_B(\mu_B,\mu)} \Bigl(\frac{Q}{\mu_S e^{\gamma_E} \nu^{\m,\B}}\Bigr)^{2\eta_S(\mu_S,\mu)}\,.
\end{split}
\ee
Taking the inverse Laplace transform with respect to $\nu^{\m,\B}$ and taking the cumulant in \eq{cumulant}, we easily obtain in momentum space,
\be
\label{appx:resummedtaumB}
\begin{split}
\sigmac(x,Q^2,\taumB) &= H(Q^2,\mu_H) \Bigl( \frac{Q}{\mu_H}\Bigr)^{\eta_H(\mu_H,\mu)} \Bigl(\frac{Q^2 \taumB}{\mu_J^2}\Bigr)^{\eta_J(\mu_J,\mu)} \Bigl(\frac{Q^2 \taumB}{\mu_B^2}\Bigr)^{\eta_B(\mu_B,\mu)} \Bigl(\frac{Q \taumB}{\mu_S}\Bigr)^{2\eta_S(\mu_S,\mu)-\Omega} \\
&\times \biggl[L_q^\m(x,Q^2)\tilde b_q^{\m,\B}\Bigl(\partial_\Omega - \ln\frac{\mu_B^2}{Q\mu_S},x,\mu_B\Bigr) + L_{\bar q}^\m(x,Q^2)\tilde b_{\bar q}^{\m,\B}\Bigl(\partial_\Omega - \ln\frac{\mu_B^2}{Q\mu_S},x,\mu_B\Bigr)\biggr] \\
&\times \tilde j\Bigl(\partial_\Omega - \ln\frac{\mu_J^2}{Q\mu_S},\mu_J\Bigr) \tilde s \bigl(\partial_\Omega,\mu_S\bigr) \Bigl(\frac{Q\taumB}{\mu_S e^{\gamma_E}}\Bigr)^\Omega \frac{\theta(\taumB)}{\Gamma(1+\Omega)} e^{\mathcal{K}(\mu_H,\mu_J,\mu_B,\mu_S,\mu)} \,,
\end{split}
\ee
with a sum over quark and antiquark flavors $q,\bar q$,  and where the sums of evolution kernels $\mathcal{K},\Omega$ are given by
\begin{subequations}
\begin{align}
\mathcal{K}(\mu_H,\mu_J,\mu_B,\mu_S,\mu) &= K_H(\mu_H,\mu) + K_J(\mu_J,\mu) + K_B(\mu_B,\mu) + 2K_S(\mu_S,\mu) \\
\Omega  \equiv \Omega(\mu_J,\mu_B,\mu_S,\mu) &= \eta_J(\mu_J,\mu) + \eta_B(\mu_B,\mu) + 2\eta_S(\mu_S,\mu)\,,
\end{align}
\end{subequations}
where the individual evolution kernels $K_{H,J,B,S},\eta_{J,B,S}$ are defined in \appx{rge}.

The fixed-order operators $\tilde j,\tilde b_{q,\bar q},\tilde s$ in \eq{appx:resummedtaumB} each take the form \eq{foperator}, which in this paper we will truncate to $\cO(\as)$, working to NNLL accuracy.
In \eq{foperator}, $\Gamma_F^n,\gamma_F^n,\beta_n$ are the coefficients in the fixed-order expansions \eq{anomdimexpansions} of the anomalous dimensions and beta function, and where $j_G = 2$ for the jet function and $j_G=1$ for the soft function, and the constants $c_G^1$ are given by
\be
c_J^1 = (7-\pi^2)C_F - \frac{\Gamma_J^0}{4}\frac{\pi^2}{6}\,,\quad c_S^1 = \frac{\pi^2}{3} C_F - \Gamma_S^0 \frac{\pi^2}{3}\,.
\ee
Note that the cusp parts of the hard, jet/beam, and soft anomalous dimensions are related to the cusp anomalous dimension in \eq{anomdimexpansions} by
\be
\Gamma_H = 2\Gamma_{\text{cusp}}^q \,, \quad \Gamma_{J,B} = -2\Gamma_{\text{cusp}}^q \,,\quad \Gamma_S = 2\Gamma_{\text{cusp}}^q\,.
\ee
Meanwhile the beam function operators $\tilde b_q^{\m,\B}$ in the $\taumB$ cross sections are given by
\be
\label{appx:beamoperators}
\begin{split}
\tilde b_q^{\m,\B}(\partial_\Omega,x,\mu_B) &= f_{q}(x,\mu_B) \biggl\{ 1 + \frac{\as(\mu_B)}{4\pi} \biggl( -C_F \frac{\pi^2}{3} - \frac{\Gamma_J^0}{4}\Bigl(\partial_\Omega^2 + \frac{\pi^2}{6}\Bigr)  + \frac{\gamma_J^0}{2} \partial_\Omega\biggr)\biggr\} \\
&\quad+ \frac{\as(\mu_B)}{2\pi}\int_x^1 \frac{dz}{z} \biggl[ C_F f_q\Bigl(\frac{x}{z},\mu_B\Bigr) F_q(z) + T_F f_g\Bigl(\frac{x}{z},\mu_B\Bigr) F_g(z)\biggr] \\
&\quad+ \frac{\as(\mu_B)}{2\pi} \int_x^1 \frac{dz}{z} \biggl[ C_F P_{qq}(z)f_q\Bigl(\frac{x}{z},\mu_B\Bigr) + T_F P_{qg}(z)f_g\Bigl(\frac{x}{z},\mu_B\Bigr)\biggr] \bigr[\partial_\Omega + \delta b^{\m,\B}(z)\bigr]\,,
\end{split}
\ee
where $\tilde b_q^{\m,\B}$ differ only in the last term,
\be
\label{appx:deltab}
\delta b^\m(z)  = 0\,,\quad \delta b^\B(z) = \ln z\,,
\ee
and the functions $F_{q,g}$ are given by
\begin{subequations}
\begin{align}
F_q(z) &\equiv (1+z^2) \left[\frac{\theta(1-z)\ln(1-z)}{1-z}\right]_+ + \theta(1-z) \Bigl(1 - z - \frac{1+z^2}{1-z}\ln z\Bigr) \\
F_g(z) &\equiv P_{qg}(z) \Bigl(\ln\frac{1-z}{z} - 1\Bigr) + \theta(1-z)\,,
\end{align}
\end{subequations}
and $P_{qq,qg}$ are given by \eq{splittingfunctions}. The additional term $\delta b^\B(z) = \ln z$ that appears in the final integrand in \eq{appx:beamoperators} for $\tilde b^\B$ is due to the nontrivial $\vect{k}_\perp^2$ dependent terms in \eq{genbeamcoeffs} for the generalized beam function, which generate the $\delta b^\B(z) = \ln z$ term upon integration over the transverse momentum in \eq{tauBcs}. Thus the difference the $\taum$ and $\tauB$ cross sections will become more pronounced at smaller $x$, when the $\delta b^\B(z) = \ln z$ term inside the integrand of \eq{appx:beamoperators} can grow larger.

To evaluate the action of the fixed-order operators given by \eqs{foperator}{appx:beamoperators} in the resummed cross section \eq{appx:resummedtaumB}, it is useful to tabulate the following relations:
\begin{align} \label{appx:Omegaderivatives}
\mathcal{G}(\Omega) &\equiv \Bigl(\frac{Q\taun}{\mu e^{\gamma_E}}\Bigr)^\Omega \frac{1}{\Gamma(1+\Omega)}\,, \nn \\
 \partial_\Omega\mathcal{G}(\Omega)&  = \Bigl[ - \ln\frac{\mu}{Q\taun} -
H(\Omega)\Bigr] \cG(\Omega) \,, \\
 \partial_\Omega^2\mathcal{G}(\Omega)  &= \Bigl[ \Bigl(\ln\frac{\mu}{Q\taun}+ H(\Omega)\Bigr)^2 - \psi^{(1)}(1+\Omega)\Bigr] \cG(\Omega) \,, \nn
\end{align}
where $H$ is the harmonic number function, $H(\Omega) = \gamma_E + \psi^{(0)}(1+\Omega)$ and $\psi^{(n)}(x)  = (d^n/dz^n)[\Gamma'(z)/\Gamma(z)]$ is the polygamma function. The result of taking these derivatives in the expression \eq{appx:resummedtaumB} is equivalent to the results of convolving logs in the fixed-order jet, beam, and soft functions with the momentum-space evolution kernels in deriving the expression \eq{resummedtaumB}. The two formalisms yield equivalent expressions for the resummed cross section.

\subsection{$\tauCM$ cross section}

The resummed $\tauCM$ cross section obtained from RG evolution of the hard, jet, beam, and soft functions in \eq{tauCMcs} is given by
\begin{align}
\label{appx:resummedtauCM}
\sigmac(x,Q^2,\tauCM) &= H(Q^2,\mu_H)   \Bigl( \frac{Q}{\mu_H}\Bigr)^{\eta_H(\mu_H,\mu)} \Bigl(\frac{Q^2}{\mu_J^2}\Bigr)^{\eta_J(\mu_J,\mu)}   \Bigl(\frac{xQ^2 }{\mu_B^2}\Bigr)^{\eta_B(\mu_B,\mu)} \Bigl(\frac{\sqrt{x}Q }{\mu_S}\Bigr)^{2\eta_S(\mu_S,\mu)-\Omega}  \tilde j\Bigl(\partial_\Omega - \ln\frac{\sqrt{x}\mu_J^2}{Q\mu_S},\mu_J\Bigr) \nn \\
&\quad\times \biggl[L_q^\CM(Q^2)\tilde b_q^\CM\Bigl(\partial_\Omega - \ln\frac{\mu_B^2}{\sqrt{x}Q\mu_S},x,y,\tauCM,\mu_B\Bigr) + L_{\bar q}^\CM(Q^2)\tilde b_{\bar q}^\CM\Bigl(\partial_\Omega - \ln\frac{\mu_B^2}{\sqrt{x}Q\mu_S},x,y,\tauCM,\mu_B\Bigr)\biggr]  \nn \\
&\quad\times \tilde s \bigl(\partial_\Omega,\mu_S\bigr) \biggl(\frac{\sqrt{x}Q\abs{\tauCM - 1 +y}}{\mu_S e^{\gamma_E}}\biggr)^\Omega \frac{e^{\mathcal{K}(\mu_H,\mu_J,\mu_B,\mu_S,\mu)}}{\Gamma(1+\Omega)} 
\,,
\end{align}
where the operator $\tilde b_{q}^\CM$ is given by
\begin{align}
\label{appx:beamCMoperator}
\tilde b_q^\CM(\partial_\Omega,x,y,\tauCM,\mu_B) &=  \theta(\tauCM - 1 + y) \tilde b_q^\m(\partial_\Omega,x,\mu_B) + \frac{\as(\mu_B)}{2\pi} \int_x^1 \frac{dz}{z} \biggl[ C_F P_{qq}(z)f_q\Bigl(\frac{x}{z},\mu_B\Bigr) + T_F P_{qg}(z)f_g\Bigl(\frac{x}{z},\mu_B\Bigr)\biggr] \nn \\
&\quad\times \biggl\{ \theta(\tauCM -1+y) \biggl[ \ln\biggl( \frac{z}{1-z} \frac{\tauCM - 1 + y}{x(1-y)}\biggr) - H(-\Omega) - \frac{1}{\Omega}\biggr] - \theta(1-y-\tauCM)\frac{\pi}{\sin\pi\Omega} \nn \\
&\qquad  + \frac{1}{\Omega}\biggl( \frac{(1-y)X}{\abs{\tauCM - 1 + y}}\biggr)^\Omega \hyp\biggl( - \Omega,-\Omega,1-\Omega; - \frac{\tauCM -1+y}{(1-y)X}\biggr)\biggr\}  \theta\Bigl( \tauCM - (1 - y)(1-X)\Bigr) 
\,,
\end{align}
and similarly for $\tilde b_{\bar q}^\CM$. Here $X \equiv x(1-z)/(x+z-xz)$.
The additional more complicated terms in $\tilde b_q^\CM$ are due to the nontrivial $\vect{p}_\perp$ integral in \eq{tauCMcs} which convolves the terms in the generalized beam function with nontrivial $\vect{p}_\perp^2$ dependence with the dependence of the jet function on $(\vect{q}_\perp+\vect{p}_\perp)^2$, with $\vect{q}_\perp\neq0$ when $y<1$. Note that the apparent singularities as $\Omega\to 0$ (the fixed-order limit) cancel in the sum of all terms. The result \eq{appx:resummedtauCM} is equivalent to the expression \eq{resummedtauCM} derived from RG evolution directly in momentum space.

\subsection{Generic $\taun$ cross section}

In similar fashion we can form the resummed $\taun$ cross section for an arbitrary definition \eq{tau1def} of the 1-jettiness. Using the generic factorization theorem \eq{taucsfinal}
\be
\label{appx:resummedtau}
\begin{split}
&\sigmac(x,Q^2,\taun) = H(Q^2,\mu_H)   \Bigl( \frac{Q}{\mu_H}\Bigr)^{\eta_H(\mu_H,\mu)} 
\Bigl(\frac{s_J}{\mu_J^2}\Bigr)^{\eta_J(\mu_J,\mu)}   
\Bigl(\frac{s_B}{\mu_B^2}\Bigr)^{\eta_B(\mu_B,\mu)} 
\Bigl(\frac{Q_R}{\mu_S}\Bigr)^{2\eta_S(\mu_S,\mu)-\Omega}   \\
&\times 
\biggl[L_q(\qJ,\qB,Q^2)\tilde b_q\Bigl(\partial_\Omega - \ln\frac{\mu_B^2 Q_R}{s_B\mu_S},\qJ,\qB,\taun,\mu_B\Bigr) 
     + L_{\bar q}(\qJ,\qB,Q^2)\tilde b_{\bar q}\Bigl(\partial_\Omega - \ln\frac{\mu_B^2 Q_R}{s_B \mu_S },\qJ,\qB,\taun,\mu_B\Bigr)
\biggr]   \\
&\times 
\tilde j\Bigl(\partial_\Omega - \ln\frac{\mu_J^2 Q_R}{s_J\mu_S },\mu_J\Bigr) 
\tilde s \bigl(\partial_\Omega,\mu_S\bigr) 
\biggl(\frac{Q_R\abs{\taun - \tau_q}}{\mu_S e^{\gamma_E}}\biggr)^\Omega 
\frac{e^{\mathcal{K}(\mu_H,\mu_J,\mu_B,\mu_S,\mu)}}{\Gamma(1+\Omega)} 
\,,
\end{split}
\ee
where the operator $\tilde b_{q}$ is given by
\be
\label{appx:genbeamoperator}
\begin{split}
\tilde b_{q}(\partial_\Omega,\qJ,\qB,\taun,\mu_B) &=  \theta(\taun - \tau_q) \tilde b_q^\m(\partial_\Omega,x,\mu_B) + \frac{\as(\mu_B)}{2\pi} \int_x^1 \frac{dz}{z} \biggl[ C_F P_{qq}(z)f_q\Bigl(\frac{x}{z},\mu_B\Bigr) + T_F P_{qg}(z)f_g\Bigl(\frac{x}{z},\mu_B\Bigr)\biggr] \\
&\quad\times \biggl\{ \theta(\taun - \tau_q) \biggl[ \ln\biggl( \frac{1-X_q}{X_q} \frac{\taun - \tau_q}{\tau_q}\biggr) - H(-\Omega) - \frac{1}{\Omega}\biggr] - \theta(\tau_q -\taun)\frac{\pi}{\sin\pi\Omega} \\
&\qquad  + \frac{1}{\Omega}\biggl( \frac{\tau_q X_q}{\abs{\taun - \tau_q}}\biggr)^\Omega \hyp\biggl( - \Omega,-\Omega,1-\Omega; - \frac{\taun - \tau_q}{\tau_q X_q}\biggr)\biggr\}  \theta\Bigl( \taun - \tau_q(1-X_q)\Bigr) 
\,,
\end{split}
\ee
and similarly for $\tilde b_{\bar q}^\CM$. In \eqs{appx:resummedtau}{appx:genbeamoperator}, $\tau_q$ and $X_q$ are given by
\be
\tau_q \equiv \frac{\qperp^2}{Q_J\bn_J\mcdot q}=\frac{\qperp^2}{s_J}\,, \qquad X_q \equiv \frac{-\qJ\mcdot q (1-z)}{[z \qB - (1-z)\qJ]\cdot q}\,.
\ee

\section{$\cO(\as)$ fixed-order cross sections}
\label{app:NLO}

\subsection{$\tauCM$ cross section}

The fixed-order $\tauCM$ cross section at $\cO(\as)$ is easily obtained from \eq{appx:resummedtauCM} by taking the limit $\mu_{H ,J,B,S} = \mu$, which turns off all the resummation. We plug the $\cO(\as)$ hard function \eq{hardNLO}, the $\cO(\as)$ jet and soft operators given by \eq{foperator}, and the $\cO(\as)$ beam function operator \eq{appx:beamCMoperator} into the expression \eq{appx:resummedtauCM}. We use \eq{appx:Omegaderivatives} to evaluate the action of these operators in \eq{appx:resummedtauCM}, and finally take the $\cK,\Omega,\eta_{H,J,B,S}\to 0$ limit. The result is:
\be
\label{NLOtauCMcs}
\begin{split}
&\sigmac(x,Q^2,\tauCM) =  \theta(\tauCM- 1+y) \int_x^1\frac{dz}{z}\bigl[ L_q^\CM(Q^2)f_q(x/z,\mu) + L_{\bar q}^\CM(Q^2) f_{\bar q}(x/z,\mu)\bigr] \\
&\qquad \times   \Biggl\{ \delta (1-z) \biggl[ 1 - \frac{\as(\mu) C_F}{4\pi}\Bigl(9+\frac{2\pi^2}{3} + 3\ln[x(\tauCM -1 +y)^2]  + 4\ln [x(\tauCM - 1+y)]\ln(\tauCM-1+y) \Bigl) \biggr]\\
&\qquad \qquad + \frac{\as(\mu) C_F}{2\pi} \biggl[ P_{qq}(z) \ln\frac{xQ^2(\tauCM - 1+y)}{\mu^2} + F_q(z) \biggr] \Biggr\} \\
&\quad + \frac{\as(\mu) T_F}{2\pi} \bigl( L_q^\CM+ L_{\bar q}^\CM\bigr)(Q^2)   \theta(\tauCM- 1+y)   \int_x^1 \frac{dz}{z} f_g\Bigl(\frac{x}{z},\mu\Bigr)  \Bigl[ P_{qg}(z) \ln\frac{xQ^2(\tauCM - 1+y)}{\mu^2} + F_g(z)\Bigr] \\
&\quad + \frac{\as(\mu)}{2\pi} \int_x^1\frac{dz}{z} \Bigl\{ C_F P_{qq}(z) \bigl[L_q^\CM(Q^2) f_q(x/z,\mu) +  L_{\bar q}^\CM(Q^2) f_{\bar q}(x/z,\mu) \bigr]  + T_F P_{qg}(z) (L_q + L_{\bar q})(Q^2) f_g(x/z,\mu)\Bigr\} \\
&\qquad \qquad\quad  \times \biggl[ \theta(\tauCM - 1 + y) \ln\frac{z}{x+z-xz}  + \theta(1-y-\tauCM) \theta\Bigl( \tauCM - \frac{z(1-y)}{x+z-xz}\Bigr) \ln\frac{(1-y)X}{1-y-\tauCM} \biggr]\,.
\end{split}
\ee
In the last line we used that in the $\Omega\to 0$ limit, the hypergeometric function in \eq{appx:beamCMoperator} behaves like \cite{Huber:2005yg,Huber:2007dx}:
\be
\hyp (-\Omega,-\Omega,1-\Omega; -T) = 1 + \Omega^2 \Li_2(-T) + \cdots\,,
\ee
In the $\Omega\to 0$ limit in \eq{appx:beamCMoperator}, only the first term in this expansion survives.

\subsection{Generic $\taun$ cross section}

The fixed-order $\cO(\as)$ cross section is similarly obtained from \eq{appx:resummedtau} by taking the limit of equal scales $\mu=\mu_H=\mu_J=\mu_B=\mu_S$, and thus $\cK,\Omega,\eta_{H,J,B,S}\to 0$. 
For the cumulant to $\cO(\as)$, we obtain:

\be
\label{NLOtaucs}
\begin{split}
&\sigmac(x,Q^2,\taun) =  \theta(\taun- \tau_q) \int_x^1\frac{dz}{z}\bigl[ L_q(\qJ,\qB,Q^2)f_q(x/z,\mu) + L_{\bar q}(\qJ,\qB,Q^2) f_{\bar q}(x/z,\mu)\bigr] \\
&\qquad \times   \Biggl\{ \delta (1-z) \biggl[ 1 - \frac{\as(\mu) C_F}{4\pi}\Bigl(9+\frac{2\pi^2}{3} + 3\ln\frac{Q_R^2 (\taun-\tau_q)^2}{Q^2}  + 4\ln \frac{Q_R^2(\taun-\tau_q)}{s_B} \ln \frac{Q_R^2(\taun-\tau_q)}{s_J} \Bigl)\biggr]\\
&\qquad \qquad + \frac{\as(\mu) C_F}{2\pi} \biggl[ P_{qq}(z) \ln\frac{s_B(\taun-\tau_q)}{\mu^2} + F_q(z) \biggr] \Biggr\} \\
&\quad + \frac{\as(\mu) T_F}{2\pi} \bigl( L_q+ L_{\bar q}\bigr)(\qJ,\qB,Q^2)   \theta(\taun- \tau_q)   \int_x^1 \frac{dz}{z} f_g\Bigl(\frac{x}{z},\mu\Bigr)  \Bigl[ P_{qg}(z)\ln\frac{s_B (\taun-\tau_q)}{\mu^2} + F_g(z)\Bigr] \\
&\quad + \frac{\as(\mu)}{2\pi} \int_x^1\frac{dz}{z} \Bigl\{ C_F P_{qq}(z) \Bigl[L_q f_q\Bigl(\frac{x}{z},\mu\Bigr) +  L_{\bar q} f_{\bar q}\Bigl(\frac{x}{z},\mu\Bigr) \Bigr]  + T_F P_{qg}(z) (L_q + L_{\bar q}) f_g\Bigl(\frac{x}{z},\mu\Bigr)\Bigr\} \\
&\qquad \qquad\quad  \times \biggl[ \theta(\taun - \tau_q ) \ln(1-X_q)  + \theta(\tau_q-\taun) \theta\Bigl( \taun - \tau_q(1-X_q)\Bigr) \ln\frac{\tau_q X_q}{\tau_q -\taun} \biggr]\,,
\end{split}
\ee
where we have used the relation in \eq{ssQR2},  $s_J s_B/Q_R^2 = Q^2$ to leading order in $\lambda$, to simplify the arguments of the logs on the second line.

\end{widetext}

\bibliography{DIS}

\end{document}